\documentclass[11pt, a4paper]{article}
\usepackage{url}
\usepackage{a4wide}
\usepackage{graphicx}
\usepackage{natbib}
\usepackage[shortlabels]{enumitem}
\usepackage{hyperref}
\usepackage{setspace}
\usepackage{multirow, booktabs}
\usepackage{amsmath, amsthm, amssymb, amsfonts}
\usepackage{bm, bbm}
\usepackage{longtable}
\usepackage{color}
\usepackage{mathtools}
\usepackage{appendix}
\usepackage{pdflscape}
\usepackage{ulem}

\DeclareMathOperator*{\argmax}{arg\,max}

\usepackage[ruled,vlined]{algorithm2e}

\allowdisplaybreaks

\DeclareMathAlphabet\mathbfcal{OMS}{cmsy}{b}{n}

\newcommand{\mc}{\mathcal}
\newcommand{\wh}{\widehat}

\title{\bf MOPED: A moving sum method for change point detection in pairwise extremal dependence}
\author{Euan T.~McGonigle$^1$, Matthew Pawley$^2$,\\ Jordan Richards$^{3*}$ and Christian Rohrbeck$^4$}

\begin{document}

\maketitle

\footnotetext[1]{School of Mathematical Sciences, University of Southampton, UK.}

\footnotetext[2]{ATASS Sports, Exeter, UK.}

\footnotetext[3]{School of Mathematics and Maxwell Institute for Mathematical Sciences, University of Edinburgh, UK. $\;^*$Corresponding author: jordan.richards@ed.ac.uk.} 

\footnotetext[4]{Department of Mathematical Sciences, University of Bath, UK.}

\bigskip
\begin{abstract}
It is increasingly the case with modern time series that many data sets of practical interest contain abrupt changes in structure. These changes may occur in complex characteristics such as the extremal dependence structure, and identifying such structural breaks remains a challenging problem. Many existing change point detection algorithms focus on changes in dependence across the entire distribution, rather than the tails, and approaches that are tailored to extremes typically make strict parametric assumptions or they are only applicable to bivariate data. We propose a nonparametric MOving sum-based approach for detecting multiple changes in the Pairwise Extremal Dependence (MOPED) of multivariate regularly varying data. To avoid the classical problem of threshold selection in the study of multivariate extremes, we further propose a multiscale, multi-threshold variant of MOPED that pools change point estimates across choices of the threshold and the bandwidth used in local estimation. Good performance of MOPED is illustrated in a simulation study, and we showcase its ability to identify subtle changes in tail dependence class in the absence of correlation changes. We further demonstrate the usefulness of MOPED by identifying changes in the extremal connectivity of electroencephalogram (EEG) signals of seizure-prone neonates.
\end{abstract}

\noindent%
{\it Keywords: Electroencephalograms; Extreme value theory; Multivariate regular variation;  Seizures; Structural breaks } 
\vfill

\section{Introduction}
\label{sec:Introduction}

In many applications, interest lies in analysing the tail behaviour of a random vector \citep{Coles1999, cooley2019decompositions, simpson2020determining,Rohrbeck2023, murphy2024deep, de2025generative}. This has given rise to a wide range of methods in multivariate extreme value theory; we refer the reader to \citet{Engelke2021} for a review article on classical methods and a summary of recent developments. Due to the scarcity of extreme events, one assumption commonly made in practice is that of stationarity in the tail dependence, with any change in the distribution of extremes being captured by a non-stationary model for the marginal distributions \citep[see, e.g.,][]{chavez2005generalized,youngman2019generalized,Richards2024}. For instance, in environmental studies, such a framework allows extreme events at spatial sites to become more severe and frequent over time, while the overall spatial structure of extreme events is assumed to remain unchanged. However, it is increasingly commonplace for data to exhibit changing tail dependence structure: \cite{Poon2003} identified complex non-stationarity in the tail dependence of stock market indices; flood events in the UK tend to be more widespread in winter than in summer \citep{Rohrbeck2021}; \cite{zhong2022modeling} modelled temporal trends in the spatial extent of European heatwaves; \cite{Kakampakou2024} found that the spatial dependence in sea surface temperature extremes across the Red Sea has changed over the past decades; \cite{murphy2024modelling} identified increasing trends in the strength of extremal dependence between UK temperatures and dryness.

There has been a growing number of methods in extreme value theory which aim to detect non-stationarity in the tail dependence of data. An exploratory approach is to estimate and compare the coefficient of tail dependence $\chi$ \citep{Coles1999} for non-overlapping time windows \citep[see, e.g.,][]{Poon2003, Kakampakou2024}. Formally, for random variables $X$ and $Y$ with cumulative distribution functions $F_X$ and $F_Y$ respectively,
\begin{equation}
\chi = \lim_{u\to 1}~\Pr\left\{ F_X(X)>u \mid F_Y(Y)>u \right\}. 
\label{eq:chi}
\end{equation}
We say that $X$ and $Y$ are asymptotically independent when $\chi=0$, and asymptotically dependent when $\chi>0$. Beyond this diagnostic tool, several statistical tests for detecting structural changes in extremal dependence are available. \cite{Buecher2015} introduced a test for a change in tail dependence of a sequence of independent bivariate random vectors, and this framework was generalized to $\beta$-mixing time series by \cite{Hoga2018}. \cite{hazra2025estimating} used a parametric extremal dependence model and likelihood ratio tests to identify a single change point in bivariate extremal dependence, whilst \cite{Hoga2022} introduce a modelling framework which allows for a transition between asymptotic dependence and asymptotic independence. \cite{drees2023} introduced another testing procedure for a structural change in the tail dependence which is, however, limited to lower dimensions due to its computational burden. An alternative has been by proposed by \cite{Pawley2025} who focus on testing for changes in summaries of pairwise extremal dependence, which decreases the computational burden and permits application in higher dimensions at the cost of reduced power in certain cases. Finally, we note that changes in tail dependence may also refer to a shift in temporal dependence. For instance, \cite{Dupuis2019} considered testing for a change point in serial dependence of a single time series of extremes. 

Whilst the above methods directly consider nonstationary tail dependence, another class of methods -- so-called nonparametric change point detection approaches -- may also be used. These approaches can discriminate between general distributional changes and can be applied to the problem of detecting change points, or structural breaks, in extremal dependence. For example, there are approaches based on density functions \citep{padilla2022optimal2} and kernel transforms of the data \citep{ celisse2018new, arlot2019kernel}. Another common technique relies on energy-based notions of distance \citep{matteson2014, mcgonigle2025nonparametric}, whilst graph-based methods applicable to non-Euclidean data may also be employed \citep{chen2015graph}. However, nonparametric methods are less informative in practical situations, where it is often the aim to study a particular type of change point present in the data based on the application of interest.

Working in a framework similar to \cite{Buecher2015} and \cite{drees2023}, we consider the problem of detecting abrupt changes in the tail dependence structure amongst a subset, or all, of the variables of a serially independent multivariate time series. In this work, we constrain our focus to identifying structural changes in only the tail dependence, and hereafter assume that the marginal distributions are stationary. For identifying structural changes in marginal tail behaviour, we instead direct the reader to, for example, \cite{dierckx2010change,dupuis2015detecting,hoga2017testing,kojadinovic2017detecting,e2020change,castillo2022distribution}, and \cite{girard:hal-05044135}. The key contribution of this paper is to introduce a computationally efficient algorithm which identifies structural changes in tail dependence. We summarise tail dependence in the vector over a time period using the tail pairwise dependence matrix (TPDM) of \cite{cooley2019decompositions} which makes the assumption of asymptotic dependence between variables. The TPDM acts analogously to the covariance matrix in classical Gaussian statistics, which has made it an increasingly popular tool in the study of multivariate extremes; examples of its usage include clustering \citep{elsom2024extreme, richards2025modern}, causal inference \citep{jiang2025separation}, asymmetric dependence modelling \citep{jiang2024efficient}, graph learning \citep{lee2022partial,gong2024partial}, principal component analysis \citep{jiang2020principal}, linear prediction \citep{lee2021transformed}, and time series modelling \citep{mhatre2024transformed}.

We propose a moving sum \citep[MOSUM; see, e.g.,][]{eichinger2018} procedure for multiple change point detection in the extremal dependence of a multivariate time series, using a detector statistic that is carefully devised to detect changes in the TPDM; we refer to this method as MOPED (MOving sum method for changes in Pairwise Extremal Dependence). We showcase the efficacy of MOPED, relative to an existing state-of-the-art nonparametric change point detection algorithm, via a simulation study and real data application. In our application to multivariate EEG signals from seizure-prone neonatal subjects, we find that MOPED identifies significant structural changes in the extremal dependence of the signals when the subjects undergo seizures, and these changes typically occur close to the point of seizure.

The remainder of the paper is organised as follows. In Section~\ref{sec:method}, we provide background on multivariate regular variation and the tail pairwise dependence matrix, which we then use to define our extremal dependence change point algorithm, MOPED. In Section~\ref{sec:numerical}, we provide a discussion of the numerical implementation of our change point algorithm and a simulation study highlighting its efficacy and empirical properties. In Section~\ref{sec:application},  we apply our algorithm to identify change points and structural changes in the extremal dependence of multivariate EEG signals for seizure-prone neonates. We provide concluding remarks and a discussion of extensions in Section~\ref{sec:discussion}. Accompanying software implementing
the method is available as the R package \verb!moped! at \url{https://github.com/EuanMcGonigle/moped}.

\section{Methodology}

In this section, we introduce the necessary background concepts, define the multiple change point model for the TPDM, and discuss our proposed change point detection methodology.

\label{sec:method}
\subsection{Multivariate regular variation}
    A $d$-dimensional random vector $\mathbf{X}\in \mathbbm{R}^d_+$ is regularly varying with tail index $\alpha>0$, denoted by $\mathbf{X}\in {\rm RV}^d_+(\alpha)$, if a sequence $b_n\to \infty$ exists such that $n\Pr(b_n^{-1}\mathbf{X} \in\cdot)\xrightarrow{v} v(\cdot)$ as $n\to\infty$, where $v(\cdot)$ is a Radon measure on the cone $\mathbbm{E}^d_+:=[0,\infty]^d\setminus{\{\mathbf{0}\}}$ and $\xrightarrow{v}$ denotes vague convergence; see, e.g., \cite{resnick2007heavy}. The limit measure $v(\cdot)$ satisfies $(-\alpha)$-homogeneity, $v(r\mathcal{B})=r^{-\alpha}v(\mathcal{B})$ for any $r>0$ and any Borel subset $\mathcal{B}\subset \mathbbm{E}^d_+$. Define by $H(\cdot)$ the angular mass measure on the positive part of the unit $(d-1)$-sphere $\mathcal{S}^{d-1}_+:=\{\mathbf{x}\in \mathbbm{R}_+^d: ||\mathbf{x}||_2=1\},$ where $||\cdot||_2$ denotes the $L_2$-norm. Then, we can decompose $v(\cdot)$ into a product measure: for $r>0$, we have $${v(\{\mathbf{x}\in \mathbbm{E}^d_+: ||\mathbf{x}||_2\geq r, \mathbf{x}/||\mathbf{x}||_2\in \mathcal{B}_H \})}=r^{-\alpha}H(\mathcal{B}_H),$$ where $\mathcal{B}_H \subset\mathcal{S}^{d-1}_+ $ is a Borel subset.  Note that $H(\cdot)$ is not a valid probability measure, but can be normalised by dividing by its total mass,  $H(\mathcal{S}^{d-1}_+)$, to give a valid probability measure, denoted by $N(\cdot):=H(\cdot)/H(\mathcal{S}^{d-1}_+)$. In the case where $\alpha = 2$ and $\mathbf{X}$ has standardised margins such that $X_i \in \mbox{RV}^1_+(\alpha),i=1,\dots,d,$ then we have $H(\mathcal{S}^{d-1}_+) = d$; hereafter we assume that this holds and note that, in practice, this is without loss of generality \citep[with respect to marginal transformations; see, e.g.,][for discussion]{jiang2024efficient}. A marginal distribution that satisfies $X\in \mbox{RV}^1_+(\alpha)$ is the type 1 Pareto distribution with unit scale and shape two, and distribution function $F(x)= 1-x^{-2},x \geq 1$; we hereafter denote this as Pareto(2).

\subsection{Tail pairwise dependence matrix}

The tail pairwise dependence matrix \citep[TPDM;][]{cooley2019decompositions} is a matrix of pairwise summaries of the extremal dependence in $\mathbf{X}=(X_1,\dots,X_d) \in \mbox{RV}^d_+(\alpha)$. The TPDM is denoted by $\Sigma:=(\sigma_{ij})_{i,j=1,\ldots,d}$, with pairwise entries
\begin{equation}
    \label{eq:edm}
    \sigma_{ij}:=\int_{\mathcal{S}^{d-1}_+ }\theta_i\theta_j{\rm d}H(\bm{\theta})=d\int_{\mathcal{S}^{d-1}_+ }\theta_i\theta_j{\rm d}N(\bm{\theta})=d\lim_{r\to \infty} \mathbb{E}\left[\frac{X_i X_j}{R^2}\mid R>r\right],
\end{equation}
with $R=||\mathbf{X}||_2$. The strength of asymptotic dependence between $X_i$ and $X_j$ increases with $\sigma_{ij}$, with $\sigma_{ij}=0$ if and only if $X_i$ and $X_j$ are asymptotically independent. When $d=2$, the pairwise $\sigma_{ij}$ is often referred to as the Extremal Dependence Measure \citep[EDM;][]{resnick2004extremal, larsson2012extremal}. 

We can estimate $\sigma_{ij}$ empirically using $n$ samples $\left\{ \mathbf{X}_t = \left( X_{1,t} , \ldots , X_{d,t}\right)^\mathsf{T} \right\}^n_{t=1}$. Assuming that the limit in~\eqref{eq:edm} holds exactly for sufficiently large $R$, the empirical estimator of $\sigma_{ij}$ is 
\begin{equation}
    \label{eq:estimationSigma}
    \wh{\sigma}_{ij}:=\frac{d}{k}\sum_{t=1}^{n}\frac{X_{i,t}X _{j,t}}{R^2_t}\mathbbm{1}(R_t>r_0),
\end{equation}
where $r_0>0$ is a suitably-chosen high radial threshold, $R_t:=||\mathbf{X}_t||_2$, and $k=\sum_{t=1}^n \mathbbm{1}(R_t>r_0)$ is the number of radial threshold exceedances, where $\mathbbm{1}(\cdot)$ is the indicator function. Choice of the exceedance threshold $r_0$ is an ubiquitous problem in extreme value analysis and involves the classical bias-variance trade-off: larger choices of $r_0$ reduce bias in estimation of $\sigma_{ij}$ at the cost of reduced threshold exceedances and higher estimation uncertainty. Following, e.g., \cite{cooley2019decompositions}, we take $r_0$ to be the $k$-th largest order statistic of $\{R_t\}_{t=1}^n$. Similarly to the bandwidth in moving sum-based change point detection methods (to be discussed in Section~\ref{sec:mosum}), the order $k$ is a hyper-parameter that must be specified. In Section~\ref{sec:multiscale}, we discuss a technique for pooling over a sequence of $k$ values, in order to improve our change point detection algorithm and to reduce sensitivity to the selected threshold.

\subsection{Change point model}

We consider a model where the tail dependence in a multivariate time series $\left\{\mathbf{X}_t:t \geq 1\right\}$ undergoes abrupt changes. In particular, we consider the setting
\begin{equation}
\mathbf{X}_t = \sum_{l=1}^{q+1} \mathbf{X}^{(l)}_t \mathbb{I} ( \tau_{l-1} < t \leq \tau_{l} )  , \qquad  1 \leq t \leq n,
\end{equation}
where each of the sequences $\left\{ \mathbf{X}_t^{(l)} = \left( X_{1,t}^{(l)} , \ldots , X_{d,t}^{(l)} \right)^\mathsf{T} : t \geq 1 \right\}, l = 1,\dots,q+1,$ are independent, identically distributed random vectors in $\mbox{RV}^d_+(2)$ (with standardised $\mbox{RV}^1_+(2)$ margins and) with corresponding angular measure $H^{(l)}(\cdot)$, such that consecutive sequences differ in their extremal dependence measure characterised by their TPDM $\Sigma^{(l)} := ( \sigma^{(l)}_{ij} )_{i,j = 1, \ldots , d}$:
\begin{equation}
    \label{eq:edm-cpt}
    \sigma_{ij}^{(l)} := \int_{\mathcal{S}^{d-1}_+ } \theta_i \theta_j {\rm d}H^{(l)}(\bm{\theta})=d\lim_{r\to \infty} \mathbbm{E}\left[\frac{X_{i,1}^{(l)} X^{(l)}_{j,1}}{R^{(l)^2}}\mid R^{(l)} > r \right],
\end{equation}
where $R^{(l)} := || \mathbf{X}_1^{(l)} ||_2$. That is, we consider a $d$-variate time series model containing $q$ change points (structural breaks), with the TPDM of the observed time series, $\{\mathbf{X}_t: t \geq 1\}$, differing across the change points. The change points are collected in $\mathcal C := \{\tau_l \}_{l=0}^{q+1}$ with the convention that $\tau_0 := 0$ and $\tau_{q+1} := n$, and neighbouring segments differ in their TPDM such that $\Sigma^{(l)} \neq \Sigma^{(l-1)}$ for all $l=2,\dots,q+1$.

\subsection{Moving sum-based method for detecting changes in the TPDM}\label{sec:mosum}

We here define a moving sum (MOSUM) detector statistic for estimating multiple change points in the TPDM of the multivariate time series $\{\mathbf{X}_t\}^n_{t=1}$; we refer to this method as MOPED (MOving sum method for changes in Pairwise Extremal Dependence). For a fixed bandwidth $G \in \mathbb{N}$, the method scans the data using a detector statistic, $T (G, t)$, computed on neighbouring moving windows of length $G$ located either side of a a candidate change point location $t$. The detector statistics measures the discrepancy between local estimates of the TDPM of the corresponding windows measured via their matrix norm. In particular, denoting the Frobenius norm by $|| \cdot ||_F$, we use the detector statistic
\begin{equation}\label{eq:test-stat}
T (G, t)=  || D(G, t) ||_F, \qquad G \leq t \leq n - G, 
\end{equation}
where $D(G,t)$ is the $d \times d$ matrix whose $(i,j)$-th entry is given by
\begin{equation}\label{eq:mosum}
D (G, t)_{i,j} =  \frac{d}{k} \sum_{s= t- G+1}^{t}  \frac{X_{i,s} X_{j,s}}{R_s^2}  \mathbb{I} (R_s > r^{(-)}_{0,t})   - \frac{d}{k} \sum_{s= t+1}^{t+G} \frac{X_{i,s} X_{j,s}}{R_{s}^2} \mathbb{I} (R_s > r^{(+)}_{0,t}) , 
\end{equation}
 for $t \in [G,n-G]$, and where $r_{0,t}^{(-)}$ and $r_{0,t}^{(+)}$ are radial exceedance thresholds for the left and right $G$-windows, respectively, of the candidate change point $t$, such that
$$ \sum_{s= t- G+1}^{t} \mathbbm{1}(R_s>r_{0,t}^{(-)})=k=\sum_{s= t+1}^{t+G} \mathbbm{1}(R_s>r_{0,t}^{(+)}).$$
The entry $D(G,t)_{i,j}$ measures the local difference, in a $G$-window to the left and right of the candidate change point location $t$, in the pairwise tail dependence between the $i$-th and $j$-th components of the random vector $\mathbf{X}_t$. Then, $T(G,t)$ is the aggregated measure of discrepancy across all components of $\mathbf{X}_t$. Note that, asymptotically, the quantity $\sigma_{i,j}$ for $i = j$ is equal to one. Therefore, in practice, we set $D(G,t)_{ii} = 0$ for $1 \leq i \leq d$ when computing $T(G,t)$.

When the data in the left and right $G$-windows, i.e., $\{\mathbf{X}_s : |t - s | \leq G \},$ does not undergo a change point, the local estimates of the TDPM to the left and right of $t$ are expected to be similar and thus the detector statistic $T(G, t)$ is expected to be close to zero. On the other hand, if $|t - \tau_j|<G$ where $\tau_j$ is a true change point, then $T(G, t)$ is expected to increase and then decrease around $\tau_j$ with a local maximum at the change point $t = \tau_j$.

\begin{figure}[h!]
    \centering
    \includegraphics[width = 0.8\linewidth]{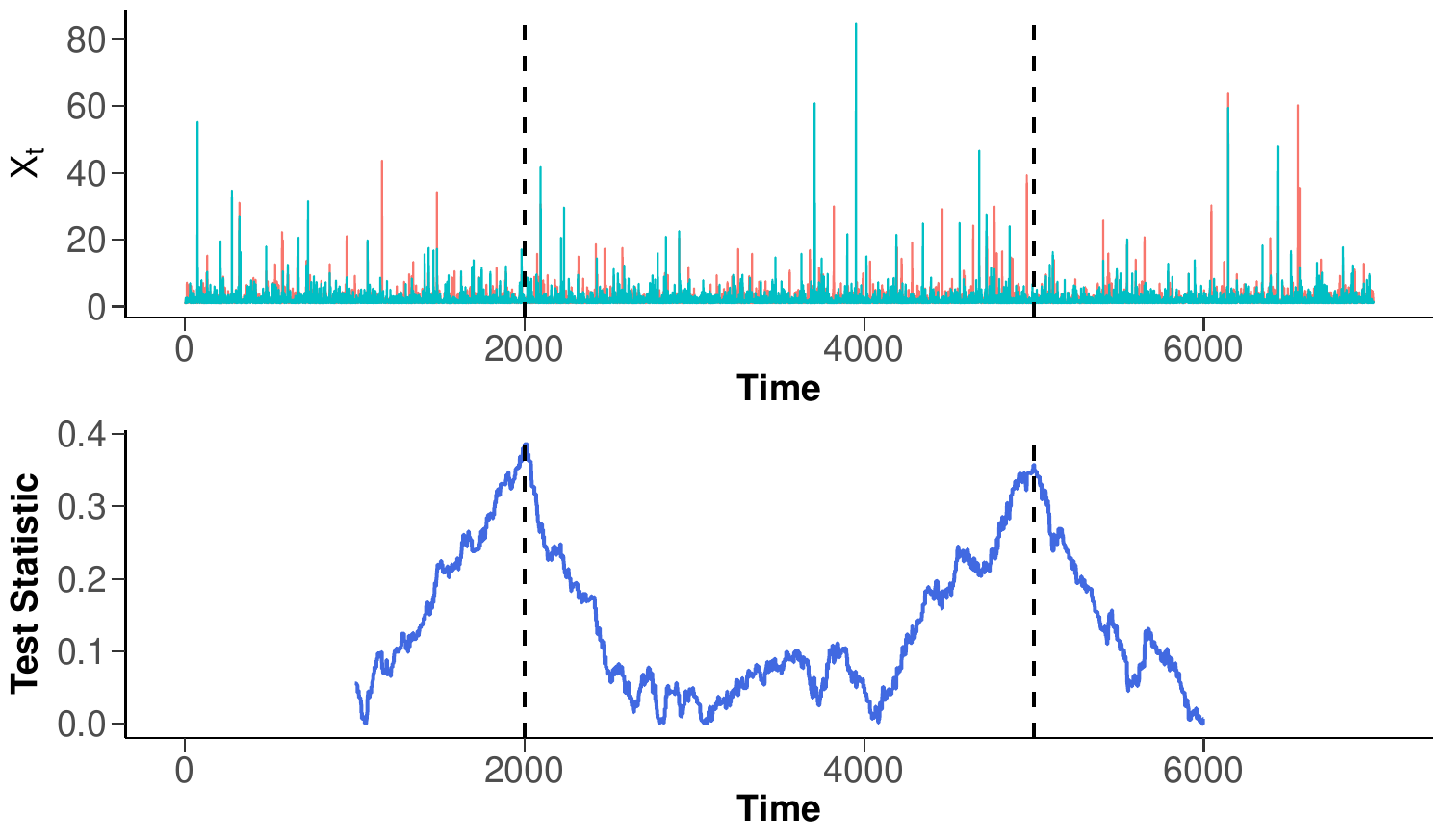}   
    \caption{Top: bivariate time series of length $n = 7000$ with change points $\tau_1 = 2000$ and $\tau_2 = 5000$ (vertical dashed lines). Bottom: corresponding detector statistic $T(G,t)$.}
    \label{fig:example-est-stat}
\end{figure}

We illustrate this behaviour using the following example. A bivariate time series of length $n = 7000$ is generated with two change points in the tail dependence at locations $\tau_1 = 2000$ and $\tau_2 = 5000$ (for more details on the data generating process, see Scenario 1 in Section~\ref{sec:numerical}). The top panel of Figure~\ref{fig:example-est-stat} shows a realisation of this process (on Pareto(2) margins) whilst the bottom panel gives the detector statistic $T(G,t)$, $G \leq t \leq n - G$, computed with $G = 1000$ and $k = G/10$. The detector statistic forms prominent peaks around the two change points, whilst it is otherwise generally flat.

Based on these observations, we detect and locate the change points in the TPDM by using local maximisers of the detector series $\{T(G, t)\}_{t=G}^{n-G}$ that are significantly large. We adopt the so-called $\eta$-criterion, first considered by \cite{eichinger2018}, for simultaneous estimation of multiple change points. For some fixed constant $\eta \in (0, 1)$ and a threshold $C > 0$ (to be specified in Section~\ref{sec:threshold}), we identify any local maximiser(s) of $T(G, t)$, denoted $\widehat\tau$, which satisfies
\begin{align}\label{eq:mosum:est} 
T( G,\widehat \tau ) > C \quad \text{and} \quad
\widehat \tau = \underset{\{t: \, \vert t - \widehat \tau \vert \leq \eta G\}}{\argmax} T(G, t).
\end{align}
That is, $\widehat{\tau}$ is declared a change point if it is a local maximiser of $T (G, t)$ over an $\eta G$-neighbourhood, at which the threshold $C$ is exceeded. In practice, local maximisers satisfying~\eqref{eq:mosum:est} may not be unique, in which case we use the mid-point of those such local maximisers instead. We denote the set of such estimators fulfilling~\eqref{eq:mosum:est} by $\widehat{\mathcal{C}}=\{\wh{\tau}_1,\dots\}$ with total number of estimated change points $\widehat q := \vert \widehat{\mathcal{C}} \vert$.

\subsection{Threshold selection via permutation testing}\label{sec:threshold}

The choice of threshold $C$ influences the finite sample performance of the method, which will depend on unknown quantities involved in the underlying extremal dependence structure of the time series. Resampling is a widely-used method for the calibration of change point detection algorithms, including threshold selection \citep{matteson2014, bucher2016dependent}. Under the assumption of independence of the observations, we utilise a permutation testing procedure in order to approximate the quantiles of $\max_{G \leq t \leq n - G} \{ T(G,t) \}$ in the absence of any change point, from which we can then select the threshold $C$ in a data-driven manner.

We perform $M$ permutations of the timesseries. For the $m$-th permutation, we first randomly permute the times series $\{\mathbf{X}_t \}_{t=1}^n$ to form a new set of observations $\{\mathbf{X}^{[m]}_t \}_{t=1}^n$. Using these permuted observations, the MOPED detector statistics, $\{ T^{[m]} (G, t) \}_{t=G}^{n-G},$ are obtained using Equation~\eqref{eq:test-stat}. Finally, we record the maximum value of the MOPED detector statistics as $T^{[m]} := \max_{G \leq t \leq n-G} \{ T^{[m]}(G,t)\}$.

The permutation test would be exact if all $n!$ possible permutations are computed. Since this is not computationally feasible, we instead perform an approximate test by performing the above procedure using $M\ll n!$ random permutations. From the $M$ permutation test statistic values $\{T^{[m]} \}_{m=1}^M$, we set the threshold $C=Q_{1-\alpha} ( \{ T^{[m]}  \}_{m=1}^{M})$ as the empirical $(1- \alpha)$-th quantile of $\{T^{[m]} \}_{m=1}^M$ for the chosen approximate significance level $\alpha \in (0,1)$. Having selected the set of change points $\wh{\mc C}$ using the rule in Equation~\eqref{eq:mosum:est}, we can compute an approximate $p$-value for each $\wh{\tau} \in \wh{\mathcal{C}}$ as
\begin{equation}\label{eq:p-val}
p (\wh{\tau} ) := \frac{1 + \left| \{ 1 \leq m \leq M : T^{[m]} \geq T(G, \wh{\tau} ) \}  \right| }{M + 1}.    
\end{equation}
Combining the procedures outlined in Sections~\ref{sec:mosum} and~\ref{sec:threshold}, we have the following full algorithmic description of MOPED in Algorithm~\ref{alg1}.
\normalem
\begin{algorithm}[t!]
\DontPrintSemicolon
 \KwIn{Multivariate time series $\{\mathbf{X}_{t} \}_{t=1}^n$, bandwidth $G$, order $k$, selection parameter $\eta$, permutation parameters $\alpha$ and $M$ \;}
 \BlankLine

    \For{$t \in \{G,\ldots , n-G\}$}{
    \BlankLine
       Compute MOSUM statistic $T(G,t)$ according to Equation~\eqref{eq:test-stat} \;
    }
      \For{$m \in \{1, \ldots , M \}$}{
       Compute permutation statistic $T^{[m]}$ \;
     }
    Set threshold $C\leftarrow Q_{1-\alpha} ( \{ T^{[m]}  \}_{m=1}^{M}) $ \;
                $\widehat{\mc C} \leftarrow$ Set of change point estimators obtained with bandwidth $G$ and threshold $C$ according to~\eqref{eq:mosum:est} \;
      \For{$\widehat \tau \in  \widehat{ \mc C}$}{
      \BlankLine
       Compute $p$-values $p(\widehat{\tau} ) $ according to Equation~\eqref{eq:p-val}  \;
     }
 \KwOut{Estimated change locations $\widehat{\mc C}$, approximate $p$-values $\{ p ( \widehat{\tau} ) : \ \widehat{\tau} \in \widehat{\mc C} \}$}
 \caption{MOPED algorithm}
 \label{alg1}
\end{algorithm}

\subsection{Multiscale, multiple threshold change point detection}
\label{sec:multiscale}

In moving window-based change point detection methods, one of the key considerations is the choice of detecting bandwidth $G$, whilst in extreme value statistics, a crucial aspect is the choice of order $k$ used in the extremal threshold. In this section, we extend the single-scale, single extremal threshold MOPED algorithm described in Section~\ref{sec:mosum} to incorporate multiple bandwidths and multiple extremal thresholds.  

Inspired by methods proposed by \cite{messer2014} and \cite{mcgonigle2025nonparametric}, we first describe an approach that allows us to perform the MOPED procedure for a fixed $G$ but multiple values of $k$, and combine the resulting change point estimates into a single set. We then describe a similar method to combine these results across multiple values of $G$.

First, we denote a set of $J > 1$ ranks $\mc K := \{k_j, \ 1 \leq j \leq J : k_1 <  \ldots < k_J \}$. We run MOPED for each $k \in \mc{K}$, denoting the corresponding change point estimates as $\widehat{\mathcal{C}} (k)$, and then combine the change point estimates across different values of $k$ using the `bottom-up' merging method proposed by \cite{messer2014}. We first add all estimates in $\wh {\mc C} (k_1)$, for $k_1$ the smallest rank, to the set of final estimates $\wh {\mc C}$. Then, sequentially for $j = 2, \ldots , J$, we accept $\wh{\tau} \in \wh {\mc C} (k_j)$ as a final change point estimate if and only if $\min_{t \in \wh {\mc C}} | t - \wh{\tau} | \geq \eta G$ with $\eta$ as before. That is, we only accept the estimates that are not close to previously detected change point estimates using a smaller rank/larger extremal threshold; we expect that close change point estimates will correspond to the same true change point, and so enforce that the final change point estimates $\wh {\mc C}$ are sufficiently far apart. The algorithmic description of the multiple threshold MOPED procedure is summarised in Algorithm~\ref{alg2} below.

\begin{algorithm}[h]
\DontPrintSemicolon
\SetKwData{edMOSUM}{MOPED}
\SetKwData{sort}{SORT}
 \KwIn{Multivariate time series $\{X_{t} \}_{t=1}^n$, bandwidth $G$, set of ranks $\mc K$, selection parameter $\eta$, permutation parameters $\alpha$ and $M$ \;}
 \BlankLine
Initialise $\wh {\mc C} \leftarrow {\mc C} (\mc K) \leftarrow  \emptyset$ \;
    \For{$k \in \mc K$}{
    \BlankLine
       $ \{ \wh{\mc C} (k) , \mc P (k) \}  \leftarrow \edMOSUM(\{X_{t} \}_{t=1}^n, G, k, \eta , \alpha, M  ) $
        \BlankLine
    $\mc{C} (\mc K) \leftarrow \mc{C} (\mc K) \cup \wh{\mc {C}} (k)$ \;
    }
    \BlankLine
\For{$\wh{\tau} \in \mc C (\mc K)$ \underline{in increasing order with respect to} $k$}{
    \lIf{\textup{$\min_{{t} \in \wh{\mc C} }| {t}- \wh{\tau}  | \geq \eta G$}}{Add $\wh{\tau}$ to $\wh{\mc C}$}
}
 \KwOut{Estimated change locations $\wh{\mc C}$, estimated number of changes $\wh{q} = \vert \wh{\mc C} \vert$}
 \caption{Multi-threshold MOPED algorithm}
 \label{alg2}
\end{algorithm}

The above approach outlines how to combine MOPED estimates with multiple threshold orders $k$ into a single set of change point estimates, for a single fixed bandwidth $G$. Appropriate choice of this $G$ is necessary to achieve good practical eprformance: a larger $G$ is suitable for detecting smaller changes over large time intervals, whilst a smaller $G$ can detect larger changes over shorter time periods. A multiple-bandwidth moving sum procedure can be particularly well-suited to so-called multiscale change points where both of the change point types described above occur within the same time series. Using multiple bandwidths, and then merging the results, can therefore improve the adaptivity of our proposed moving window-based procedure.

Here, we again combine the MOPED method with the bottom-up merging procedure, but now with respect to the different bandwidths, $G$. This is motivated by the observation that, with larger bandwidths, the MOPED detector statistic may be contaminated by several changes, resulting in the corresponding estimates being potentially unreliable. Therefore, it is reasonable to keep all change point estimates from the smallest bandwidth and, moving on to the next bandwidth in an iterative manner, only keep new estimates that cannot be accounted for by the previously accepted estimators. Denote by $\mc{G} = \{ G_h, 1 \leq h \leq H : G_1 < \ldots < G_H\}$ the set of $H>1$ bandwidths, and let $\widehat{\mathcal{C}}(G)$ denote the set of change point estimates detected using the multi-threshold MOPED procedure, as in Algorithm~\ref{alg2}, with single bandwidth $G$ and ranks $\mc K$. Then, as with the multiple threshold merging, we add all estimates in $\wh {\mc C} (G_1)$ returned with the first bandwidth $G_1$ to the set of final estimates $\wh {\mc C}$ and, sequentially for $h = 2, \ldots , H$, accept $\wh{\tau} \in \wh {\mc C} (G_h)$ as a final change point estimate if and only if $\min_{t \in \wh {\mc C}} | t - \wh{\tau} | \geq \eta G_h$ with $\eta$ as before. That is, we only accept the estimates that do not correspond with the change points which have previously been detected at a finer scale. The full multiscale, multi-threshold version of MOPED is presented in Algorithm~\ref{alg3} for completeness.

\begin{algorithm}[h]
\DontPrintSemicolon
\SetKwData{multithresh}{MULTI-THRESH-MOPED}
\SetKwData{sort}{SORT}
 \KwIn{Multivariate time series $\{X_{t} \}_{t=1}^n$, set of bandwidths $\mc G$, set of ranks $\mc K$, selection parameter $\eta$, permutation parameters $\alpha$ and $M$\;}
Initialise $\widehat{\mc C} \leftarrow \mc C ( {\mc G)} \leftarrow \emptyset$ 
\BlankLine
\For{$G \in \mc G$}{
    $\widehat{\mc C} (G) \leftarrow \multithresh(\{X_{t} \}_{t=1}^n, G, \mc K , \eta , \alpha, M)$
\BlankLine

\lFor{$\wh{\tau} \in \wh{\mc C}(G)$}{ Add $(\wh{\tau},G)$ to $\mc C (\mc G)$}
}
\BlankLine

\For{$\wh{\tau} \in \mc C (\mc G)$ \emph{in increasing order with respect to} $G$}{
    \lIf{\textup{$\min_{{t} \in \wh {\mc C}}| t- \wh{\tau}  | \geq \eta G$}}{Add $\wh{k}$ to $\wh{\mc C}$}
}
\BlankLine
 \KwOut{Estimated change point locations $\wh{\mc C}$, estimated number of changes $\wh{q} = \vert \wh {\mc C} \vert$}
 \caption{Multiscale, multi-threshold MOPED algorithm}
 \label{alg3}
\end{algorithm}

\section{Numerical results}
\label{sec:numerical}

In this section, we provide practical recommendations for the implementation of our approach, and conduct numerical studies to investigate the efficacy of the method.

\subsection{Practical implementation}

\textbf{Computational complexity.} Fast evaluation of the test statistic is paramount for reduced computational complexity of MOPED. Note that, for a single bandwidth $G$ and order $k$, a naive approach to computing the test statistic by recalculating the sums in Equation~\eqref{eq:test-stat} would yield a method with computational complexity $O(M n G)$. Owing to the MOSUM-based approach, we can instead sequentially update some of the quantities involved in the computation of Equation~\eqref{eq:test-stat}, reducing the computational cost to $O(M n \log (G))$ as follows.

The test statistic $D(G,t)_{i,j}$ in Equation~\eqref{eq:mosum} can be updated in a sequential fashion by first computing the time series of products $\{ X_{i,s} X_{j, s} /R_s \}_{s = 1}^n$, as well as the rolling sorted values $\{R^{(s)} \}_{s = t - G+1}^t$ of $\{ R_s \}_{s = t - G+1}^t$ for $G \leq t \leq n$. Then, since only the largest $k$ values of $\{R^{(s)} \}_{s = t - G+1}^t$ make a contribution in the first term in Equation~\eqref{eq:mosum} (and likewise for $\{R^{(s)} \}_{s = t+1}^{t+G}$ and the second term), the summation terms for computing $D(G, t+1)_{i,j}$ can be updated using the associated rolling sorted values, by keeping track of deletions and insertions that occur in updating the rolling sorts from time $t$ to time $t+1$, which can be achieved in $O(\log(G))$ complexity (see e.g., \cite{jula2022multiscale}). Therefore we require $O(n \log (G))$ total computations to compute the detector statistic, and hence $O(M n \log (G))$ when including the permutation testing.

\smallskip

\textbf{Bandwidth.} 
Due to the nonparametric nature of MOPED, in combination with the difficulty of detecting changes in the extremal dependence, it is advised to use larger bandwidths than that shown to work well for the MOSUM procedure for univariate mean change detection \citep{eichinger2018}. 
In practice, the practitioner may have prior knowledge that aids the choice of $G$.
In our simulation studies and data applications, we consider $G$ in multiples of 500. 
\smallskip

\textbf{Parameter for change point estimation and detection.} We set $\eta = 0.4$ in~\eqref{eq:mosum:est} following the recommendation in \cite{meier2021mosum}. For the permutation testing, we use $M = 200$ permutations with approximate significance levels of $\alpha=0.05$ or $\alpha = 0.1$.
\smallskip

\textbf{Extremal threshold.} The choice of threshold level $r_0$, and the corresponding order $k$ that defines the $k$-th order statistic, depend on the practitioner's interest and domain knowledge. This is a problem commonly faced both in extreme value theory \citep[see, e.g.,][]{scarrott2012review, murphy2025automated} and by general-purpose change point detection methods, such as the choice of the quantile level in \cite{jula2022multiscale} or the choice of lag in \cite{mcgonigle2025nonparametric}. In the absence of any information, we recommend setting the threshold as $k=0.05G$, such that $r_0$ corresponds to the $95$\% empirical quantile.

\smallskip

\textbf{Multiscale, multi-threshold MOPED.}
As discussed in Section~\ref{sec:multiscale},  the multiscale, multi-threshold variant of MOPED  pools change point estimates over values of the bandwidth $G$ and order $k$. We show in the following simulation study, Section~\ref{sec:sim}, that this approach can improve the accuracy of change point estimation relative to MOPED with fixed hyper-parameters ($G$ and $k$). However, the multiscale, multi-threshold version is less conservative than the ``fixed" version of MOPED, in the sense that it may produce spurious change point estimates. As observed by \cite{mcgonigle2023robust}, bottom-up merging has a propensity to produce false positives as it accepts all estimates from the finest bandwidth. Thus, we advocate use of the multiscale, multi-threshold procedure in applications where the accuracy of change point estimates is more important than testing for  structural breaks in the extremal dependence of multivariate time series (and vice versa for the fixed variant of MOPED).
\subsection{Simulation study}
\label{sec:sim}
 Our MOSUM-based extremal dependence change point algorithm is compared against two competing approaches: the state-of-the-art nonparametric method, E-divisive \citep{matteson2014}, and the parametric extreme value method of \cite{hazra2025estimating}. The latter identifies a single change point in bivariate data using a likelihood ratio test between two candidate H\"usler-Reiss max-stable distributions \citep{husler1989maxima}. 
 
 Two data generating scenarios are considered. In each, data are first generated on a uniform scale before the margins are transformed to one appropriate for the applied change point method: for MOPED, E-divisive, and the method of \cite{hazra2025estimating}, these are Pareto(2), standard Gaussian, and standard Gumbel, respectively. In Scenario 1, we generate data from a $d$-variate Student's $t$-copula with $\nu=3$ degrees of freedom and with correlation matrix, $\Omega$, varying across the change point; we switch between i) independence (i.e., $\Omega = I_{d\times d}$) and ii) all pairwise correlations equal to $\rho>0$ (i.e., where all off-diagonal elements of $\Omega$ are equal to $\rho$). Note that, in cases where we have $q=0$ change points, the second correlation matrix is used to generate the data, i.e., we consider non-zero pairwise $\rho$. 
 
 In Scenario 2, we switch between a $d$-variate Student's $t$-copula ($\nu=3$) and a Gaussian copula, both with positive-definite correlation matrix $\Omega$. The matrix $\Omega$ is generated randomly and kept equal across the change point. In this way, we design a data generating model with no change point(s) in the correlation structure, but where the theoretical value of the TPDM $\Sigma$ varies across the change point; note that the theoretical TPDM for a Gaussian copula is the identity matrix, whilst the TPDM for the Student's $t$-copula has non-zero diagonal entries (if the corresponding entry in $\Omega$ is strictly positive). In other words, the extremal dependence class of the data changes from asymptotic independence to asymptotic dependence; see discussion in Section~\ref{sec:Introduction}. Three random seeds are used for generating $\Omega$ and results are reported for each.
 

 We consider sample size $n=5000$, and two values of $\rho$: a weak and strong dependence case with $\rho=0.2$ and $\rho=0.6$, respectively. We vary dimension $d\in\{2,8,15\}$ between scenarios, and consider $q\in \{0,1,2\}$  equally-spaced change points. For MOPED, we consider two versions: a ``fixed'' version of MOPED (see Algorithm~\ref{alg1}), where the tuning parameters are $G=1500$ and $k = 0.1G$, and the multiscale, multi-threshold MOPED (hereafter referred to as MMMOPED; see Algorithm~\ref{alg3}). For MMMOPED, we merge over the sequence of bandwidths $\mathcal{H}=\{500,1000,1500\}$ and ranks $\mathcal{K}=\{0.2G, 0.1G, 0.05G\}$. 
 
 Note that the method of \cite{hazra2025estimating} is only applicable for dimension $d=2$. Moreover, it cannot test for the significance of a change point, and rather it returns exactly the most likely change point (according to a parametric likelihood ratio test). Thus, we only report the results of the \cite{hazra2025estimating} method when $d=2$ and $q=1$.
 
 Tables~\ref{tab:scenario1} and~\ref{tab:scenario2} provide the results of the simulation study when $d=2$ for Scenarios 1 and~2 respectively; similar tables for $d=8$ and $d=15$ are provided in Appendix~\ref{Sup:tables}. Across all tables, we report the distribution of $\wh{q}-q$, where $\wh{q}$ is the estimated number of change points and $q$ is the truth, alongside two measures of accuracy for the change point estimates (when $q > 0$) advocated by \cite{van2020evaluation}: the average covering metric \citep[CM;][]{arbelaez2010contour} and V-measure \citep[VM;][]{rosenberg2007v}. Define the partition $\mathcal{P}=\{\mathcal{A}_l\}_{l=1}^{q+1}$ of $\{1,\dots,n\}$ provided by the true change point locations $\{\tau_l\}^{q+1}_{l=1}$, i.e., $\mathcal{A}_j=\{\tau_{j-1}+1,\dots,\tau_j\}$, and its corresponding estimate, $\wh{\mathcal{P}}=\{\wh{\mathcal{A}}_l\}_{l=1}^{\wh{q}+1}$. The CM is then
 $$
 \rm{CM}(\wh{\mathcal{P}},\mathcal{P})=\frac{1}{n}\sum_{\mathcal{A}\in\mathcal{P}}|\mathcal{A}|\max_{\wh{\mathcal{A}}\in\wh{\mathcal{P}}}\left\{\frac{|\mathcal{A} \cap \wh{\mathcal{A}}|}{|\mathcal{A} \cup \wh{\mathcal{A}}|}\right\}.
 $$
 The VM is calculated using the conditional entropy of $\wh{\mathcal{P}}$; for brevity, we omit its formal definition. Both metrics take values in $[0,1]$, with larger values indicating better accuracy. Results are averaged over 100 repeated experiments.

 For Scenario 1, Tables~\ref{tab:scenario1}, \ref{tab:scenario1_sm1}, and \ref{tab:scenario1_sm2} give the simulation study results when $d=2$, $d=8$, and $d=15$, respectively. We observe that MOPED performs competitively with E-divisive across all considered configurations of $(d,\rho,q)$, as both methods attain similar CM and VM scores, and the distributions of $\wh{q}-q$ are broadly similar. As expected, MOPED performs better when the change in strength of extremal dependence is greater (i.e., when $\rho=0.6$). In some cases, MOPED performs markedly better than E-divisive; for example, in the case where we expect identification of a change point to be the most difficult, i.e., when $\rho=0.2$ and $d=2$, MOPED provides both higher CM/VM scores (when $q \geq 1$) and is more likely to identify the true number of change points. Moreover, Tables~\ref{tab:scenario1}, \ref{tab:scenario1_sm1}, and \ref{tab:scenario1_sm2} illustrate improvements can be made to MOPED by aggregating over sequences of bandwidths $G$ and ranks $k$; we note that, in cases where the change in extremal dependence is small, i.e., $\rho=0.2$ and $q > 0$, multiscale, multi-threshold MOPED provides more accurate estimates of the change point locations (higher CM and VM scores). Finally, we note that the parametric extreme value method of \cite{hazra2025estimating} provides the most accurate estimates of the change point when there is known to be only $q=1$. However, this method is limited to dimension $d=2$, whereas Tables~\ref{tab:scenario1_sm1} and \ref{tab:scenario1_sm2} illustrate that MOPED performs well for much higher dimensions. We note here that we have not optimised the hyper-parameters, $G$ and $k$, for MOPED or MMMOPED, and so these results represent a lower-bound on the performance of our change point detection algorithm.

\begin{table*}[t!]
\caption{Distribution of the estimated number of change points and the average CM and VM over 100 realisation for Scenario 1, $d=2$, of the simulation study. The modal value of $\wh{q}-q$ is provided in bold in each row.}
\label{tab:scenario1}
 \centering
  \begin{tabular}{cccccccccc}
\cmidrule(lr){4-8} 
&&&\multicolumn{5}{c}{$\wh{q}-q$}&&\\
 $\rho$ & q & Method & $ \leq 2$ & -1 & $\mathbf{0}$ & 1 & $\geq 2$ & CM & VM \\
\cmidrule(lr){1-2}
\cmidrule(lr){3-3}
\cmidrule(lr){4-9} 
\cmidrule(lr){9-10}
 \multirow{3}{*}{$\rho=0.2$}& \multirow{6}{*}{$0$}& MOPED & - & - & \textbf{0.897} & 0.092 & 0.011 & - & - \\ 
 && MMMOPED& 0.000 & 0.000 & \textbf{0.568} & 0.299 & 0.133 & - & - \\ 

&  & E-divisive & - & - & \textbf{0.948} & 0.013 & 0.039 & - & - \\ 
\cmidrule(lr){3-10}
 \multirow{3}{*}{$\rho=0.6$}& & MOPED  & - & - & \textbf{0.889} & 0.100 & 0.011 & - & - \\ 
  &&MMMOPED& 0.000 & 0.000 & \textbf{0.571} & 0.305 & 0.124 & - & - \\ 
&  & E-divisive & - & - & \textbf{0.953} & 0.015 & 0.032 & - & - \\ 
\cmidrule(lr){1-2}
\cmidrule(lr){3-3}
\cmidrule(lr){4-8} 
\cmidrule(lr){9-10}
 \multirow{4}{*}{$\rho=0.2$} & \multirow{8}{*}{$1$}& MOPED & 0.000 & 0.402 & \textbf{0.565} & 
 0.033 & 0.000 & 0.729 & 0.449 \\ 
 && MMMOPED & 0.000 & 0.153 & \textbf{0.542} & 0.238 & 0.067 & 0.783 & 0.608 \\ 
&  & E-divisive & 0.000 & \textbf{0.902} & 0.059 & 0.035 & 0.004 & 0.527 & 0.058 \\ 
& & \cite{hazra2025estimating} & - & -& 1.000 & - & - & 0.924 & 0.829 \\ 
\cmidrule(lr){3-10}
 \multirow{4}{*}{$\rho=0.6$}& & MOPED & 0.000 & 0.000 & 
 \textbf{0.993} & 0.006 & 0.001 & 0.973 & 0.923 \\ 
 &&MMMOPED& 0.000 & 0.000 & \textbf{0.644} & 0.284 & 0.072 & 0.928 & 0.873 \\ 
& & E-divisive & 0.000 & 0.000 & \textbf{0.947} & 0.018 & 0.035 & 0.989 & 0.975 \\ 
& & \cite{hazra2025estimating} & - & -& 1.000 & - & - & 0.995 & 0.979 \\ 
\cmidrule(lr){1-2}
\cmidrule(lr){3-3}
\cmidrule(lr){4-8} 
\cmidrule(lr){9-10}
 \multirow{3}{*}{$\rho=0.2$} & \multirow{6}{*}{$2$}& MOPED & \textbf{0.365} & 0.344 & 0.291 & 0.000 & 0.000 & 0.604 & 0.469 \\ 
  &&MMMOPED& 0.127 & 0.247 & \textbf{0.462}& 0.146 & 0.018 & 0.709 & 0.637 \\ 
&  & E-divisive & \textbf{0.932} & 0.025 & 0.039 & 0.004 & 0.000 & 0.347 & 0.031 \\ 

\cmidrule(lr){3-10}
 \multirow{3}{*}{$\rho=0.6$}& & MOPED  & 0.000 & 0.000 & \textbf{1.000} & 0.000 & 0.000 & 0.963 & 0.929 \\ 
  &&MMMOPED& 0.000 & 0.000 & \textbf{0.552} & 0.370 & 0.078 & 0.918 & 0.888 \\ 
& & E-divisive & 0.000 & 0.000 & \textbf{0.947} & 0.025 & 0.028 & 0.986 & 0.974 \\ 
\hline
 \end{tabular}
 \end{table*}

  For Scenario 2, Tables~\ref{tab:scenario2}, \ref{tab:scenario2_sm1}, and \ref{tab:scenario2_sm2} give the results for $d=2$, $d=8$, and $d=15$, respectively. In contrast to Scenario 1, the change in the distribution across the change point is more subtle; the data generating copula changes from one which exhibits asymptotic dependence to one which exhibits asymptotic independence (see Section~\ref{sec:Introduction}), but with the (sub-asymptotic) correlation structure remaining constant. Note that, for Scenario 2, we only consider the true number of change points $q>0$; the case where $q=0$ is equivalent to the similar case in Scenario 1. For $d=2$ and $d=8$, Tables~\ref{tab:scenario2} and \ref{tab:scenario2_sm1} illustrate that the fixed-parameter MOPED method generally outperforms both E-divisive and the method of \cite{hazra2025estimating}, in terms of both accuracy (higher CM/VM scores) and correct determination of $q$ (with more mass at zero for the estimated distribution of $\wh{q}-q$). We see further improvements on these results when using MMMOPED, especially in the case where the change point is most difficult to identify, i.e., $d=2$; see Table~\ref{tab:scenario2}. The performance of MOPED and E-divisive for higher dimension $d=15$ are broadly similar, and there is no clear dominant approach. 
\begin{table*}[t!]
\caption{Distribution of the estimated number of change points and the average CM and VM over 100 realisation for Scenario 2, $d=2$, of the simulation study. The modal value of $\wh{q}-q$ is provided in bold in each row.}
\label{tab:scenario2}
 \centering
  \begin{tabular}{cccccccccc}
\cmidrule(lr){4-8} 
&&&\multicolumn{5}{c}{$\wh{q}-q$}&&\\
$\Omega$ & q & Method & $ \leq 2$ & -1 & $\mathbf{0}$ & 1 & $\geq 2$ & CM & VM \\
\cmidrule(lr){1-2}
\cmidrule(lr){3-3}
\cmidrule(lr){4-8} 
\cmidrule(lr){9-10}
 \multirow{4}{*}{$1$} & \multirow{12}{*}{$1$}& MOPED & 0.000 & \textbf{0.582} & 0.391 & 0.026 & 0.001 & 0.660 & 0.316 \\ 
&&MMMOPED& 0.000 & 0.293 & \textbf{0.471} & 0.197 & 0.039 & 0.726 & 0.489 \\ 
&  & E-divisive & 0.000 & \textbf{0.949} & 0.019 & 0.029 & 0.003 & 0.508 & 0.021 \\  
&  & \cite{hazra2025estimating} & - & - & 1.000 & - & - & 0.553 & 0.195 \\ 
\cmidrule(lr){3-10}
 \multirow{4}{*}{$2$} & & MOPED   & 0.000 & \textbf{0.559} & 0.417 & 0.024 & 0.000 & 0.669 & 0.332 \\ 
&&MMMOPED & 0.000 & 0.255 & \textbf{0.505} & 0.198 & 0.042 & 0.739 & 0.517 \\ 
& & E-divisive    & 0.000 & \textbf{0.940} & 0.021 & 0.035 & 0.004 & 0.509 & 0.024 \\ 
&  & \cite{hazra2025estimating} & - & - & 1.000 & - & - & 0.554 & 0.202 \\ 
\cmidrule(lr){3-10}
 \multirow{4}{*}{$3$} & & MOPED  & 0.000 & \textbf{0.555} & 0.424 & 0.021 & 0.000 & 0.670 & 0.334 \\ 
&&MMMOPED & 0.000 & 0.272 & \textbf{0.493} & 0.194 & 0.041 & 0.734 & 0.506 \\ 
& & E-divisive   & 0.000 & \textbf{0.940} & 0.018 & 0.038 & 0.004 & 0.508 & 0.024 \\ 
&  & \cite{hazra2025estimating} & - & - & 1.000 & - & - & 0.554 & 0.202\\ 
\cmidrule(lr){1-2}
\cmidrule(lr){3-3}
\cmidrule(lr){4-8} 
\cmidrule(lr){9-10}
  \multirow{3}{*}{$1$} & \multirow{9}{*}{$2$}& MOPED    & \textbf{0.536} & 0.314 & 0.150 & 0.000 & 0.000 & 0.511 & 0.326 \\ 
 &&MMMOPED& 0.238 & \textbf{0.351} & 0.315 & 0.085 & 0.011 & 0.617 & 0.517 \\
&  & E-divisive& \textbf{0.938} & 0.013 & 0.049 & 0.000 & 0.000 & 0.345 & 0.028 \\ 
\cmidrule(lr){3-10}
 \multirow{3}{*}{$2$} & & MOPED   & \textbf{0.494} & 0.344 & 0.161 & 0.001 & 0.000 & 0.526 & 0.355 \\ 
&&MMMOPED & 0.215 & \textbf{0.357} & 0.330 & 0.091 & 0.007 & 0.626 & 0.535 \\ 
& & E-divisive & \textbf{0.941} & 0.013 & 0.046 & 0.000 & 0.000 & 0.344 & 0.027 \\ 
\cmidrule(lr){3-10}
 \multirow{3}{*}{$3$} & & MOPED & \textbf{0.495} & 0.334 & 0.170 & 
0.001 & 0.000 & 0.529 & 0.357 \\ 
&&MMMOPED& 0.219 & \textbf{0.364} & 0.324 & 0.086 & 0.007 & 0.623 & 0.530 \\ 
& & E-divisive & \textbf{0.941} & 0.013 & 0.045 & 0.000 & 0.001 & 0.344 & 0.027 \\ 

   \hline
 \end{tabular}
 \end{table*}

\section{Application}
\label{sec:application}
\subsection{Overview}
\begin{figure}[b!]
    \centering
    \includegraphics[width = 0.4\linewidth]{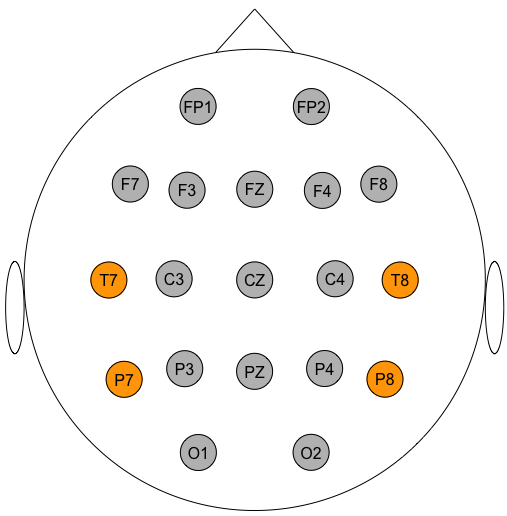}
    \caption{Standard 10-20 EEG scalp topography with 19 channels. The $d=4$ channels coloured in orange are those considered in the lower dimensional case study.}
    \label{fig:scalp}
\end{figure}

We now consider multivariate electroencephalogram (EEG) recordings from 79 human neonates, who were monitored for the occurrence of seizures in the Baby Brain Activity Center, which is a neonatal intensive care unit in Finland \citep{Stevenson2019}. The signals are observed at $d=19$ channels of the brain, which are illustrated in Figure~\ref{fig:scalp}, and the sample size $n$ differs between subjects. Alongside the observed time series, the dataset includes annotations of the occurrence of seizures provided by three expert clinicians. Extremal dependence of these data has been previously analysed by \cite{talento2025spectral}, who identified changes in the extremal dependence strength and brain connectivity for those neonates who had experienced seizures. Similarly, \cite{Guerrero2023} and \cite{Redondo2024} have previously identified changes in the pre- and post-seizure extremal behaviour of brains, albeit in adults rather than neonates. Given the abrupt nature of seizures \citep{NIH} and their links with changing extremal dependence, MOPED may be able to identify their occurrence or absence.

Following \cite{talento2025spectral}, we pre-process each time series by applying a Butterworth low-pass filter which retains only the EEG signals comprising oscillations in the Delta frequency band of $(0,4]$ Hertz (Hz). We focus on the Delta band, commonly associated with sleep patterns, as erratic behaviour in this frequency band is often linked with seizures \citep{douglass2002burst,muthuswamy2002higher}. The Butterworth low-pass filter is applied using the \texttt{R} package \texttt{eegkit} \citep{helwig2018eegkit}. To remove serial dependence in these data, we sub-sample every 256th value, where 256Hz is the sampling rate, and thus we sub-sample every second. We consider only continuous time series, i.e., neonates with no interruption in the Delta band activity across the observation period, and discard all other time series. The remaining data are marginally standardised to Pareto(2) margins by applying an empirical rank transform.

We compare change point estimates using MOPED and E-divisive. For MOPED, the hyper-parameters are set to $G=1000$ and $k=0.1G$; we report estimates for other hyper-parameter configurations in Appendix~\ref{sec:sup_figs}. As discussed in Sections~\ref{sec:multiscale} and \ref{sec:sim}, the multiscale, multi-threshold MOPED (MMMOPED) algorithm is less conservative than the fixed MOPED method, and so it always provides larger estimates $\wh{q}$ of the total number of change points. Thus, for brevity, we generally omit the results of MMMOPED from the main text, and instead provide these in Figures~\ref{fig:MMMOPED_results_sup1} and \ref{fig:MMMOPED_results_sup2} of Appendix~\ref{sec:sup_figs}; for MMMOPED, we pool over bandwidth values of $G\in\{500,1000,1500\}$ and ranks $k\in\{0.1G,0.05G,0.025G\}$. For all three methods, we consider both a low and high dimensional setting, with $d=4$ and $d=19$ signals, respectively, and a significance level of $\alpha=0.05$. When $d=4$, we follow \cite{talento2025spectral} and consider signals observed at channels T7, T8, P7, and P8, which are equally distributed across the right and left hemispheres of the brain (see Figure~\ref{fig:scalp}). 

We consider a subset of the 79 neonates and focus on three cases, using the expert clinician annotations: 1) where there are no recorded seizures; 2) where there is one contiguous time period of seizure activity; 3) where there are multiple discontiguous periods of seizure activity.
\subsection{Results}

\begin{figure}[t!]
    \centering
            \includegraphics[width = 0.45\linewidth]{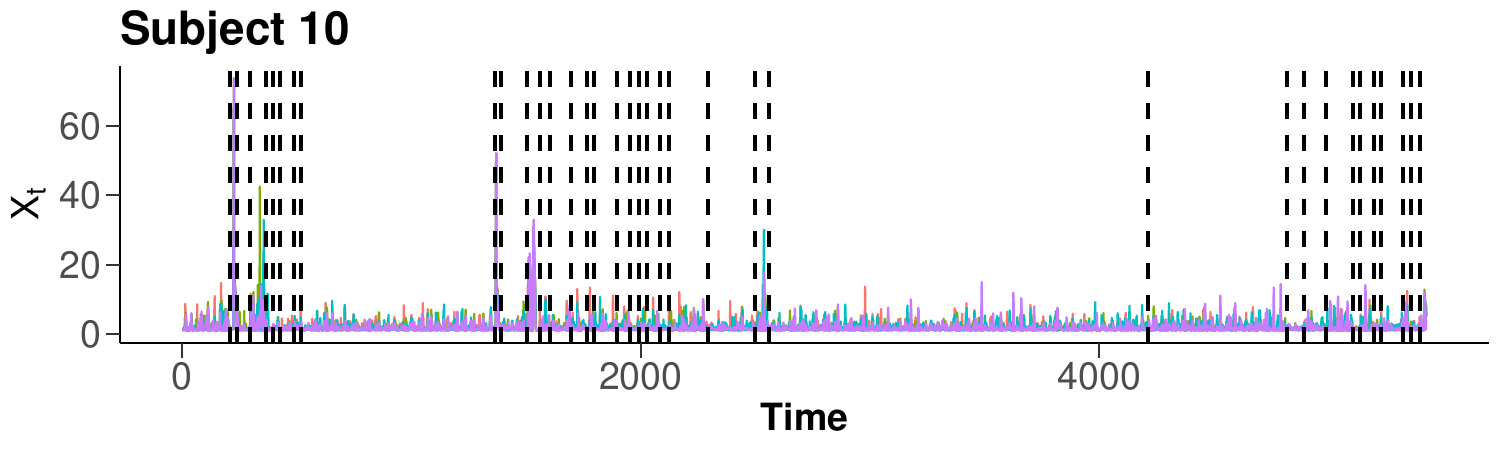}   \includegraphics[width = 0.45\linewidth]{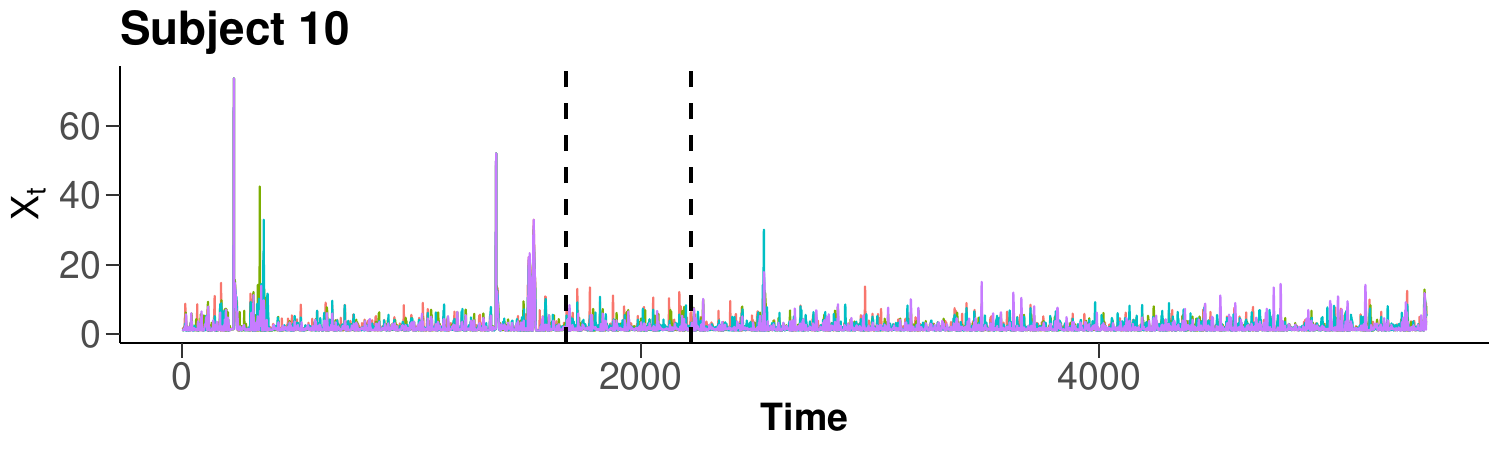}
    \includegraphics[width = 0.45\linewidth]{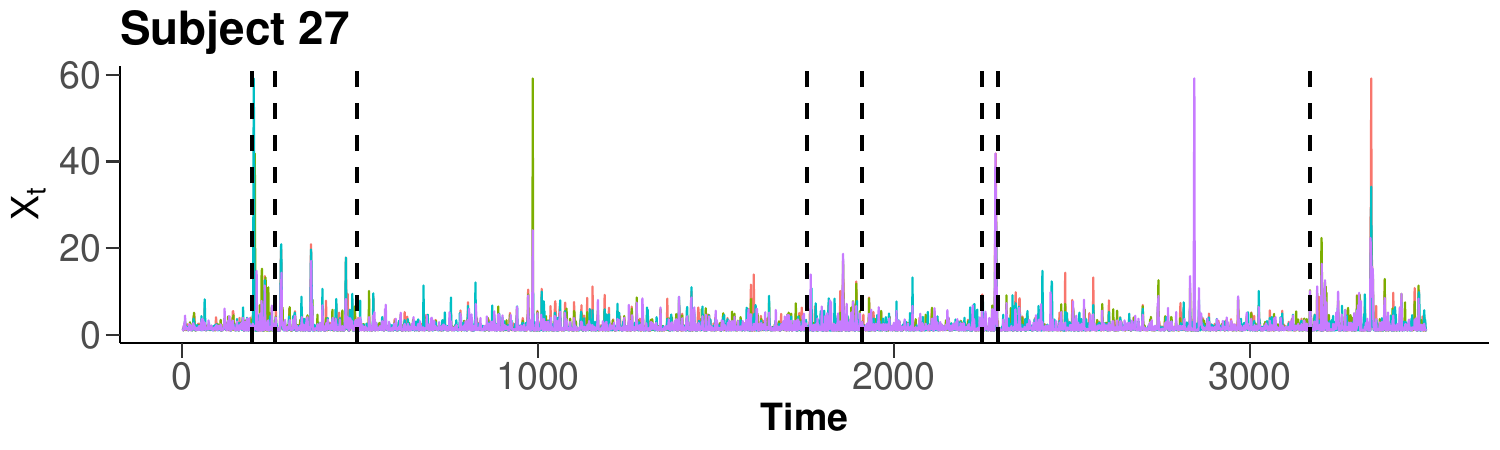}   \includegraphics[width = 0.45\linewidth]{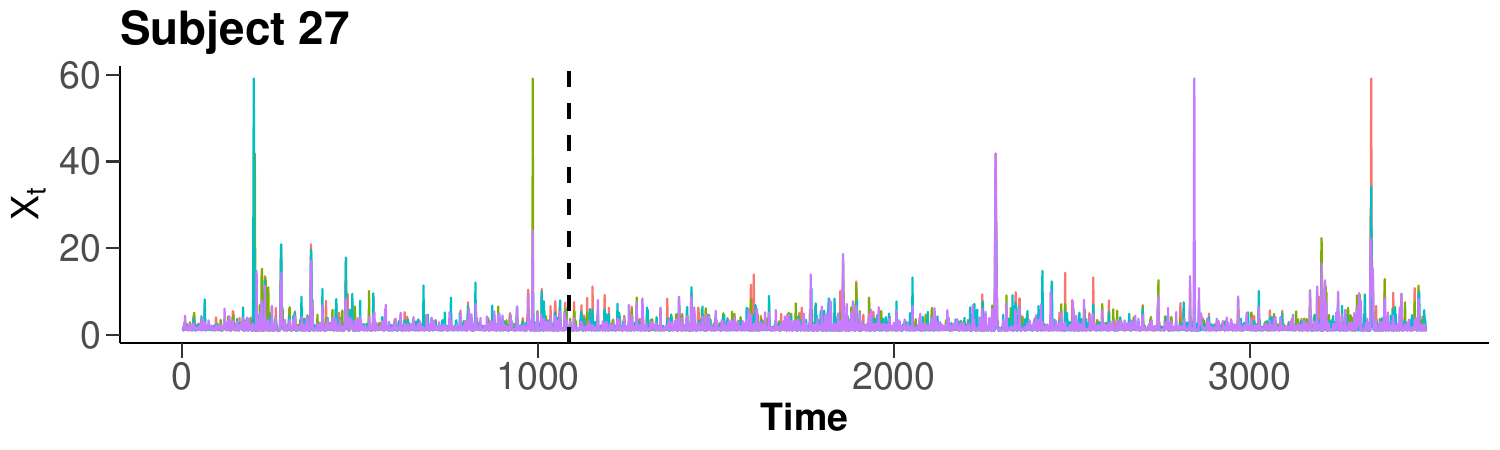}
        \includegraphics[width = 0.45\linewidth]{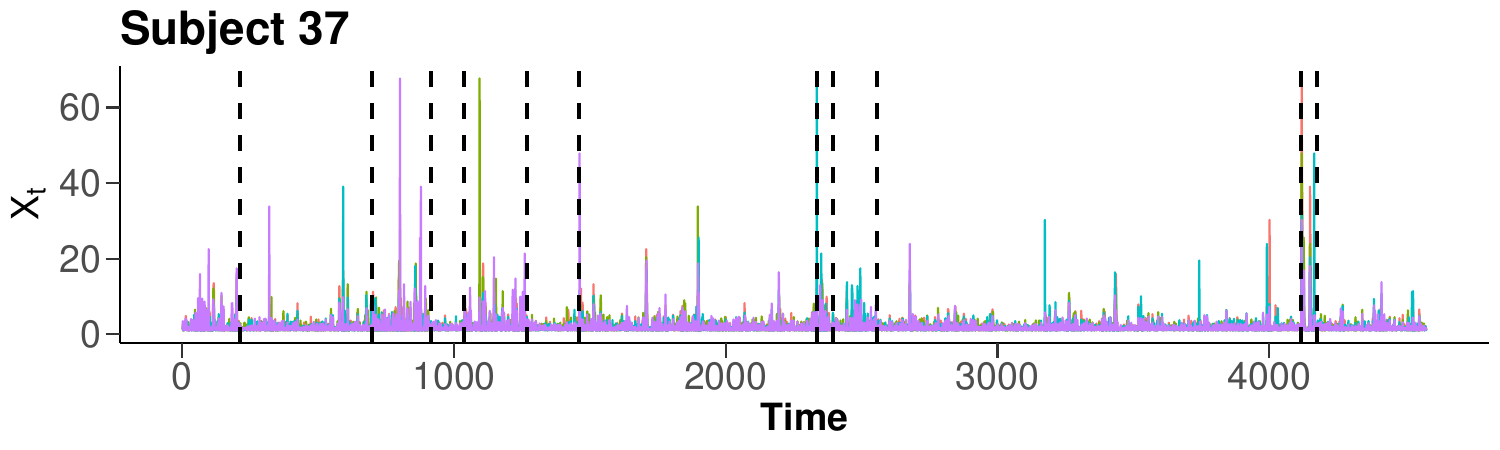}   \includegraphics[width = 0.45\linewidth]{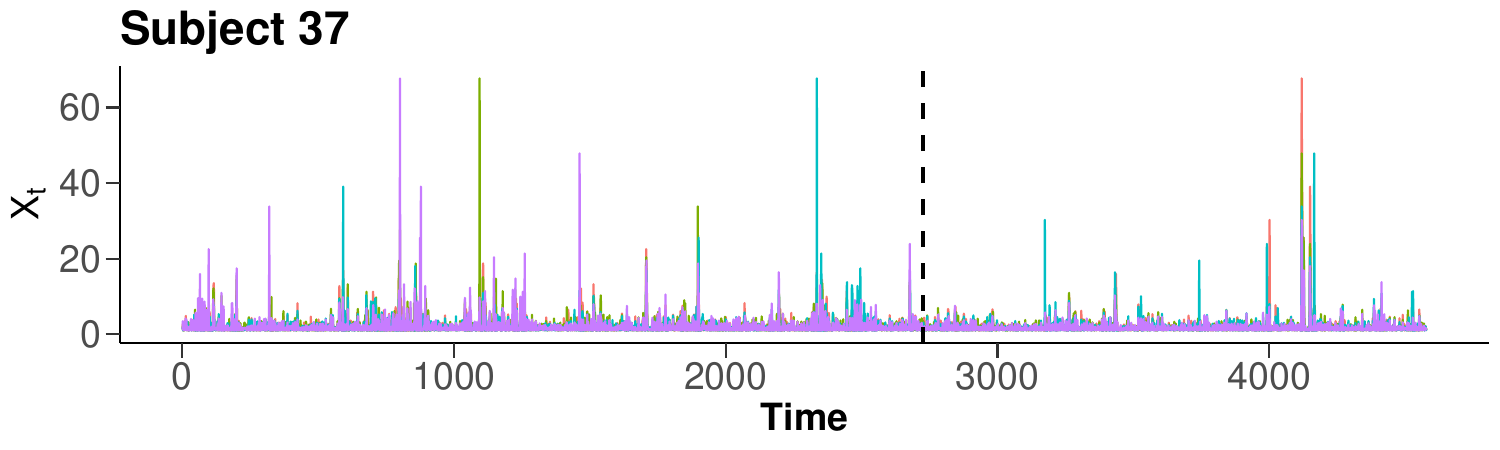}
    \caption{Change point estimates for Subjects 10, 27, and 37 (top to bottom row). Estimates (vertical dotted lines) are provided by E-divisive (left) and MOPED (right). Observations of $\{X_{it}\}^n_{t=1}$ are plotted against time $t$, with the colour corresponding to the $i$-th channel, $i=1,\dots,d$. Here, $d=4$ and no subjects were identified as having experienced a seizure.}
    \label{fig:case1_results}
\end{figure}

For Case 1, we consider $d=4$ and Subjects 10, 27, and 37 with sample sizes {$n = 5427$}, $n=3496$, and $n=4578$, respectively. We may expect that subjects that are free of seizures experience typical brain activity, and thus the ``ground truth'' is $q=0$  change points in the extremal dependence of their EEG time series. Figure~\ref{fig:case1_results} provides estimates of the change points from E-divisive and MOPED. Here we observe that E-divisive has a tendency to estimate substantially more change points than MOPED; conversely, MOPED routinely estimates $\wh{q}$ close to zero significant change points in the TPDM. A similar conclusion is drawn when considering all $d=19$ channels; see Figure~\ref{fig:case1_results_d19}. Figure~\ref{fig:case1_results_sup} in Appendix~\ref{sec:sup_figs} provides a sensitivity study of the MOPED estimates for Subject 27, when $d=4$. Across a range of $G$ and $k$ values, MOPED provides at most $\wh{q}=2$ significant change points, whilst E-divisive returns $\wh{q}=7$. As we discuss later, MOPED tends to provide higher estimates of $\wh{q}$ for subjects who have experienced seizures. \par

\begin{figure}[t!]
    \centering
        \includegraphics[width = 0.45\linewidth]{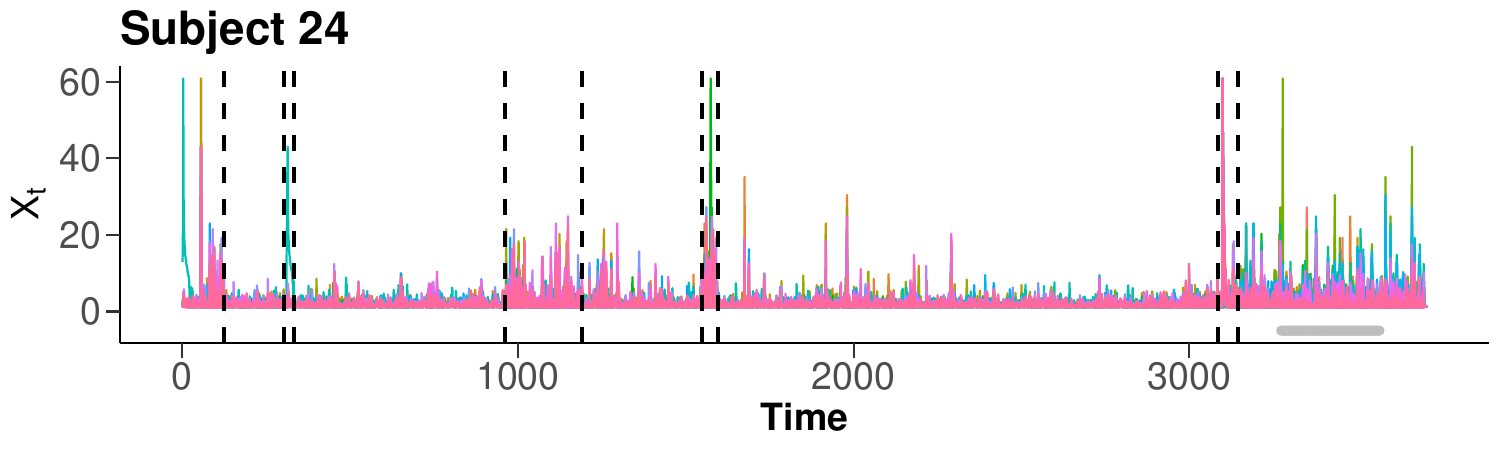}   \includegraphics[width = 0.45\linewidth]{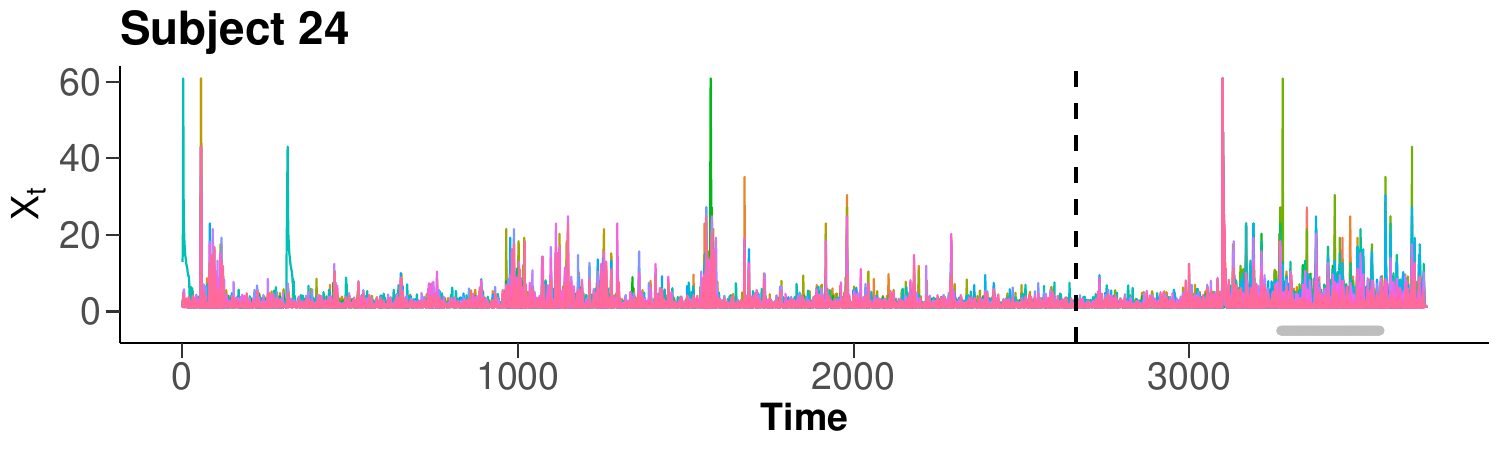}
         \includegraphics[width = 0.45\linewidth]{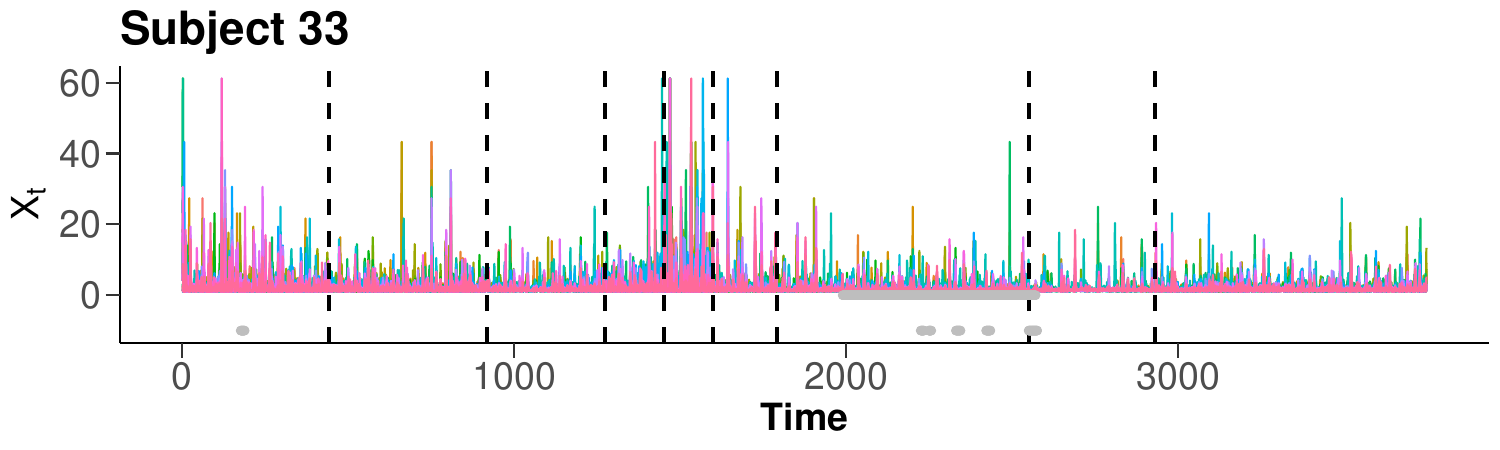}   \includegraphics[width = 0.45\linewidth]{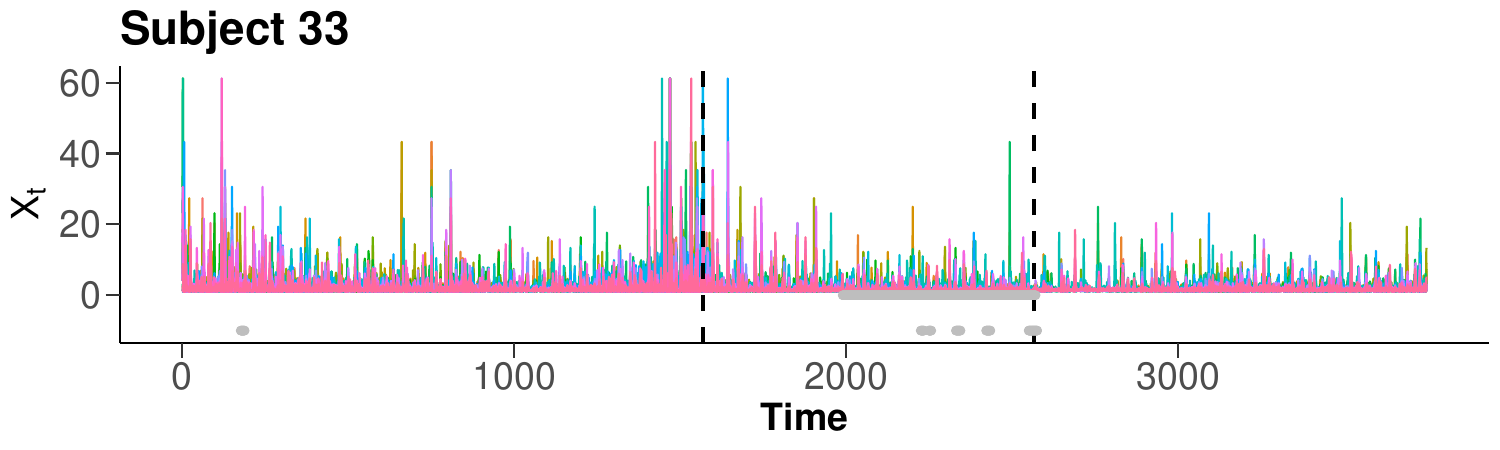}
            \includegraphics[width = 0.45\linewidth]{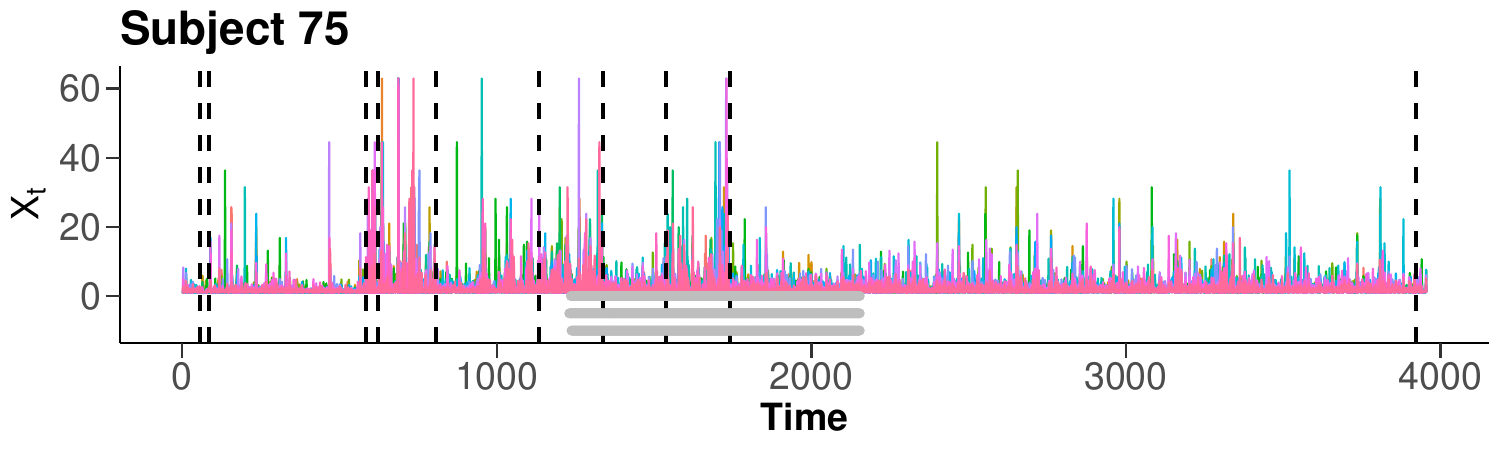}   \includegraphics[width = 0.45\linewidth]{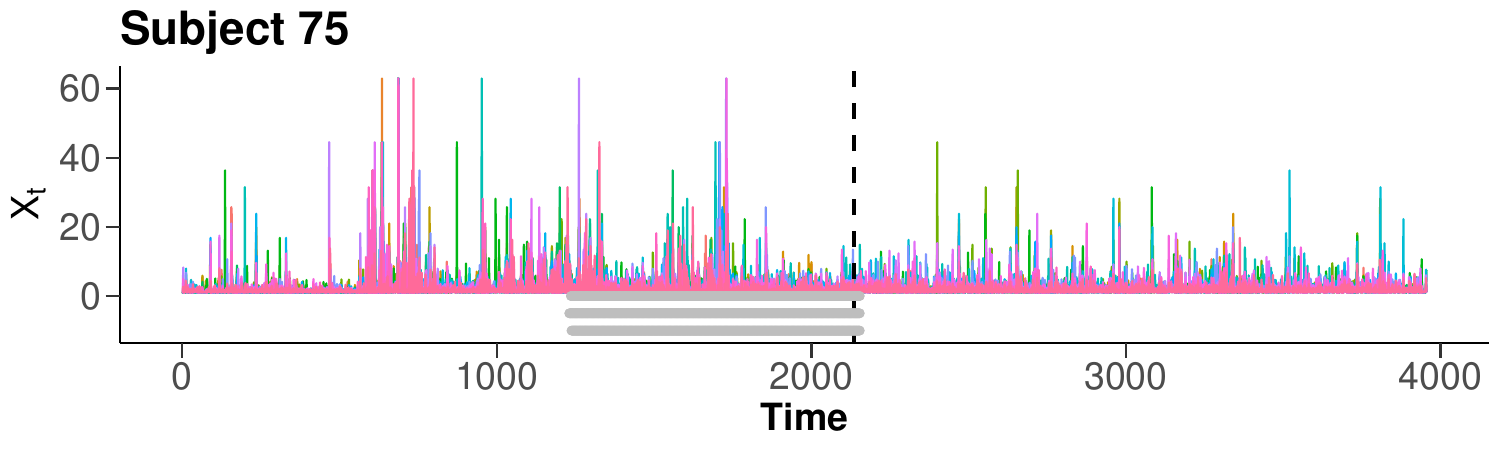}
            
    \caption{Change point estimates for Subjects 24, 33, and 75 (top to bottom row). Estimates (vertical dotted lines) are provided by E-divisive (left) and MOPED (right).  Observations of $\{X_{i,t}\}^n_{t=1}$ are plotted against time $t$, with the colours corresponding to the $i$-th channel, $i=1,\dots,d$. Here $d=19$ and the grey dots along the x-axis denote time-points identified as seizures by expert clinicians; these points are arranged in three rows, corresponding to the different expert clinicians.}
    \label{fig:case2_results}
\end{figure}

We now focus on three subjects for Case 2, i.e., they were identified as experiencing a single uninterrupted period of seizure throughout the observation period. Subjects 24, 33, and 75 ($n=3706$, $n=3747$, and $n=3955$, respectively) all experience one contiguous period of interrupted seizure; for Subjects 33 and 75, this occurs midway through the recording period, whereas for Subject 24, the seizure occurs at the end. We note that for Subject 33 there is a small period of seizure activity at the start of the observation period, but this was only identified by one of the clinical experts; hence we choose to ignore this period of activity.
Here, we consider all channels ($d=19$) with $G$ and $k$ as previously, and provide the results in Figure~\ref{fig:case2_results}. As in Figure~\ref{fig:case1_results}, Figure~\ref{fig:case2_results} shows that the E-divisive method is sensitive to changes in the EEG signals and routinely estimates a larger number $q$ of change points compared to MOPED. Although we have no ground truth in this application, it may be reasonable to assume that the true number of change points is close to two, with a change in the extremal behaviour in brain activity occurring both before and after the seizure period. Estimates from MOPED seem to agree with this hypothesis, as the method provides a single change point estimate proceeding the seizure occurrence of Subject 24 (where it would not be feasible to estimate one after the seizure event) and provides two change point estimates for Subject 33, which are either side of the annotated period of seizure activity. For Subject 75, MOPED provides a single change point estimate, at the end of the seizure activity period. Figure~\ref{fig:case2_results_sup} in Appendix~\ref{sec:sup_figs} provides a sensitivity study of the MOPED estimates for Subject 33; we generally observe one or two change point estimates, that are close to the period of seizure activity. 

\begin{figure}[t!]
    \centering
  \includegraphics[width =0.5 \linewidth]{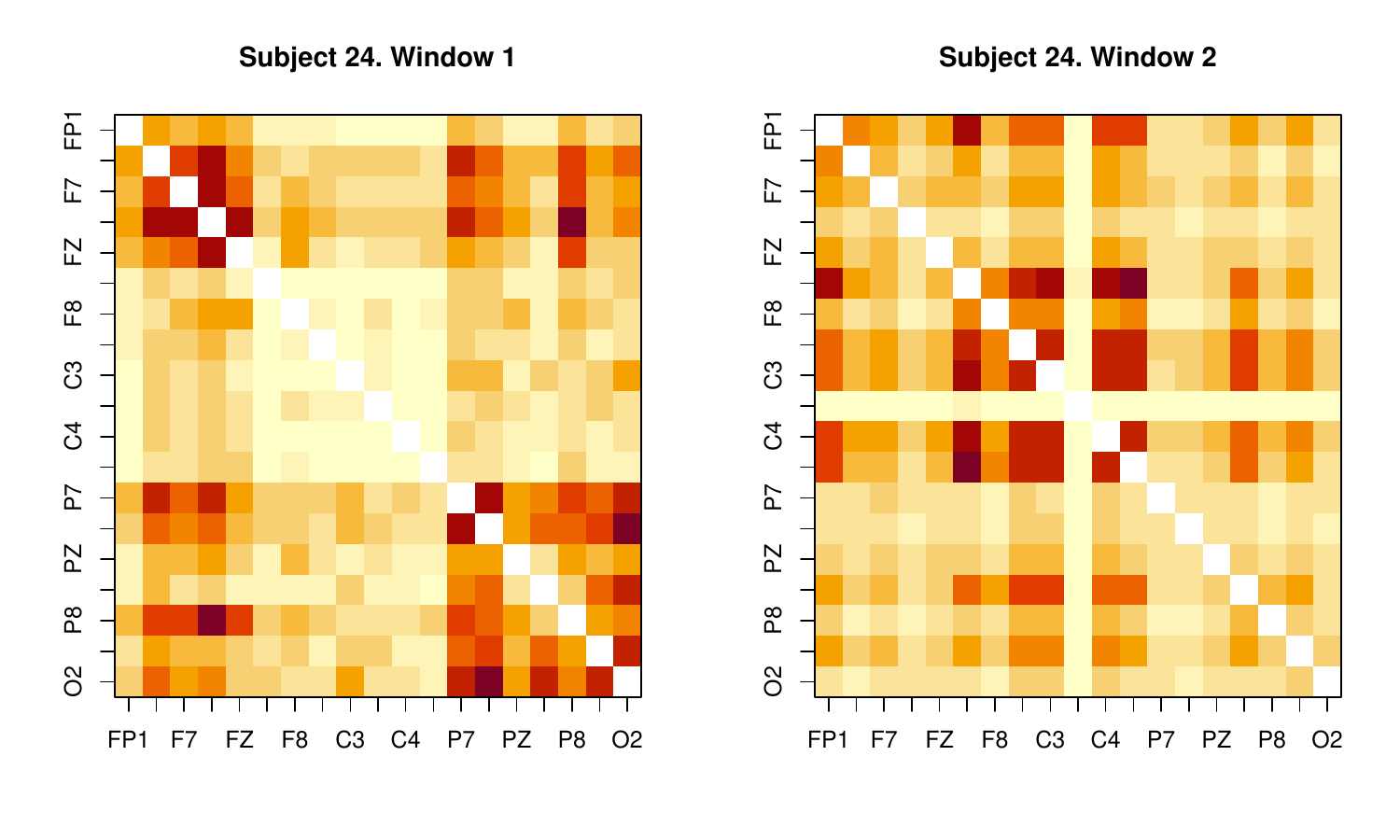}
        \includegraphics[width = \linewidth]{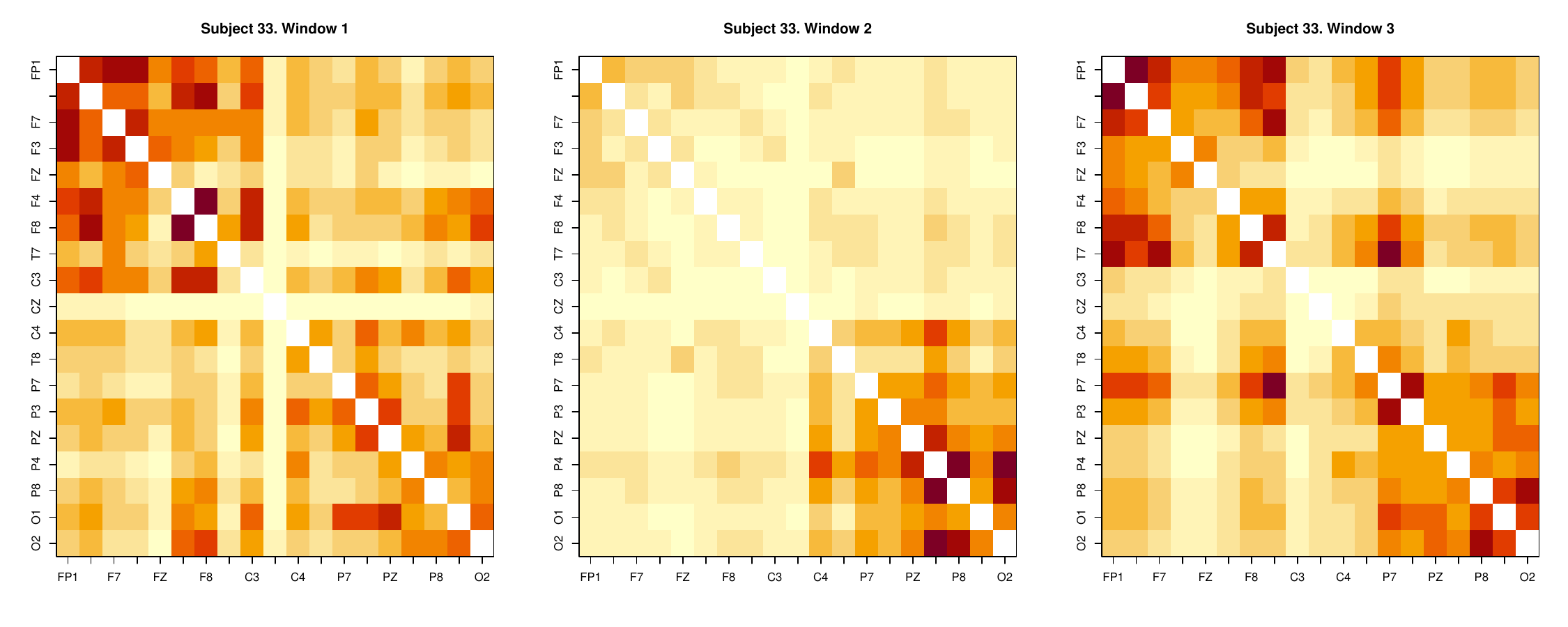}
    \includegraphics[width =0.5\linewidth]{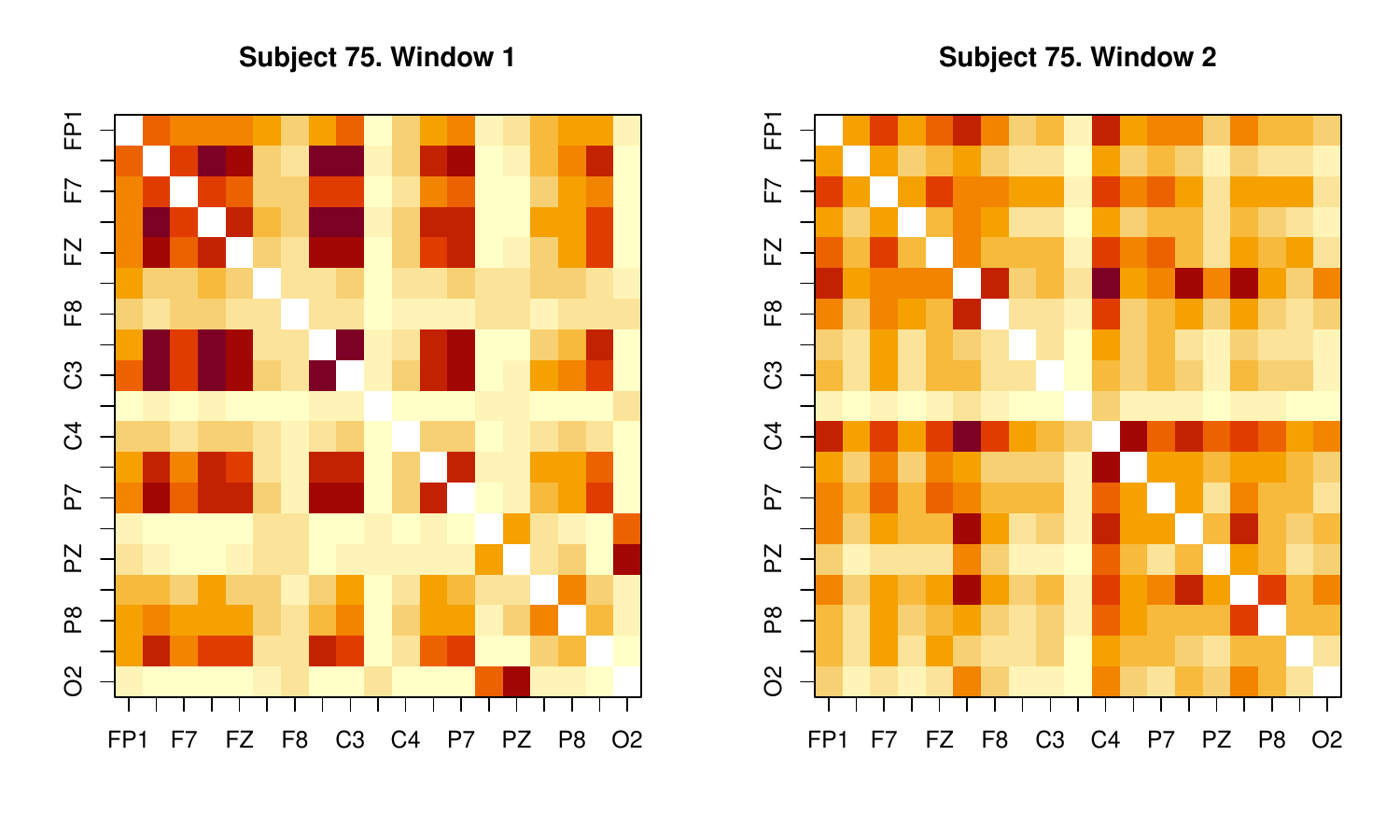}
    \caption{Heat maps of estimated TPDM for Subjects 24, 33, and 75 for the windows of observations demarcated by the MOPED change point estimates shown in Figure~\ref{fig:case2_results}. The windows are ordered from left to right, and the number of windows changes with the subject. The column and row names correspond to the channels in Figure~\ref{fig:scalp}.}
    \label{fig:case2_TPDM}
\end{figure}

Figure~\ref{fig:case2_TPDM} provides local estimates of the TPDM for the subjects shown in Figure~\ref{fig:case2_results}. For each window of data demarcated by the estimated change points provided in the right panels of Figure~\ref{fig:case2_results}, we estimate the TPDM using all data and with $r_0$ corresponding to the $0.95$-quantile of the empirical radii. Figure~\ref{fig:case2_TPDM} showcases the distinct changes in the TPDM which are being identified by MOPED. Of note are the TPDM estimates for Subject 33, which have a block-diagonal structure that corresponds to strong extremal dependence between only Channels C4--O2 (lower hemisphere; see Figure~\ref{fig:scalp}) during the period of seizure activity. This localised extremal activity may be indicative of a seizure.

\begin{landscape}
\begin{figure}[t!]
    \centering

        \includegraphics[width = 0.3\linewidth]{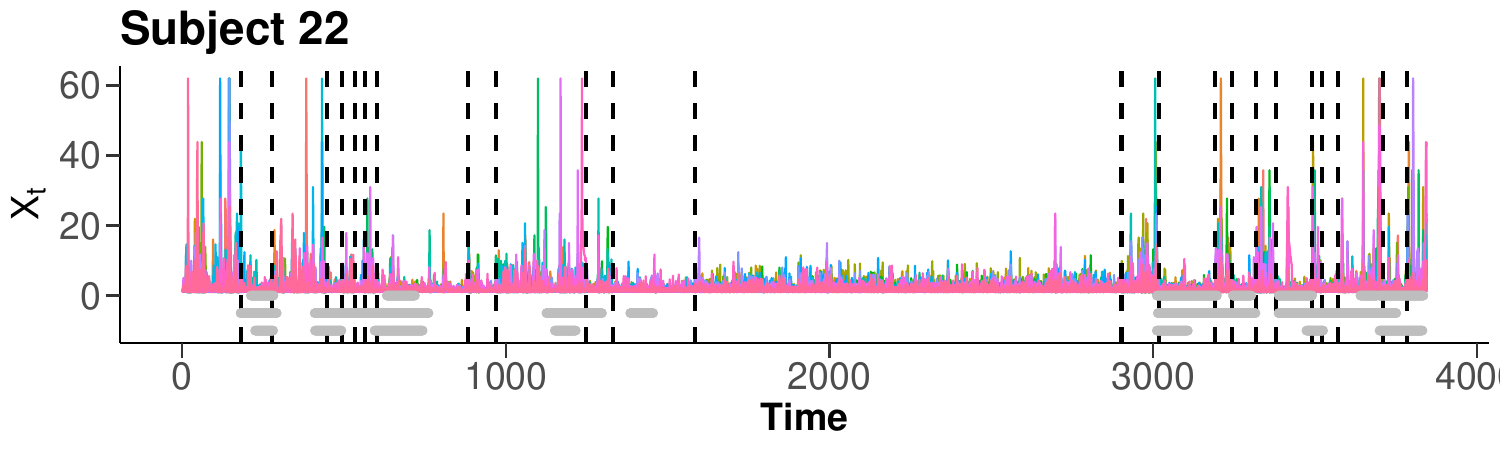}   \includegraphics[width = 0.3\linewidth]{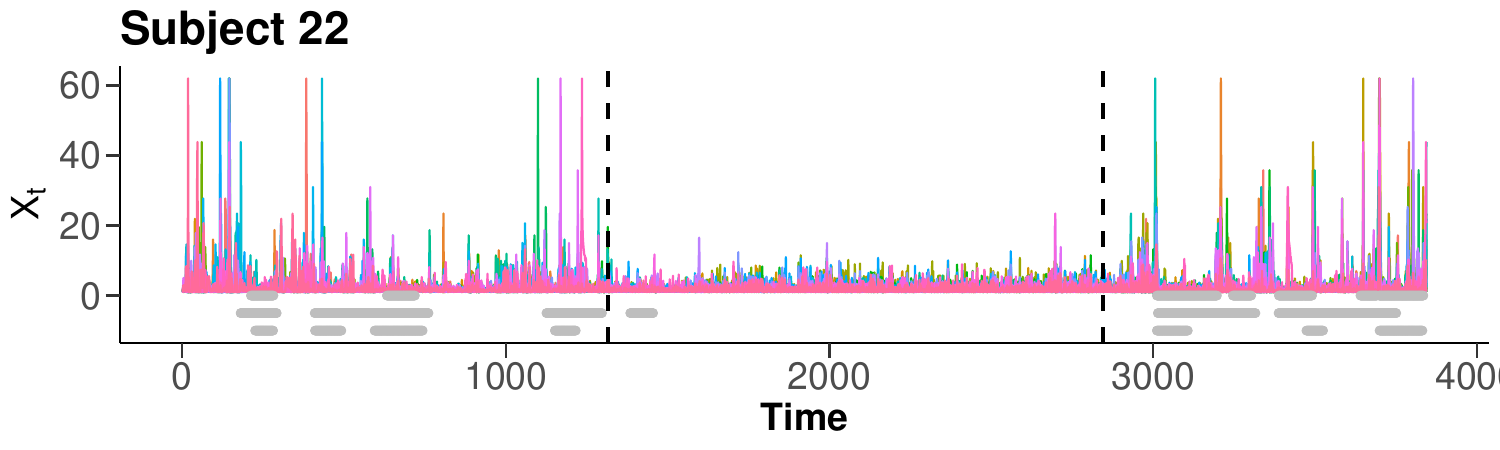}
        \includegraphics[width = 0.3\linewidth]{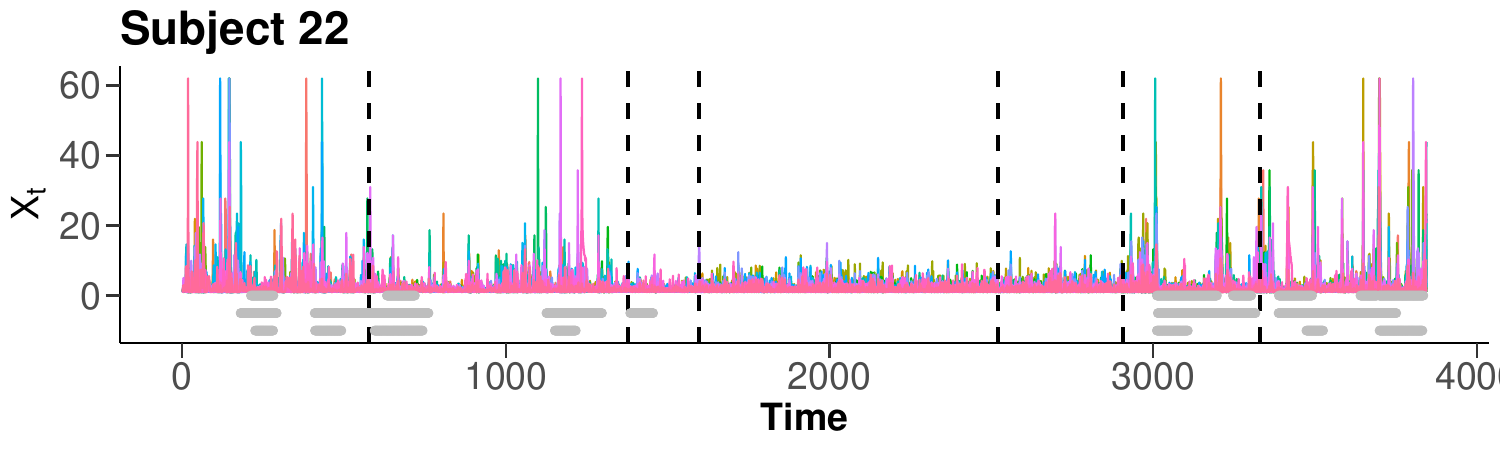}
         \includegraphics[width = 0.3\linewidth]{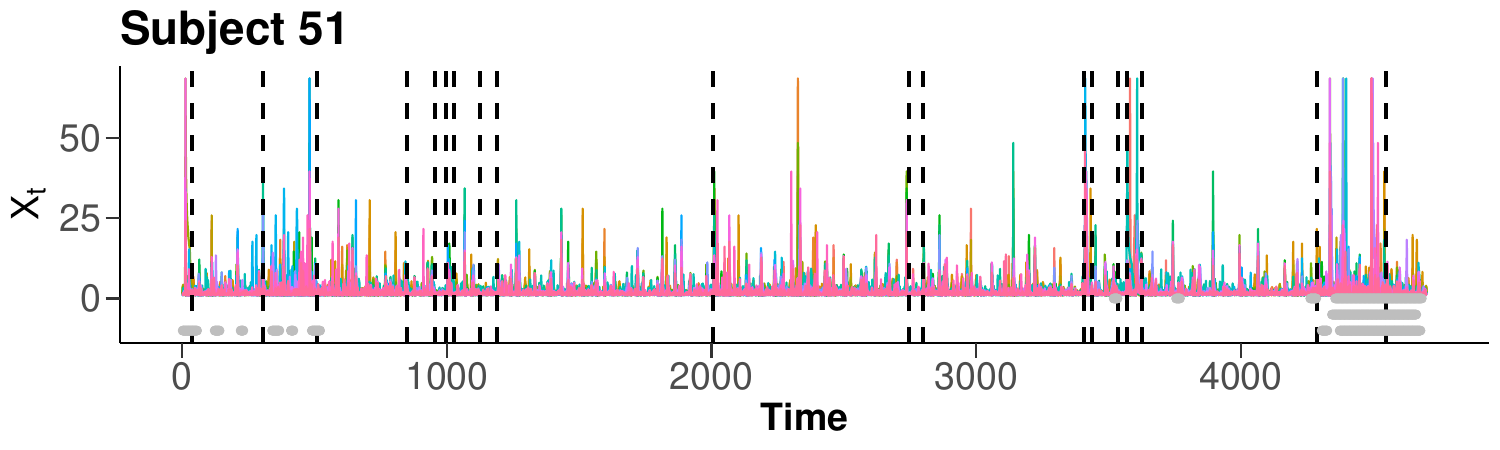}   \includegraphics[width = 0.3\linewidth]{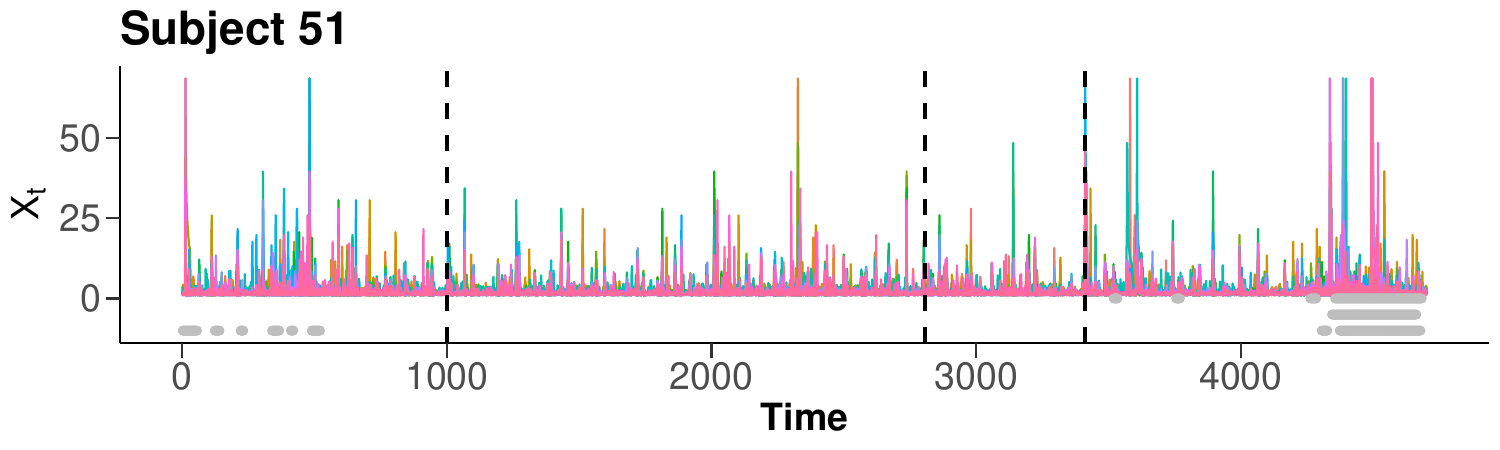}
             \includegraphics[width = 0.3\linewidth]{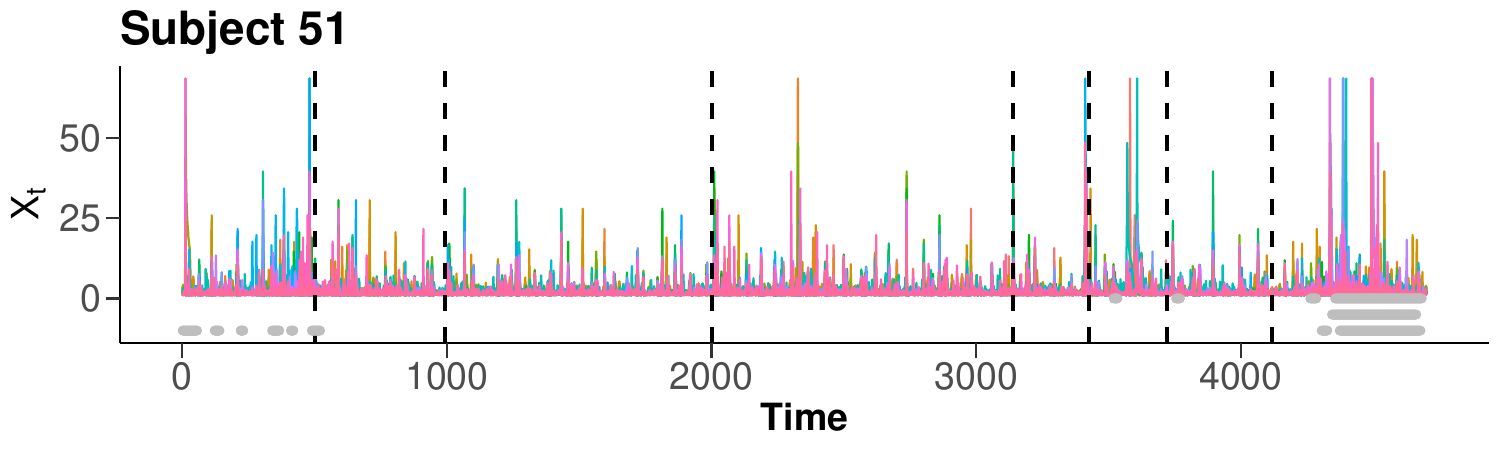}
            \includegraphics[width = 0.3\linewidth]{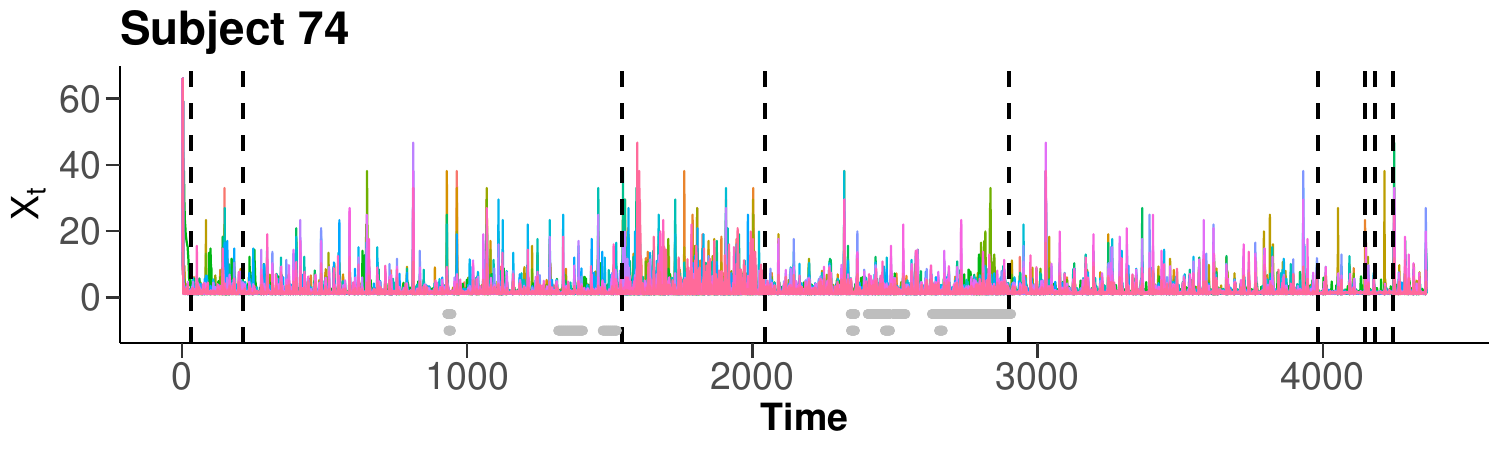}   \includegraphics[width = 0.3\linewidth]{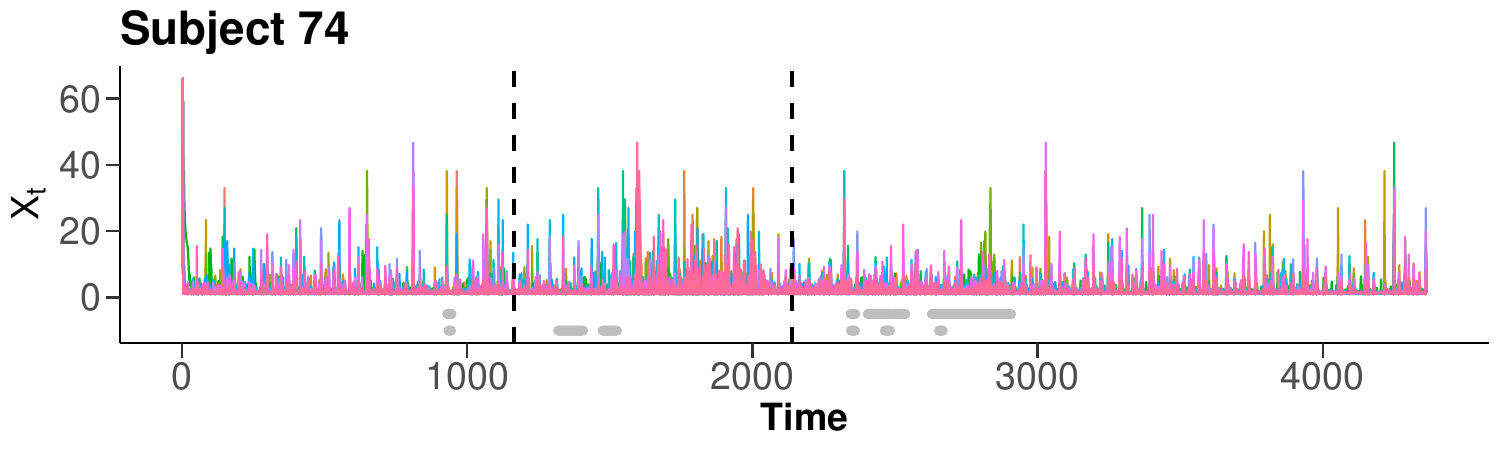}
                \includegraphics[width = 0.3\linewidth]{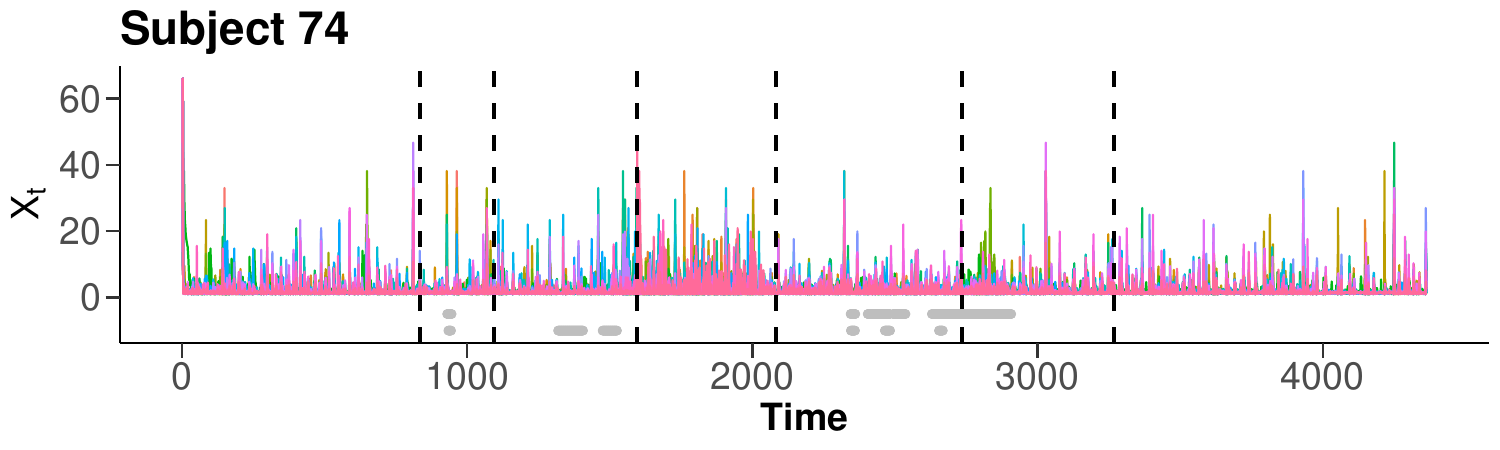}
    \caption{
    Change point estimates for Subjects 22, 51, and 74 (top to bottom row). Estimates (vertical dotted lines) are provided by E-divisive (left),  MOPED (centre), and MMMOPED (right).  Observations of $\{X_{i,t}\}^n_{t=1}$ are plotted against time $t$, with the colours corresponding to the $i$-th channel, $i=1,\dots,d$.
    Observations of $\{X_{i,t}\}^n_{t=1}$ are plotted against time $t$, with the colours corresponding to the $i$-th channel, $i=1,\dots,d$. Here $d=19$ and the grey dots along the x-axis denote time-points identified as seizures by expert clinicians; these points are arranged in three rows, corresponding to the different expert clinicians.}
    \label{fig:case3_results}
\end{figure}
\end{landscape}

For the final Case 3, we consider Subjects 22, 51, and 74 ($n=3844$, $n=4701$, and $n=4364$), which all experience multiple prolonged periods of seizure activity. Figure~\ref{fig:case3_results} provides the E-divisive, MOPED, and MMMOPED estimates of the change points. As we have seen previously for Cases 1 and 2, E-divisive provides a higher estimated number of change points, which occur throughout the times series. For Subject 22, we observe that MOPED provides two change point estimates: one at the end of the first seizure period, and one at the start of the second seizure period. We observe a similar result for Subject 51, except with an additional change point estimate close to the onset of the second seizure period. For Subject 74, MOPED provides two change point estimates, which are located between the first and second, and second and third, periods of seizure activity. As expected, MMMOPED provides more change point estimates.  For Subjects 22 and 74, Figure~\ref{fig:case3_results} illustrates that MMMOPED provides a total of $\wh{q}=6$ change point estimates, which are clustered around the periods of seizure activity.
\section{Discussion}
\label{sec:discussion}

We have proposed a computationally efficient algorithm, MOPED, for detecting structural changes in the tail dependence of sequences of regularly varying random vectors. Our approach uses a moving sum procedure to detect changes in the TPDM of \cite{cooley2019decompositions} in a nonparametric fashion. In Section~\ref{sec:multiscale}, we proposed a multiscale, multi-threshold method for pooling change point estimates across values of the bandwidth $G$ and exceedance threshold required for local estimation of the TPDM.
Our simulation studies demonstrated that MOPED performs competitively with both the current state-of-the-art approach for nonparametric change point detection, E-divisive, and the competing extremal dependence method of \cite{hazra2025estimating}. We showcased in our simulation study that, where nonparametric methods fail, MOPED is able to identify changes in the extremal dependence class of data, even when the sub-asymptotic dependence (correlation) structure remains unchanged. We also illustrated that our multiscale, multi-threshold MOPED method is able to improve accuracy of change point estimates. In our application to seizure-prone neonatal EEG signals in Section~\ref{sec:application}, we found that MOPED consistently returned a lower number of change points than the E-divisive approach in cases where practitioners identified no seizure activity in the EEG recordings. Moreover, in cases where seizures were identified by clinicians, we found that the MOPED change point estimates tended to coincide with the onset and end of seizure activity. This may suggest that seizures can be characterised by changes in the extremal dependence of brain signals, and that MOPED may be useful as a data-driven and automatic method for identifying seizures.

We illustrated cases where the MOPED algorithm outperforms the competing parametric extremal dependence change point method of \cite{hazra2025estimating}. Moreover, due to its computational efficiency, MOPED scales to higher dimensions than previously considered in other applications of extremal dependence change point models. In contrast, E-divisive performs well for high dimensional data. However, it does not enjoy the interpretability of MOPED (and other extremal change point algorithms), in the sense that it is unknown what aspects of the distribution have changed across the change points; MOPED specifically targets changes in the pairwise extremal dependence. The interpretability, efficiency, and good performance of MOPED come at the price of moderate hyperparameter tuning. While we did not optimise the hyperparameters for MOPED, in our application we found only small differences in the change point estimates when applying MOPED with different choices for $k$ and $G$. A more in-depth analysis may reveal some intricate link between the optimal choice of tuning parameters and the change in tail dependence. One may also consider performing further simulation studies to investigate the similarities and differences of E-divisive and MOPED for a wider range of tail dependence models, in particular in higher dimensions. Our multiscale, multi-threshold MOPED algorithm may alleviate the need for hyperparameter tuning, but further investigation into the optimal choice of ranks $\mathcal{K}$ and bandwidths $\mathcal{H}$, and the manner in which change point estimates are pooled across them, is required.

Although we have provided a permutation-based method for testing the existence of structural breaks in the TPDM, one could also consider an approach based on the asymptotic theory of the TPDM estimator; see, e.g., \cite{Pawley2025}. The theoretical treatment of the MOPED algorithm, i.e., the asymptotic characterisation of the MOSUM test statistic and convergence rates for the change point estimators, remains an avenue for further work.
\section*{Declarations}
\subsection*{Ethical Approval}
Not Applicable
\subsection*{Availability of supporting data}
The open access EEG data and expert annotations of \cite{Stevenson2019} are available at \url{https://doi.org/10.5281/zenodo.4940267} (last accessed 19/08/2025). Accompanying software implementing the method is available as the R package \verb!moped! at \url{https://github.com/EuanMcGonigle/moped}.
\subsection*{Competing interests}
The authors have no relevant financial or non-financial interests to disclose.

\subsection*{Funding details} 
The authors have no relevant funding details to disclose.
\subsection*{Acknowledgments}
The authors thank Mara Sherlin Talento for helpful discussions related to the dataset. This work has made use of the resources provided by the Edinburgh Compute and Data Facility (ECDF) (\url{www.ecdf.ed.ac.uk/}).

\renewcommand{\theequation}{A.\arabic{equation}}
\renewcommand{\thefigure}{A\arabic{figure}}
\renewcommand{\thetable}{A\arabic{table}}
\renewcommand{\thesection}{A\arabic{section}}

\setcounter{figure}{0}
\setcounter{table}{0}
\setcounter{equation}{0}

\pagebreak

\begin{appendix}
\section*{Appendix}
\section{Supplementary Tables}

\label{Sup:tables}

\begin{table*}[h!]
\caption{Distribution of the estimated number of change points and the average CM and VM over 100 realisation for Scenario 1, $d=8$, of the simulation study. The modal value of $\wh{q}-q$ is provided in bold in each row.}
\label{tab:scenario1_sm1}
 \centering
  \begin{tabular}{ccccccccccc}
\cmidrule(lr){5-9} 
&&&&\multicolumn{5}{c}{$\wh{q}-q$}&&\\
$d$& $\rho$ & q & Method & $ \leq 2$ & -1 & $\mathbf{0}$ & 1 & $\geq 2$ & CM & VM \\
\cmidrule(lr){1-3}
\cmidrule(lr){4-4}
\cmidrule(lr){5-9} 
\cmidrule(lr){10-11}

\multirow{18}{*}{$8$}  & \multirow{3}{*}{$\rho=0.2$} & \multirow{6}{*}{$0$}& MOPED & - & - & \textbf{0.881} & 0.114 & 0.005 & - & - \\ 
 &&&MMMOPED& - & - & \textbf{0.500} & 0.355 & 0.145 & - & - \\ 
&&  & E-divisive & - & - & \textbf{0.936} & 0.024 & 0.040 & - & - \\ 
\cmidrule(lr){4-11}
&  \multirow{3}{*}{$\rho=0.6$}& & MOPED & - & - & \textbf{0.901} & 0.096 & 0.003 & - & - \\ 
&&&MMMOPED & - & - & \textbf{0.534} & 0.343 & 0.123 & -& - \\ 
&&  & E-divisive & - & - & \textbf{0.948} & 0.013 & 0.039 & - & - \\ 
\cmidrule(lr){2-3}
\cmidrule(lr){4-4}
\cmidrule(lr){5-9} 
\cmidrule(lr){10-11}
&  \multirow{3}{*}{$\rho=0.2$} & \multirow{6}{*}{$1$}& MOPED  & 0.000 & 0.166 & \textbf{0.772} & 0.061 & 0.001 & 0.847 & 0.680 \\ 
 &&&MMMOPED& 0.000 & 0.016 & \textbf{0.537} & 0.364 & 0.083 & 0.859 & 0.765 \\  

&& & E-divisive & 0.000 & 0.000 & \textbf{0.939} & 0.026 & 0.035 & 0.988 & 0.972 \\ 
\cmidrule(lr){4-11}
&  \multirow{3}{*}{$\rho=0.6$}& & MOPED & 0.000 & 0.000 & \textbf{1.000} & 0.000 & 0.000 & 0.991 & 0.965 \\ 
&&&MMMOPED& 0.000 & 0.000 & \textbf{0.790} & 0.188 & 0.022 & 0.951 & 0.913 \\ 
&&  & E-divisive & 0.000 & 0.000 & \textbf{0.947} & 0.021 & 0.032 & 0.993 & 0.987 \\

\cmidrule(lr){2-3}
\cmidrule(lr){4-4}
\cmidrule(lr){5-9} 
\cmidrule(lr){10-11}
&  \multirow{3}{*}{$\rho=0.2$} & \multirow{6}{*}{$2$}& MOPED   & 0.129 & 0.307 & \textbf{0.562} & 0.002 & 0.000 & 0.762 & 0.697 \\ 
 &&&MMMOPED& 0.009 & 0.070 & \textbf{0.617} & 0.246 & 0.058 & 0.838 & 0.801 \\ 

&& & E-divisive & 0.000 & 0.001 & \textbf{0.930} & 0.037 & 0.032 & 0.981 & 0.966 \\ 
\cmidrule(lr){4-11}
&  \multirow{3}{*}{$\rho=0.6$}& & MOPED  & 0.000 & 0.000 & \textbf{1.000} & 0.000 & 0.000 & 0.982 & 0.961 \\ 
&&&MMMOPED& 0.000 & 0.000 & \textbf{0.765} & 0.210 & 0.025 & 0.951 & 0.926 \\ 
&&  & E-divisive & 0.000 & 0.000 & \textbf{0.937} & 0.026 & 0.037 & 0.993 & 0.989 \\ 
\hline
 \end{tabular}
 \end{table*}
 
\begin{table*}[t!]
\caption{Distribution of the estimated number of change points and the average CM and VM over 100 realisation for Scenario 1, $d=15$, of the simulation study. The modal value of $\wh{q}-q$ is provided in bold in each row.}
\label{tab:scenario1_sm2}
 \centering
  \begin{tabular}{ccccccccccc}
\cmidrule(lr){5-9} 
&&&&\multicolumn{5}{c}{$\wh{q}-q$}&&\\
$d$& $\rho$ & q & Method & $ \leq 2$ & -1 & $\mathbf{0}$ & 1 & $\geq 2$ & CM & VM \\
\cmidrule(lr){1-3}
\cmidrule(lr){4-4}
\cmidrule(lr){5-9} 
\cmidrule(lr){10-11}

\multirow{16}{*}{$15$} &  \multirow{3}{*}{$\rho=0.2$} & \multirow{6}{*}{$0$}& MOPED  & - & - & \textbf{0.908} & 0.087 & 0.005 & - & - \\ 
&&&MMMOPED& - & - & \textbf{0.556} & 0.271 & 0.174 & - & - \\ 
&&  & E-divisive & - & - & \textbf{0.957} & 0.009 & 0.034 & - & - \\ 
\cmidrule(lr){4-11}
 &  \multirow{3}{*}{$\rho=0.6$} & & MOPED  & - & - & \textbf{0.891} & 0.104 & 0.005 & - & - \\ 
&&&MMMOPED & - & - & \textbf{0.500} & 0.322 & 0.178 & - & - \\  
&& & E-divisive  & - & - & \textbf{0.958} & 0.006 & 0.036 & - & - \\ 
\cmidrule(lr){2-3}
\cmidrule(lr){4-4}
\cmidrule(lr){5-9} 
\cmidrule(lr){10-11}
&  \multirow{3}{*}{$\rho=0.2$}& \multirow{6}{*}{$1$}& MOPED  & 0.000 & 0.068 & \textbf{0.874} & 0.057 & 0.001 & 0.898 & 0.778 \\ 
 &&&MMMOPED& 0.000 & 0.004 & \textbf{0.519} & 0.393 & 0.084 & 0.873 & 0.794 \\ 
&&  & E-divisive & 0.000 & 0.000 & \textbf{0.945} & 0.020 & 0.035 & 0.990 & 0.980 \\ 
\cmidrule(lr){4-11}
&  \multirow{3}{*}{$\rho=0.6$} & & MOPED   & 0.000 & 0.000 & \textbf{1.000} & 0.000 & 0.000 & 0.992 & 0.971 \\ 
&&&MMMOPED& 0.000 & 0.000 & \textbf{0.768} & 0.204 & 0.028 & 0.948 & 0.915 \\ 
&& & E-divisive   & 0.000 & 0.000 & \textbf{0.946} & 0.018 & 0.036 & 0.992 & 0.987 \\ 
\cmidrule(lr){2-3}
\cmidrule(lr){4-4}
\cmidrule(lr){5-9} 
\cmidrule(lr){10-11}
&  \multirow{3}{*}{$\rho=0.2$}& \multirow{6}{*}{$2$}& MOPED  & 0.042 & 0.161 & \textbf{0.796} & 0.001 & 0.000 & 0.865 & 0.818 \\
 &&&MMMOPED& 0.000 & 0.007 & \textbf{0.601} & 0.299 & 0.093 & 0.875 & 0.839 \\ 
&&  & E-divisive & 0.000 & 0.000 & \textbf{0.950} & 0.025 & 0.025 & 0.990 & 0.981 \\ 
\cmidrule(lr){4-11}
&  \multirow{3}{*}{$\rho=0.6$} & & MOPED  & 0.000 & 0.000 & \textbf{1.000} & 0.000 & 0.000 & 0.986 & 0.967 \\ 
&&&MMMOPED& 0.000 & 0.000 & \textbf{0.793} & 0.175 & 0.032 & 0.951 & 0.929 \\ 
&& & E-divisive & 0.000 & 0.000 & \textbf{0.956} & 0.024 & 0.020 & 0.995 & 0.992 \\ 
\hline
 \end{tabular}
 \end{table*}

 \begin{table*}[h!]
\caption{Distribution of the estimated number of change points and the average CM and VM over 100 realisation for Scenario 2, $d=8$, of the simulation study. The modal value of $\wh{q}-q$ is provided in bold in each row.}
\label{tab:scenario2_sm1}
 \centering
  \begin{tabular}{ccccccccccc}
\cmidrule(lr){5-9} 
&&&&\multicolumn{5}{c}{$\wh{q}-q$}&&\\
$d$& $\Omega$ & q & Method & $ \leq 2$ & -1 & $\mathbf{0}$ & 1 & $\geq 2$ & CM & VM \\

\cmidrule(lr){1-3}
\cmidrule(lr){4-4}
\cmidrule(lr){5-9} 
\cmidrule(lr){10-11}
\multirow{18}{*}{$8$}  & \multirow{3}{*}{$1$} & \multirow{9}{*}{$1$}& MOPED & 0.000 & 0.056 & \textbf{0.870} & 0.073 & 0.001 & 0.909 & 0.800 \\ 
&&&MMMOPED& 0.000 & 0.008 & \textbf{0.436} & 0.421 & 0.135 & 0.847 & 0.767 \\ 
&&  & E-divisive & 0.000 & 0.000 & \textbf{0.942} & 0.025 & 0.033 & 0.974 & 0.935 \\ 
\cmidrule(lr){4-11}

&  \multirow{3}{*}{$2$} & & MOPED & 0.000 & 0.132 & \textbf{0.788} & 0.077 & 0.003 & 0.867 & 0.721 \\ 
&&&MMMOPED  & 0.000 & 0.017 & \textbf{0.456} & 0.379 & 0.148 & 0.836 & 0.746 \\ 
&&  & E-divisive& 0.000 & 0.046 & \textbf{0.899} & 0.027 & 0.028 & 0.948 & 0.879 \\ 
\cmidrule(lr){4-11}
&  \multirow{3}{*}{$3$} & & MOPED & 0.000 & 0.257 & \textbf{0.691} & 0.051 & 0.001 & 0.817 & 0.619 \\ 
&&&MMMOPED & 0.000 & 0.053 & \textbf{0.529} & 0.310 & 0.108 & 0.830 & 0.717 \\ 
&&  & E-divisive  & 0.000 & 0.023 & \textbf{0.934}& 0.018 & 0.025 & 0.959 & 0.900 \\ 

\cmidrule(lr){2-3}
\cmidrule(lr){4-4}
\cmidrule(lr){5-9} 
\cmidrule(lr){10-11}
  & \multirow{3}{*}{$1$} & \multirow{9}{*}{$2$}& MOPED & 0.012 & 0.114 & \textbf{0.872} & 0.002 & 0.000 & 0.898 & 0.857 \\ 
   &&&MMMOPED& 0.000 & 0.015 & \textbf{0.513} & 0.350 & 0.122 & 0.846 & 0.815 \\ 
&&  & E-divisive & 0.096 & 0.059 & \textbf{0.785} & 0.042 & 0.018 & 0.876 & 0.821 \\ 

\cmidrule(lr){4-11}
&  \multirow{3}{*}{$2$} & & MOPED  & 0.042 & 0.221 & \textbf{0.732} & 0.005 & 0.000 & 0.845 & 0.801 \\ 
&&&MMMOPED & 0.005 & 0.043 & \textbf{0.530} & 0.322 & 0.100 & 0.837 & 0.805 \\ 
&&  & E-divisive & \textbf{0.431} & 0.130 & 0.413 & 0.016 & 0.010 & 0.632 & 0.477 \\ 
\cmidrule(lr){4-11}
&  \multirow{3}{*}{$3$} & & MOPED & 0.144 & 0.306 & \textbf{0.546} & 0.004 & 0.000 & 0.755 & 0.687 \\ 
&&&MMMOPED & 0.023 & 0.109 & \textbf{0.600} & 0.228 & 0.039 & 0.817 & 0.778 \\ 
&&  & E-divisive & 0.363 & 0.162 & \textbf{0.450} & 0.015 & 0.010 & 0.664 & 0.531 \\

   \hline
 \end{tabular}
 \end{table*}

\begin{table*}[h!]
\caption{Distribution of the estimated number of change points and the average CM and VM over 100 realisation for Scenario 2, $d=15$, of the simulation study. The modal value of $\wh{q}-q$ is provided in bold in each row.}
\label{tab:scenario2_sm2}
 \centering
  \begin{tabular}{ccccccccccc}
\cmidrule(lr){5-9} 
&&&&\multicolumn{5}{c}{$\wh{q}-q$}&&\\
$d$& $\Omega$ & q & Method & $ \leq 2$ & -1 & $\mathbf{0}$ & 1 & $\geq 2$ & CM & VM \\
\cmidrule(lr){1-3}
\cmidrule(lr){4-4}
\cmidrule(lr){5-9} 
\cmidrule(lr){10-11}

\multirow{18}{*}{$15$}  & \multirow{3}{*}{$1$} & \multirow{9}{*}{$1$}& MOPED& 0.000 & 0.000 & \textbf{0.971} & 0.029 & 0.000 & 0.966 & 0.911 \\ 
&&&MMMOPED & 0.000 & 0.000 & 0.248 & \textbf{0.406} & 0.345 & 0.813 & 0.765 \\ 
&&  & E-divisive & 0.000 & 0.000 & \textbf{0.950} & 0.024 & 0.026 & 0.984 & 0.958 \\ 

\cmidrule(lr){4-11}
&  \multirow{3}{*}{$2$} & & MOPED & 0.000 & 0.001 & \textbf{0.966} & 0.033 & 0.000 & 0.962 & 0.903 \\ 
&&&MMMOPED & 0.000 & 0.000 & 0.363 & \textbf{0.421} & 0.216 & 0.843 & 0.785 \\ 
&&  & E-divisive& 0.000 & 0.000 & \textbf{0.954} & 0.015 & 0.031 & 0.982 & 0.955 \\ 
\cmidrule(lr){4-11}
&  \multirow{3}{*}{$3$} & & MOPED & 0.000 & 0.000 & \textbf{0.959} & 0.040 & 0.001 & 0.961 & 0.901 \\ 
&&&MMMOPED & 0.000 & 0.000 & 0.343 & \textbf{0.416} & 0.241 & 0.844 & 0.788 \\ 
&&  & E-divisive & 0.000 & 0.000 & \textbf{0.958} & 0.021 & 0.021 & 0.984 & 0.956 \\ 

\cmidrule(lr){2-3}
\cmidrule(lr){4-4}
\cmidrule(lr){5-9} 
\cmidrule(lr){10-11}
  & \multirow{3}{*}{$1$} & \multirow{9}{*}{$2$}& MOPED& 0.000 & 0.001 & \textbf{0.999} & 0.000 & 0.000 & 0.961 & 0.927 \\
&&&MMMOPED& 0.000 & 0.000 & 0.261 & \textbf{0.406} & 0.333 & 0.851 & 0.837 \\ 
&&  & E-divisive  & 0.000 & 0.000 &\textbf{0.936} & 0.039 & 0.025 & 0.977 & 0.957 \\ 

\cmidrule(lr){4-11}
&  \multirow{3}{*}{$2$} & & MOPED & 0.000 & 0.001 &\textbf{0.999} & 0.000 & 0.000 & 0.959 & 0.924 \\ 
&&&MMMOPED & 0.000 & 0.000 & 0.310 & \textbf{0.386} & 0.303 & 0.853 & 0.838 \\ 
&&  & E-divisive  & 0.000 & 0.000 & \textbf{0.950} & 0.024 & 0.026 & 0.973 & 0.951 \\ 
\cmidrule(lr){4-11}
&  \multirow{3}{*}{$3$} & & MOPED & 0.000 & 0.004 & \textbf{0.996} & 0.000 & 0.000 & 0.956 & 0.919 \\ 
&&&MMMOPED & 0.000 & 0.000 & 0.276 & \textbf{0.471} & 0.253 & 0.854 & 0.838 \\
&&  & E-divisive& 0.001 & 0.000 & \textbf{0.957} & 0.025 & 0.017 & 0.973 & 0.951 \\ 
   \hline
 \end{tabular}
 \end{table*}

\clearpage
\renewcommand{\thefigure}{B\arabic{figure}}

\setcounter{figure}{0}

\section{Supplementary Figures}
\label{sec:sup_figs}
\begin{figure}[h!]
    \centering
            \includegraphics[width = 0.49\linewidth]{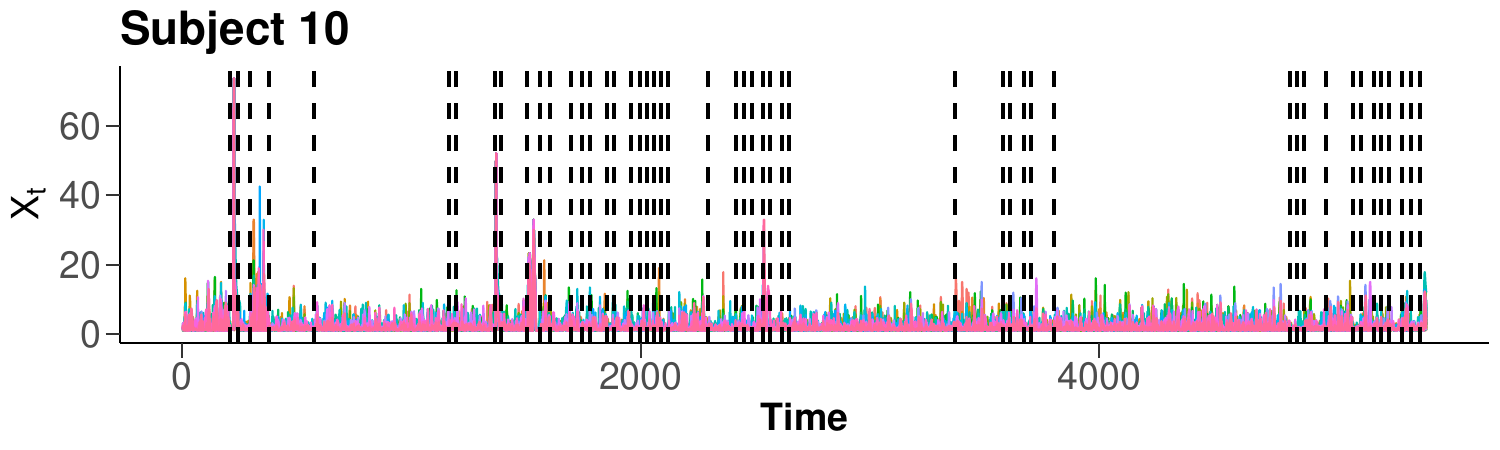}   \includegraphics[width = 0.49\linewidth]{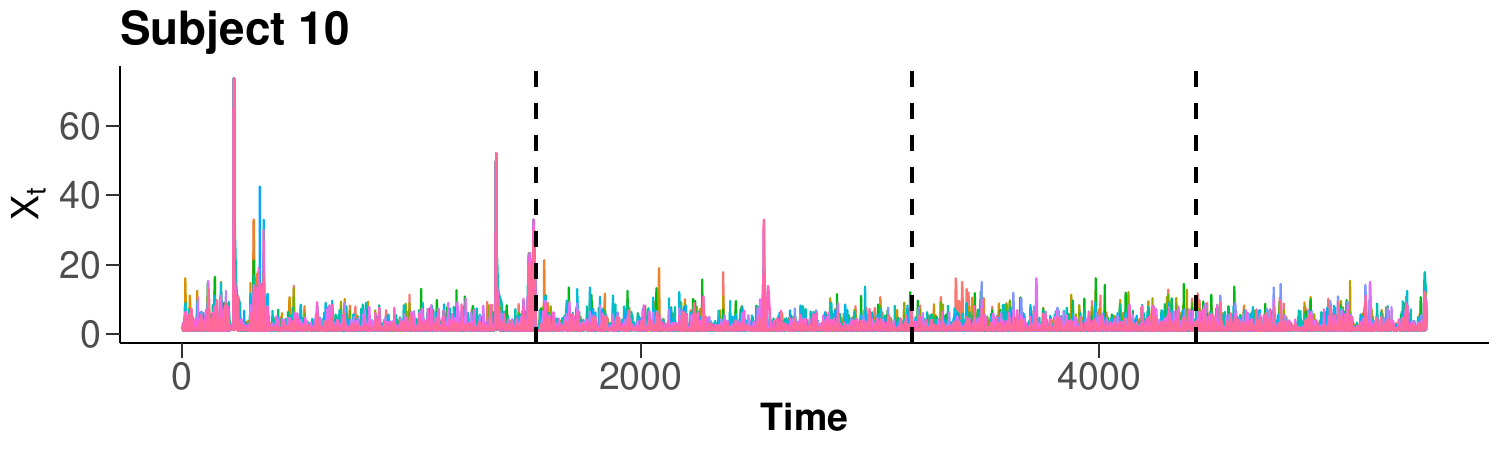}
    \includegraphics[width = 0.49\linewidth]{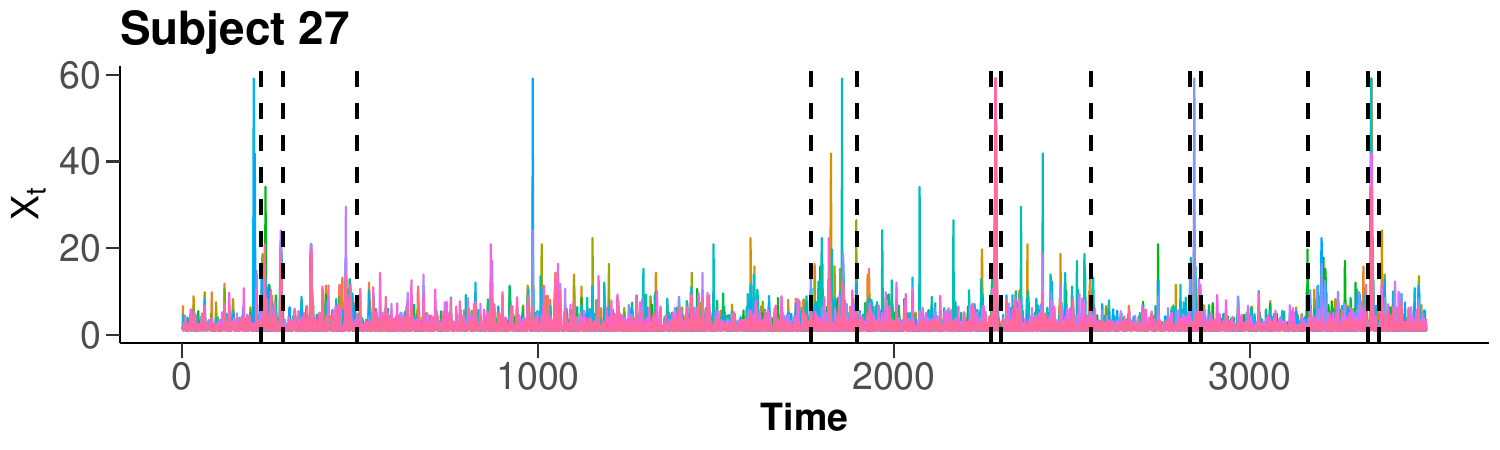}   \includegraphics[width = 0.49\linewidth]{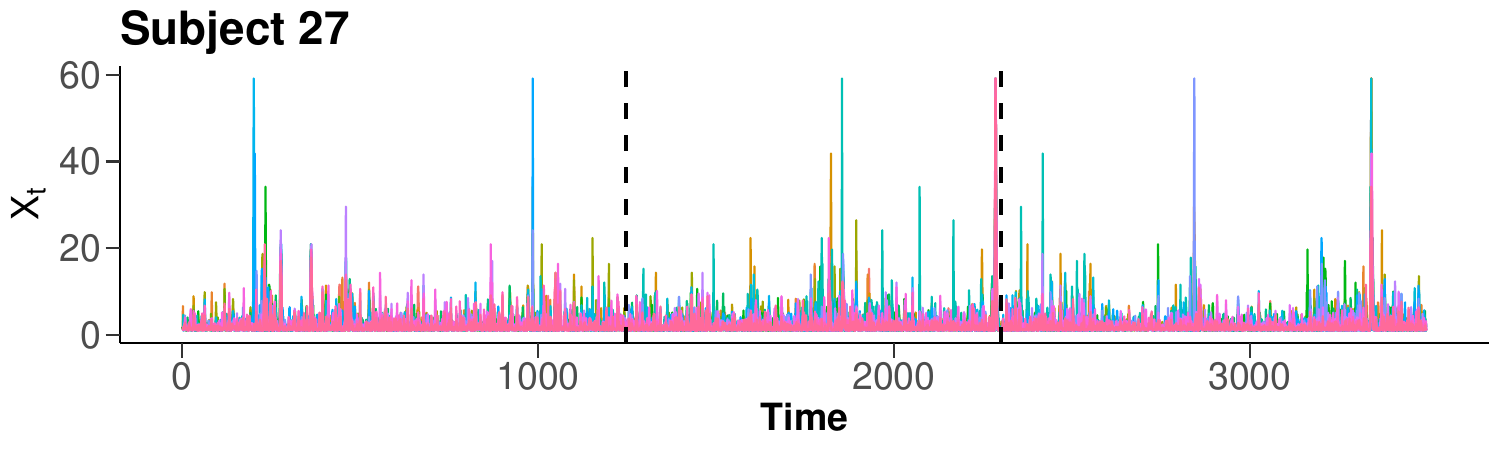}
        \includegraphics[width = 0.49\linewidth]{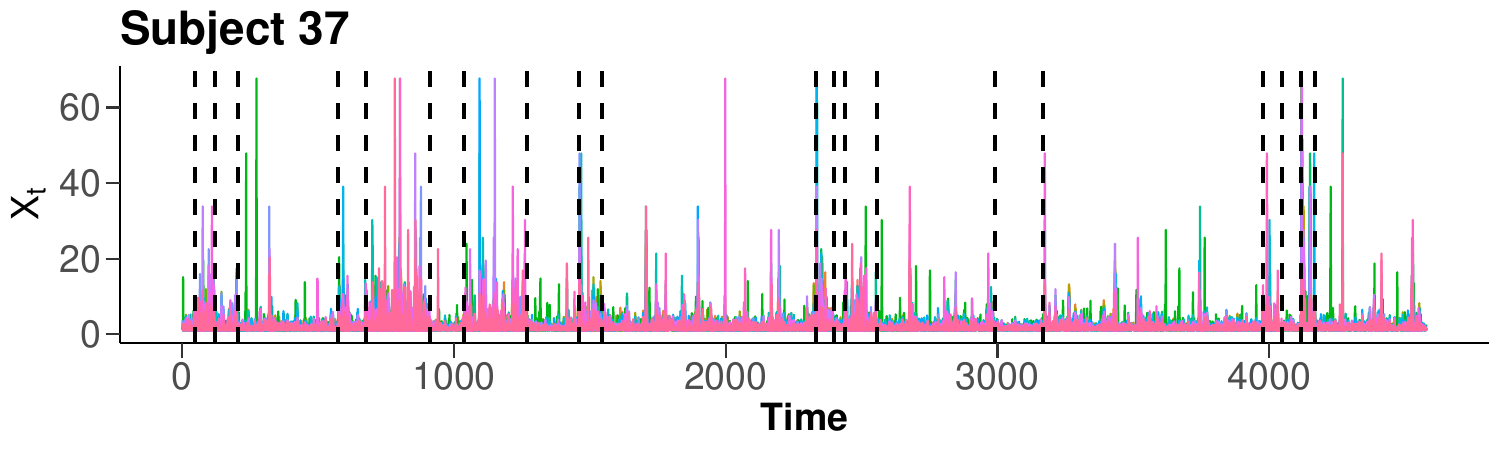}   \includegraphics[width = 0.49\linewidth]{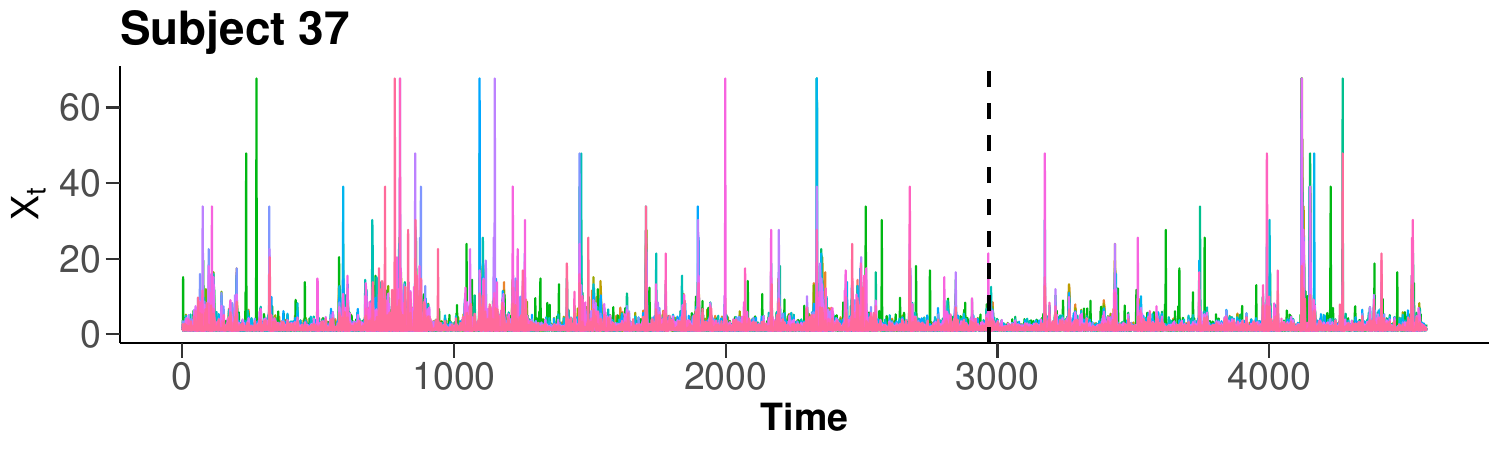}
    \caption{Change point estimates for Subjects 10, 27, and 37 (top to bottom row) of the case study. Estimates are denoted by the vertical dotted lines; the left and right columns provide estimates from E-divisive and MOPED, respectively. Observations of $\{X_{i,t}\}^n_{t=1}$ are plotted against time $t$, with the colour corresponding to the $i$-th channel, $i=1,\dots,d$. Here, $d=19$ and no subjects were identified as having experienced a seizure.}
    \label{fig:case1_results_d19}
\end{figure}

\begin{landscape}
 \begin{figure}
 \centering

    \includegraphics[width = 0.32\linewidth]{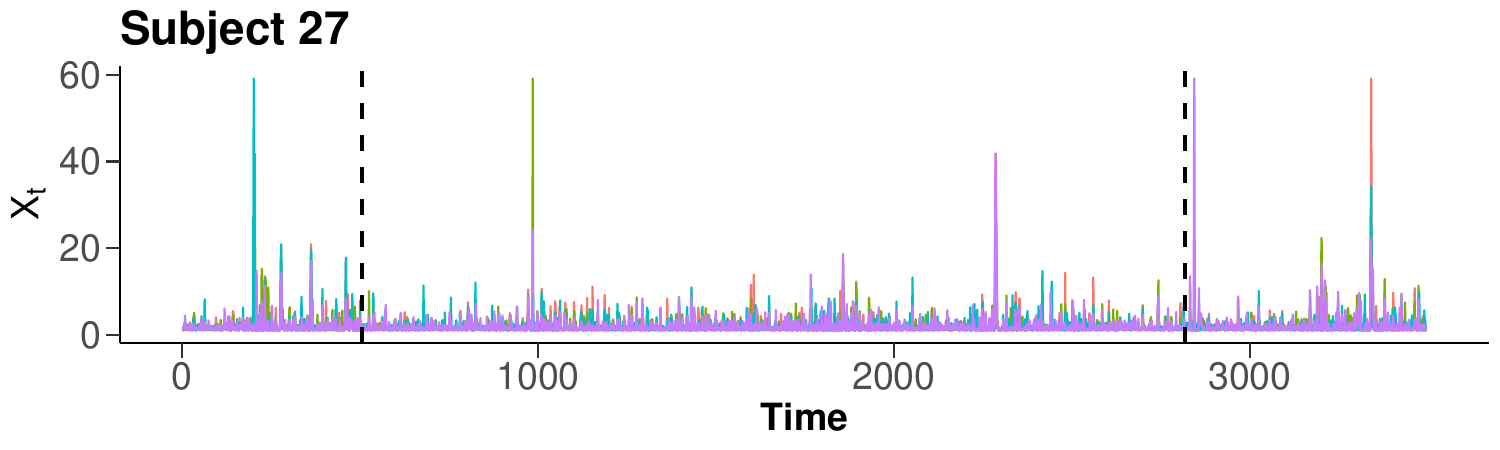}
  \includegraphics[width = 0.32\linewidth]{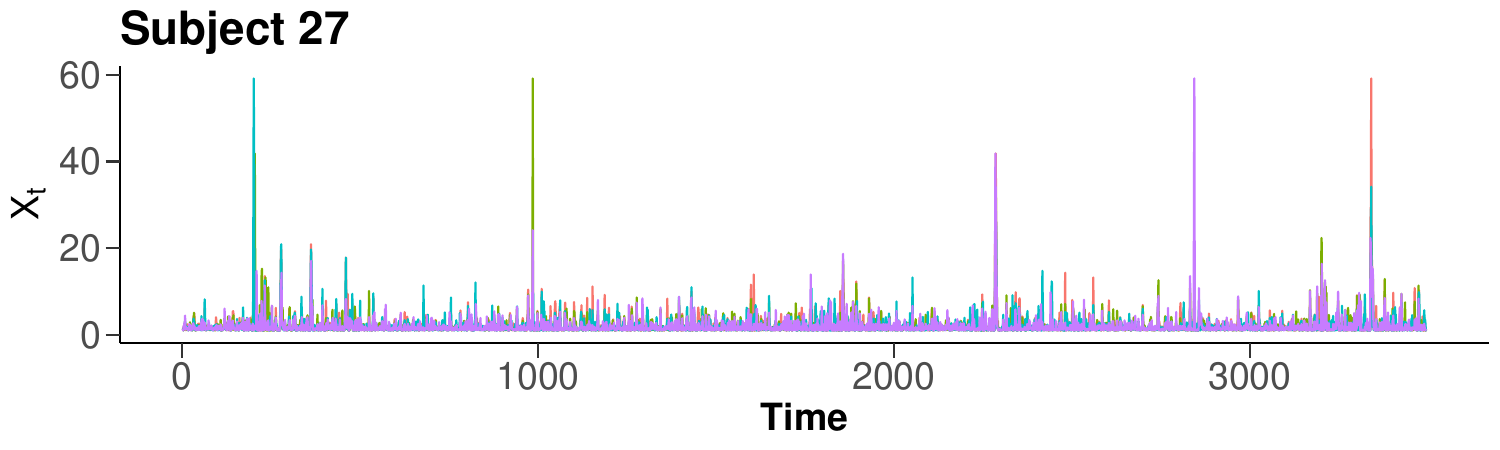}
   \includegraphics[width = 0.32\linewidth]{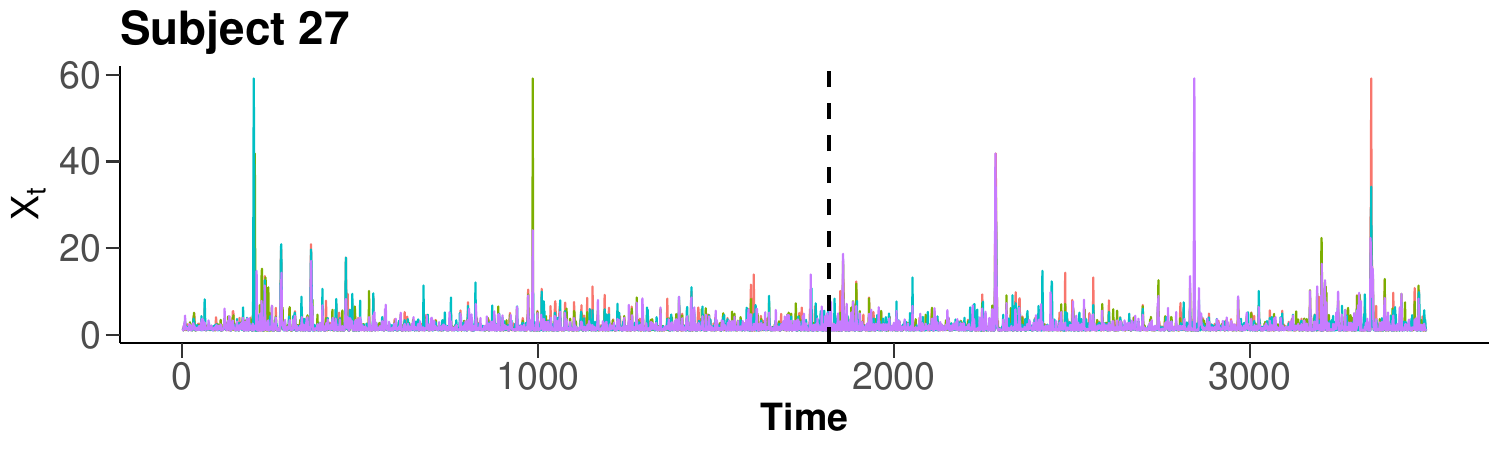}
         \includegraphics[width = 0.32\linewidth]{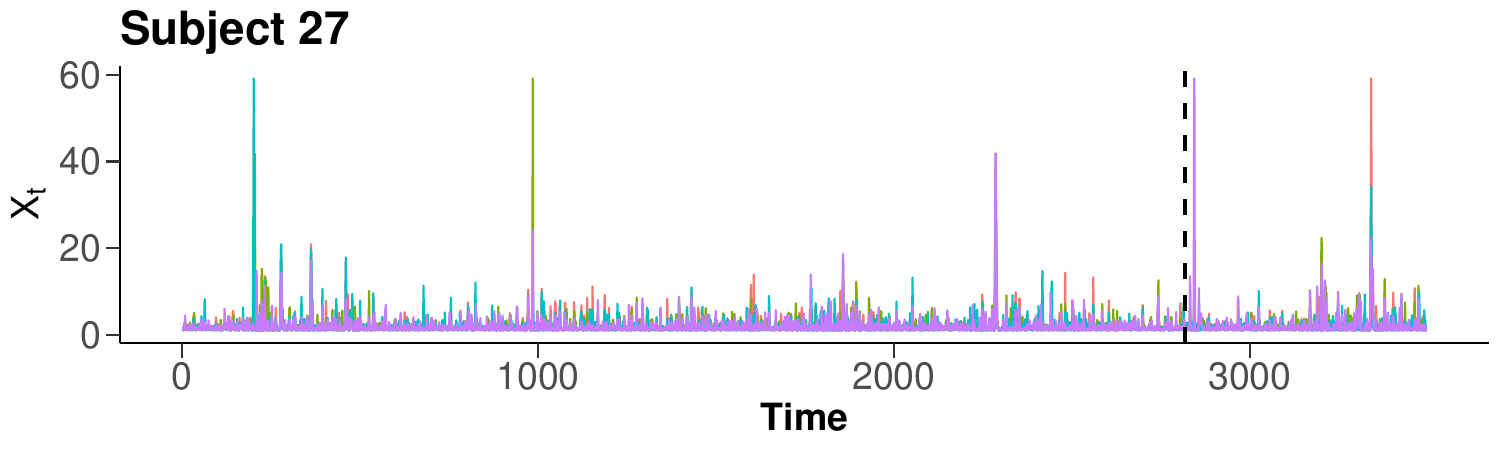}
  \includegraphics[width = 0.32\linewidth]{img/estimates_id27_EDMOSUM_G1000_q0.95_d4.pdf}
   \includegraphics[width = 0.32\linewidth]{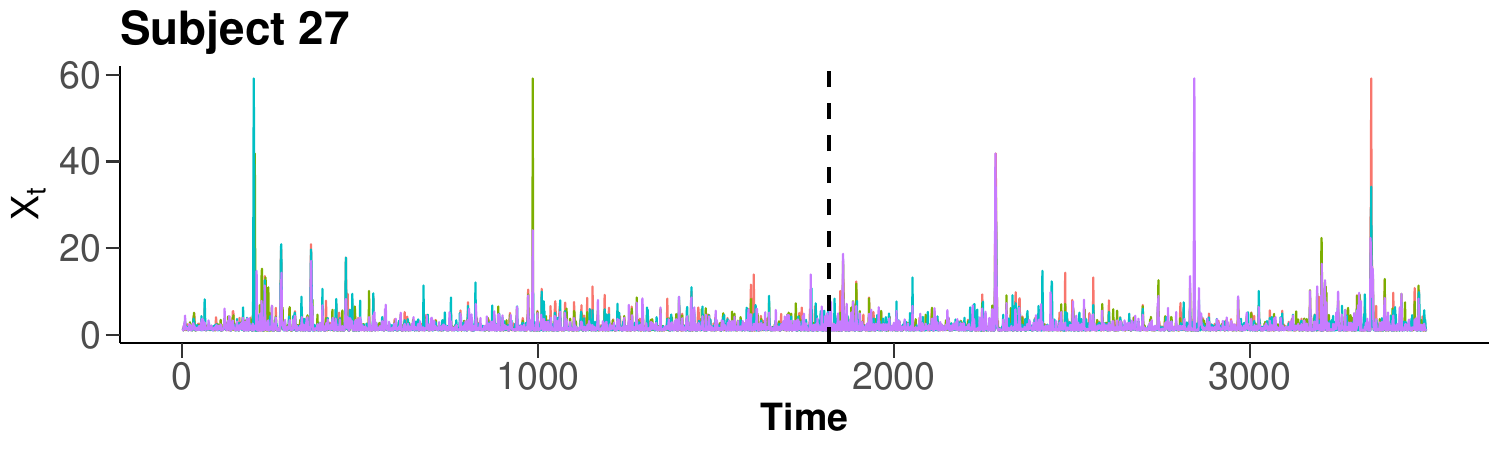}
    \includegraphics[width = 0.32\linewidth]{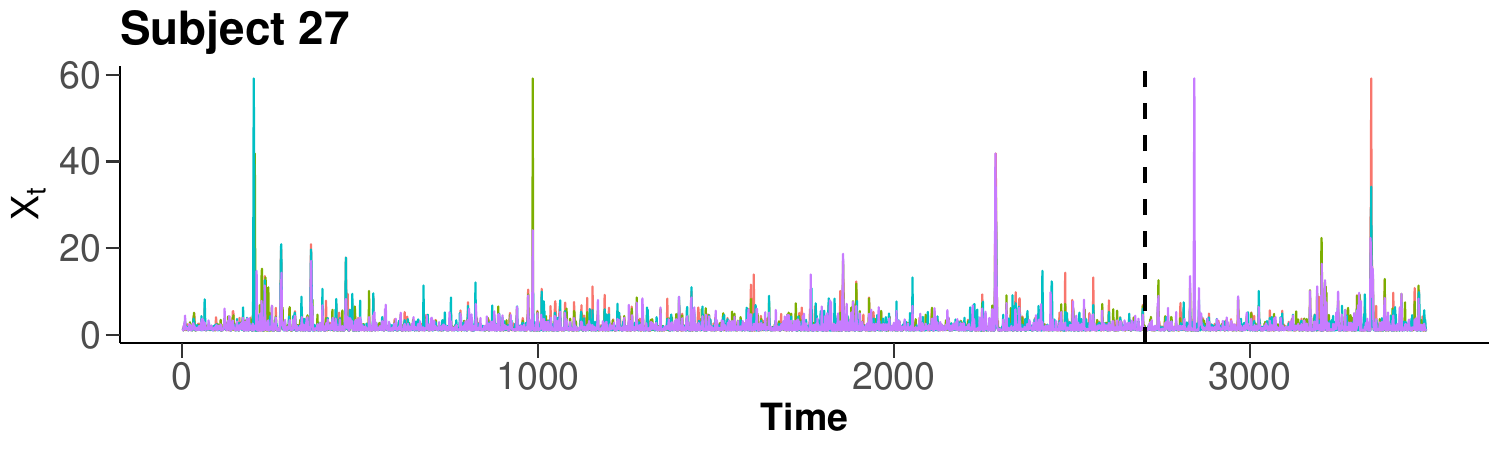}
  \includegraphics[width = 0.32\linewidth]{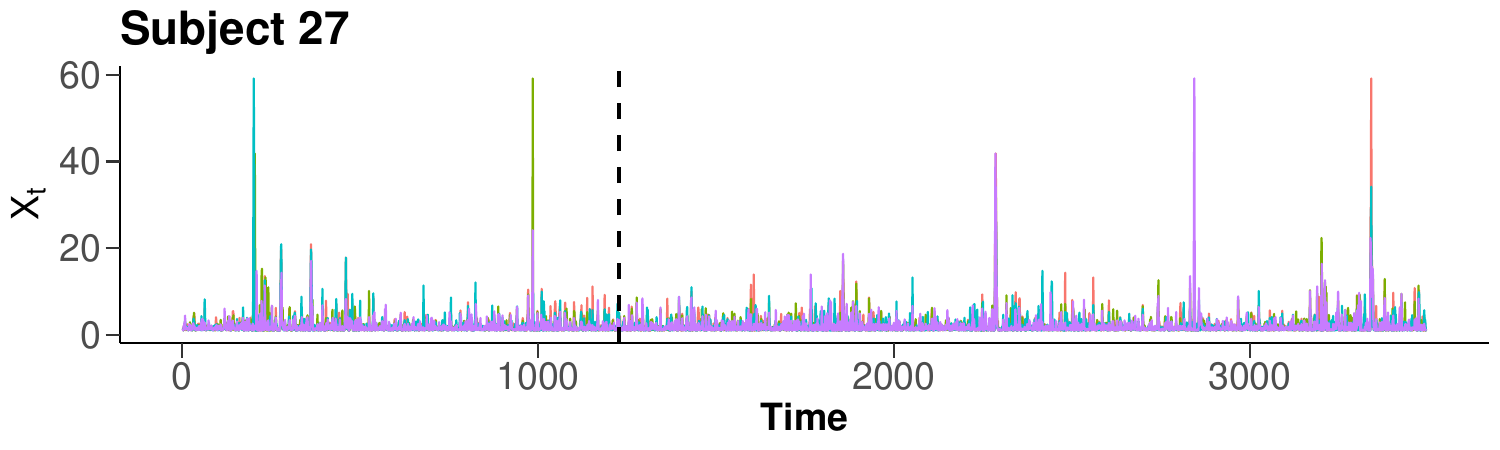}
   \includegraphics[width = 0.32\linewidth]{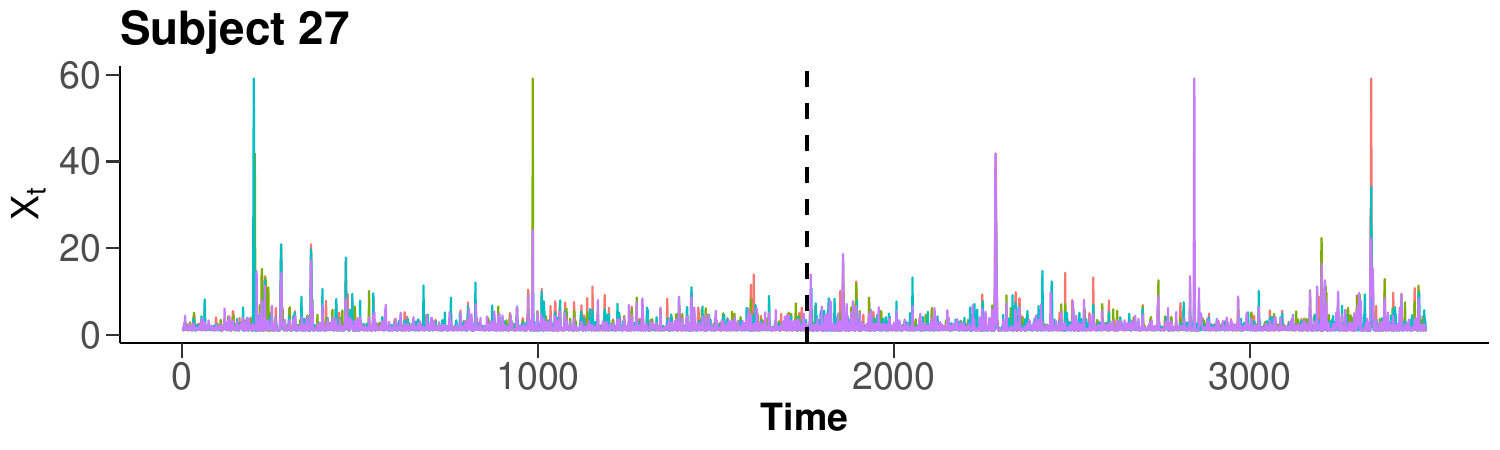}
    \caption{Change point estimates for Subject 27, using MOPED, are denoted by the vertical dotted lines. The hyper-parameters differ across panels, with $G=500$, $G=1000$, and $G=1500$ for the left, middle, and right columns, respectively, and $k$ set to $0.1G$, $0.05G$, and $0.025G$ for the top, central, and bottom rows, respectively. Observations of $\{X_{i,t}\}^n_{t=1}$ are plotted against time $t$, with the colour corresponding to the $i$-th channel, $i=1,\dots,4$.}
    \label{fig:case1_results_sup}
 \end{figure}
\end{landscape}

\begin{landscape}
 \begin{figure}
 \centering

    \includegraphics[width = 0.32\linewidth]{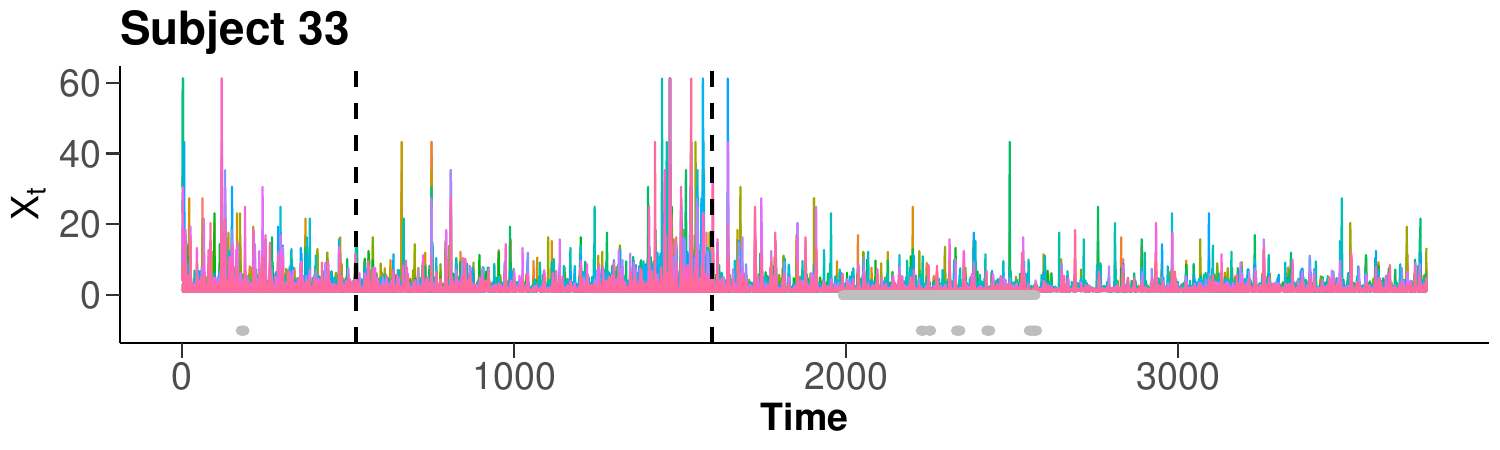}
  \includegraphics[width = 0.32\linewidth]{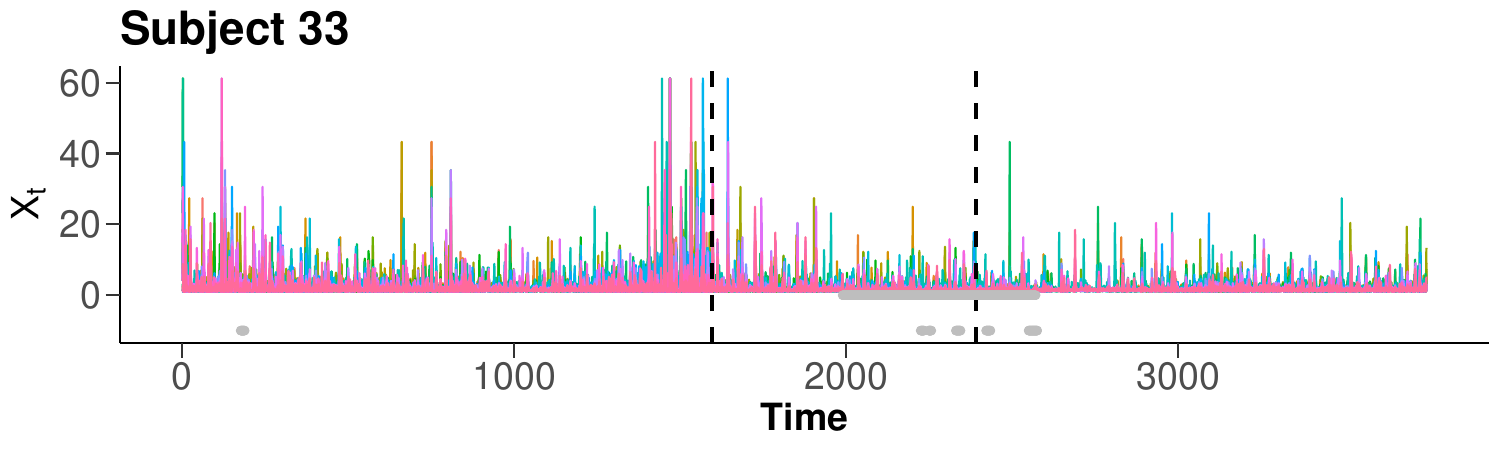}
   \includegraphics[width = 0.32\linewidth]{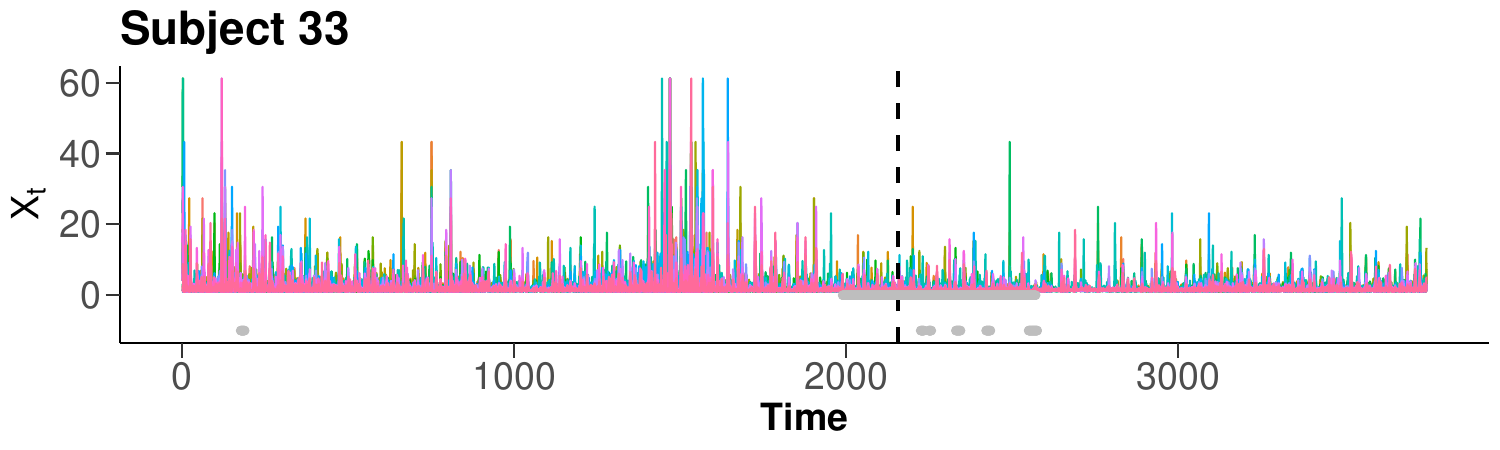}
         \includegraphics[width = 0.32\linewidth]{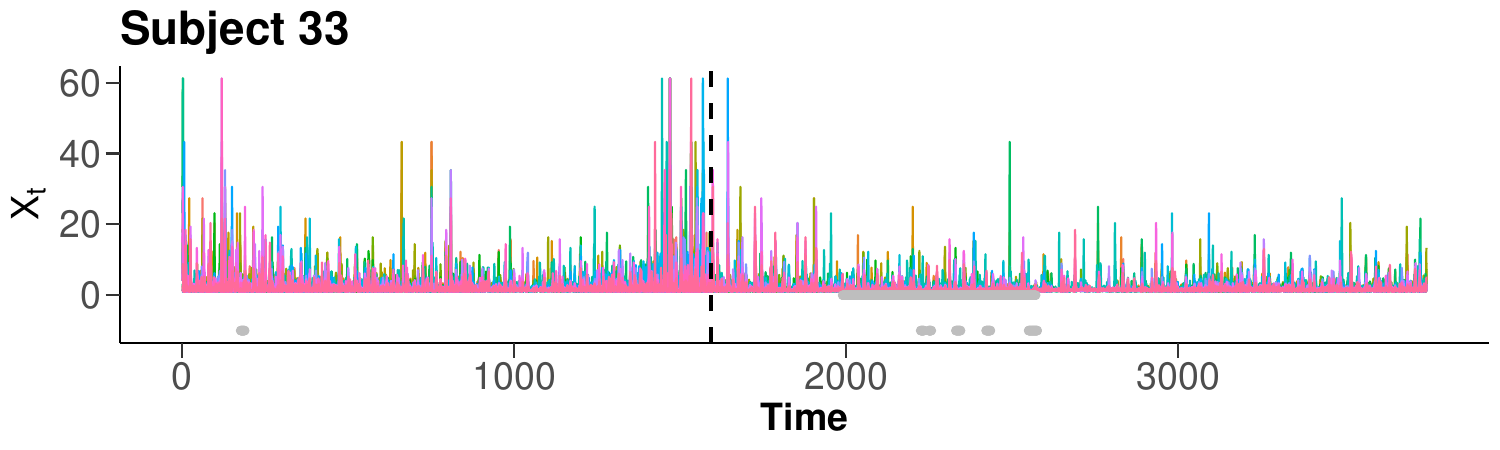}
  \includegraphics[width = 0.32\linewidth]{img/estimates_id33_EDMOSUM_G1000_q0.95_d19.pdf}
   \includegraphics[width = 0.32\linewidth]{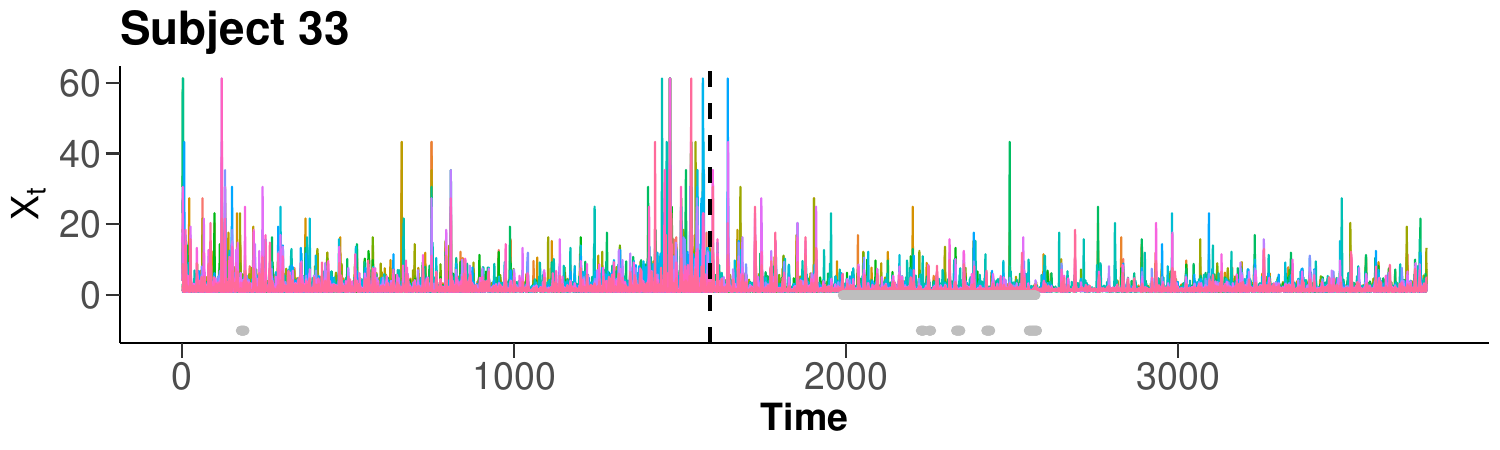}
    \includegraphics[width = 0.32\linewidth]{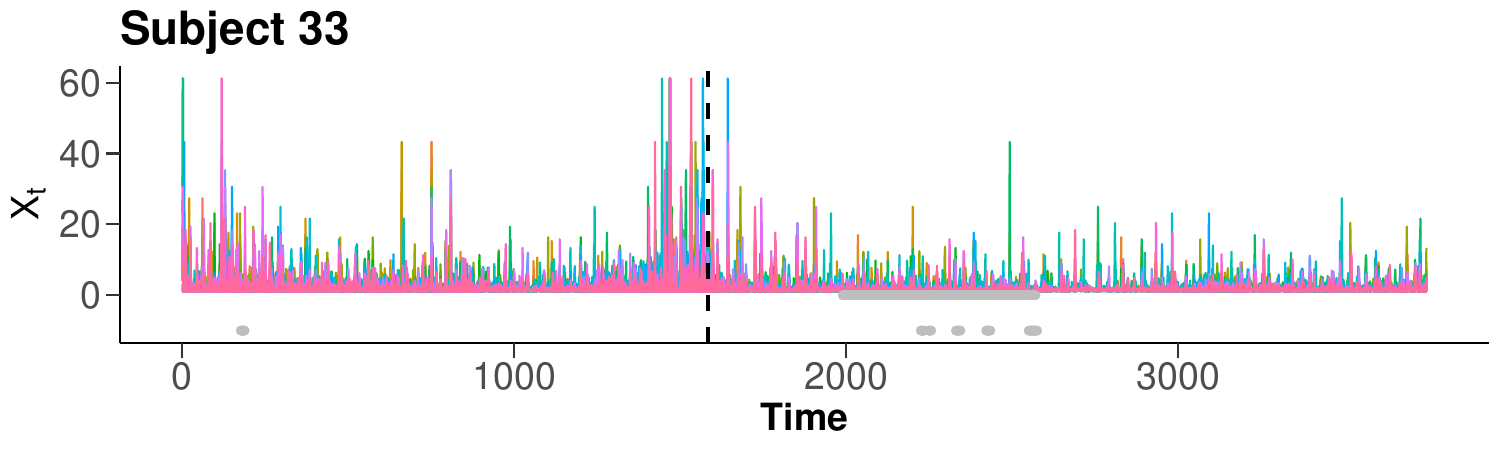}
  \includegraphics[width = 0.32\linewidth]{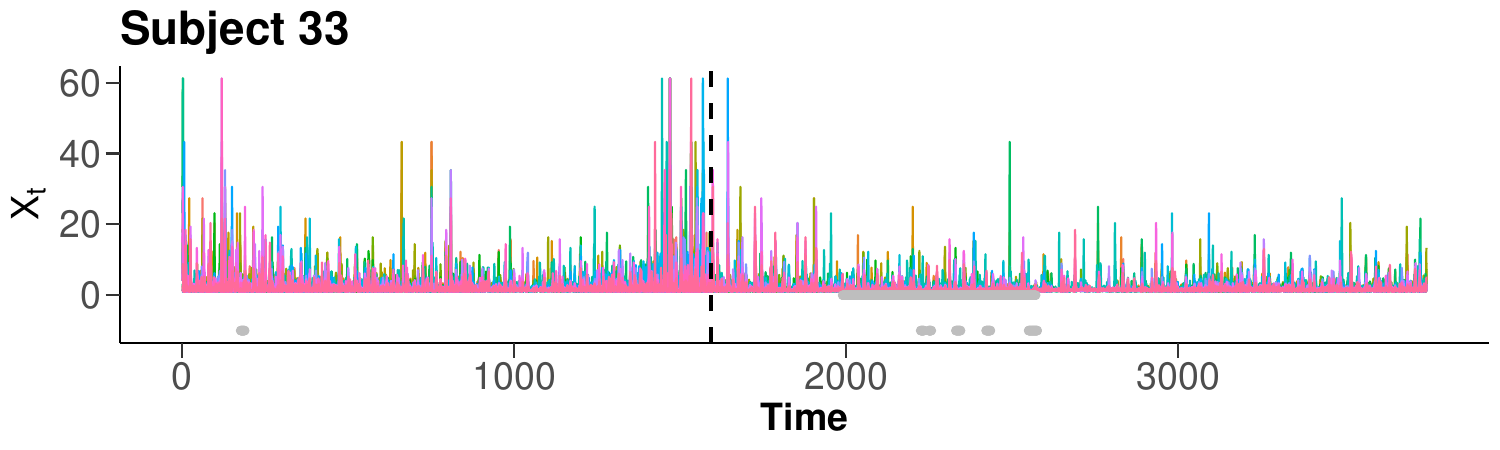}
   \includegraphics[width = 0.32\linewidth]{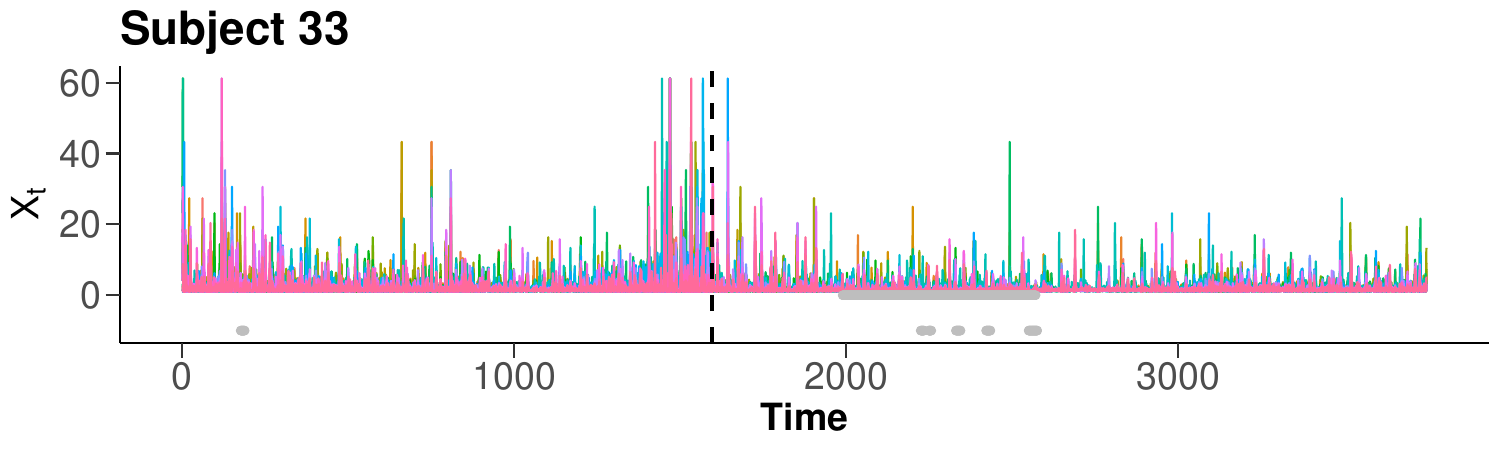}
    \caption{Change point estimates for Subject 33, using MOPED, are denoted by the vertical dotted lines. The hyper-parameters differ across panels, with $G=500$, $G=1000$, and $G=1500$ for the left, middle, and right columns, respectively, and $k$ set to $0.1G$, $0.05G$, and $0.025G$ for the top, central, and bottom rows, respectively. Observations of $\{X_{i,t}\}^n_{t=1}$ are plotted against time $t$, with the colour corresponding to the $i$-th channel, $i=1,\dots,19$.}
    \label{fig:case2_results_sup}
 \end{figure}
\end{landscape}

\begin{landscape}
 \begin{figure}
 \centering
 \includegraphics[width = 0.32\linewidth]{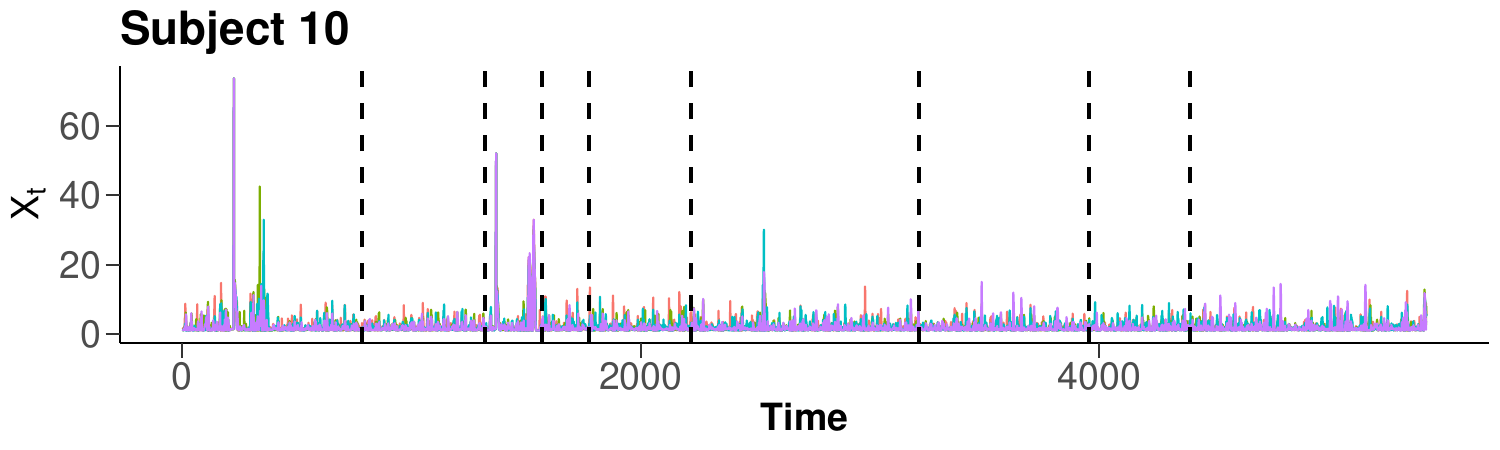}
   \includegraphics[width = 0.32\linewidth]{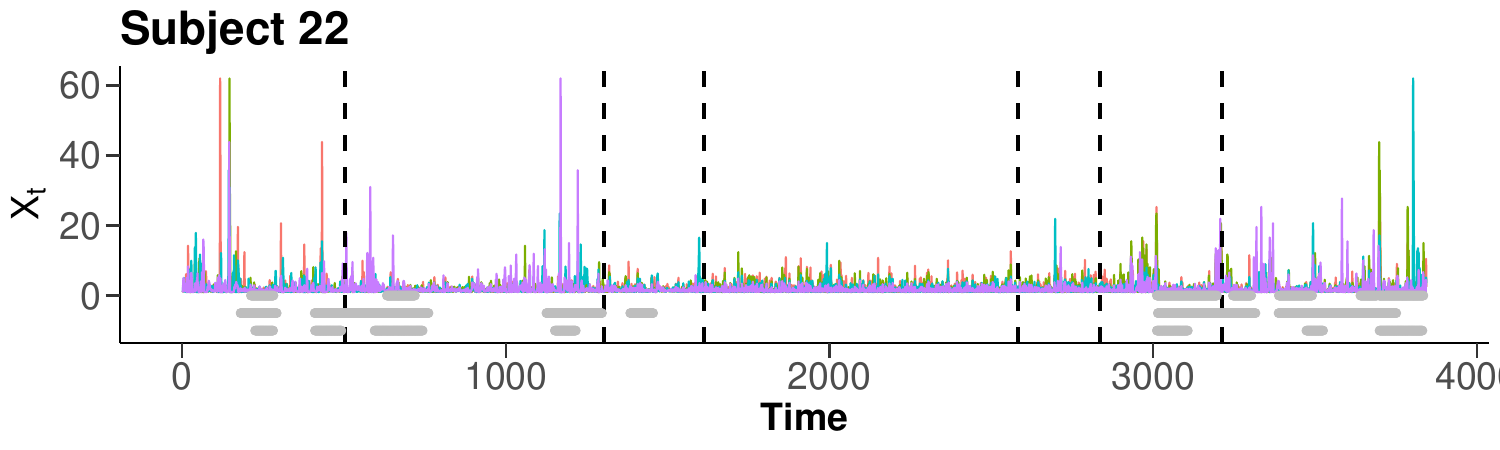}
  \includegraphics[width = 0.32\linewidth]{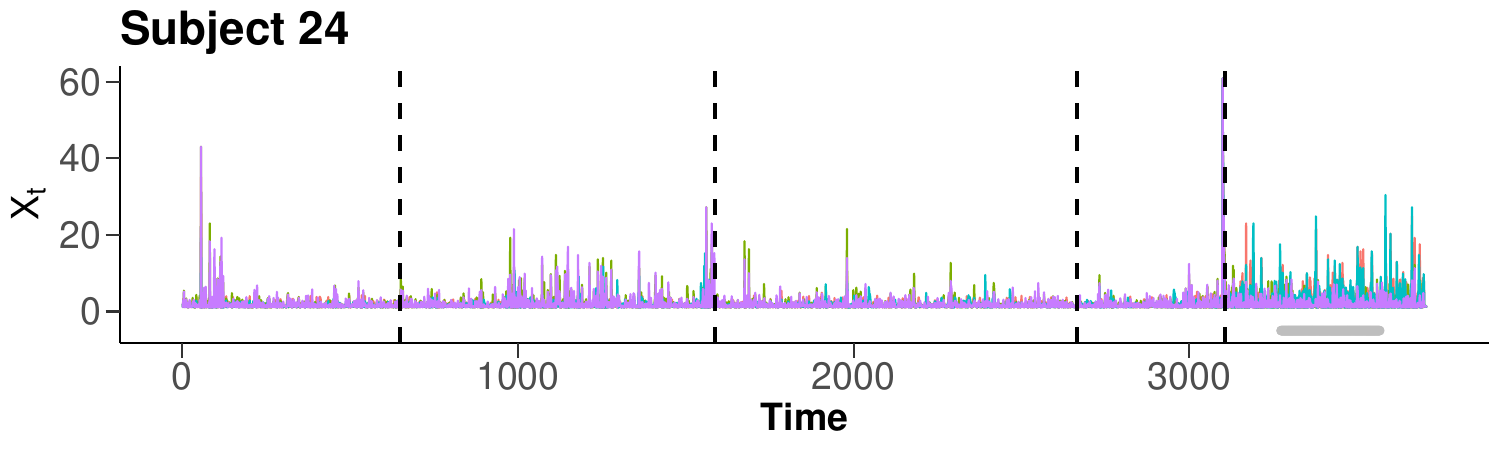}
 \includegraphics[width = 0.32\linewidth]{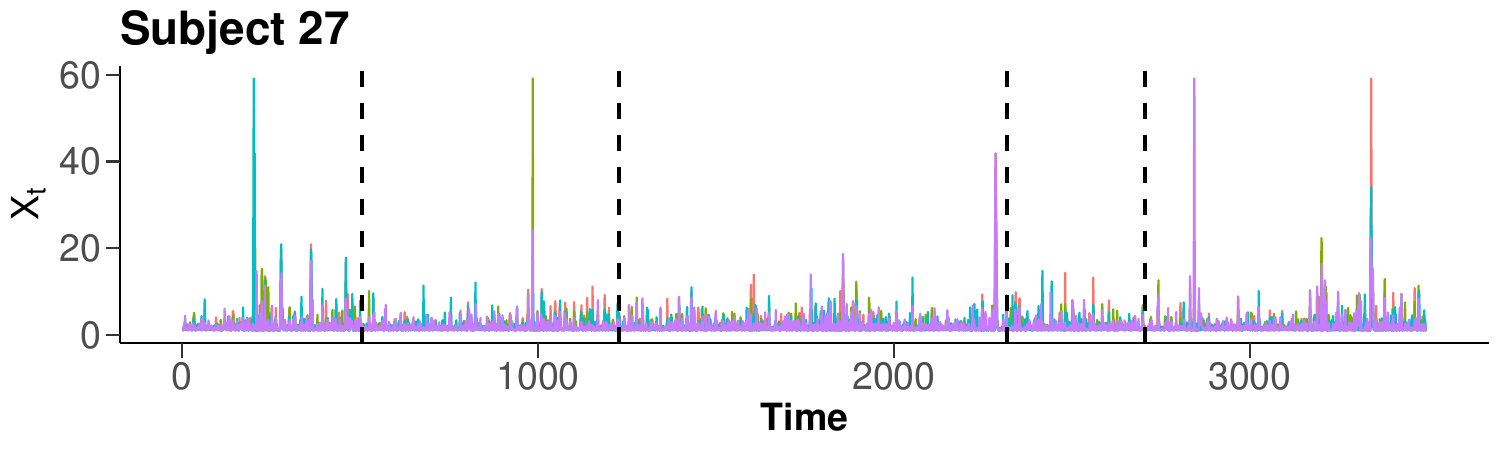}
  \includegraphics[width = 0.32\linewidth]{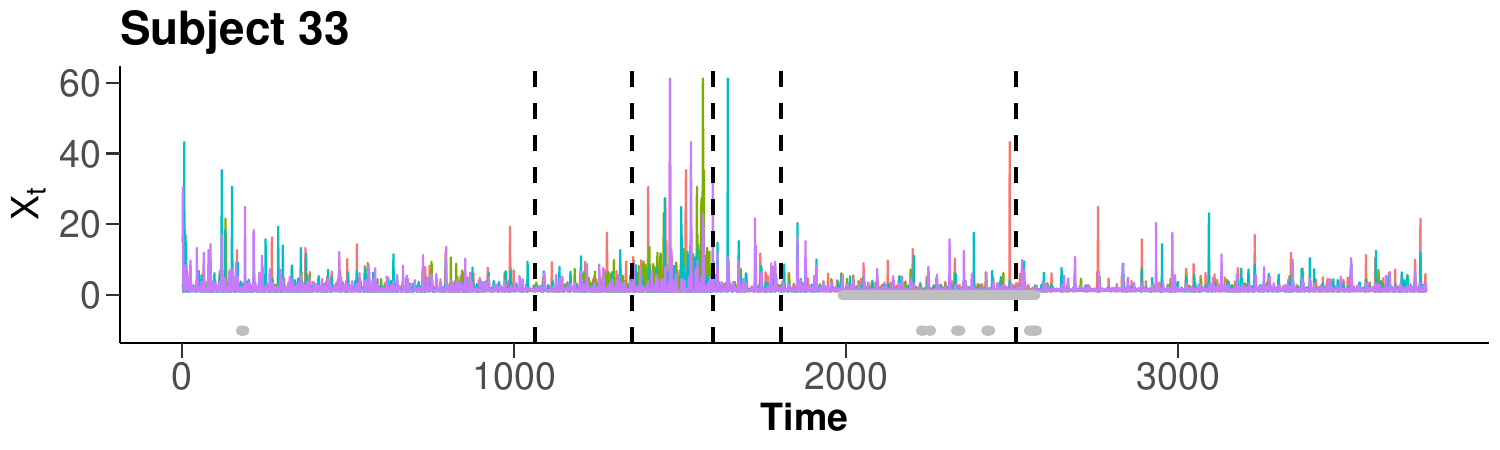}
   \includegraphics[width = 0.32\linewidth]{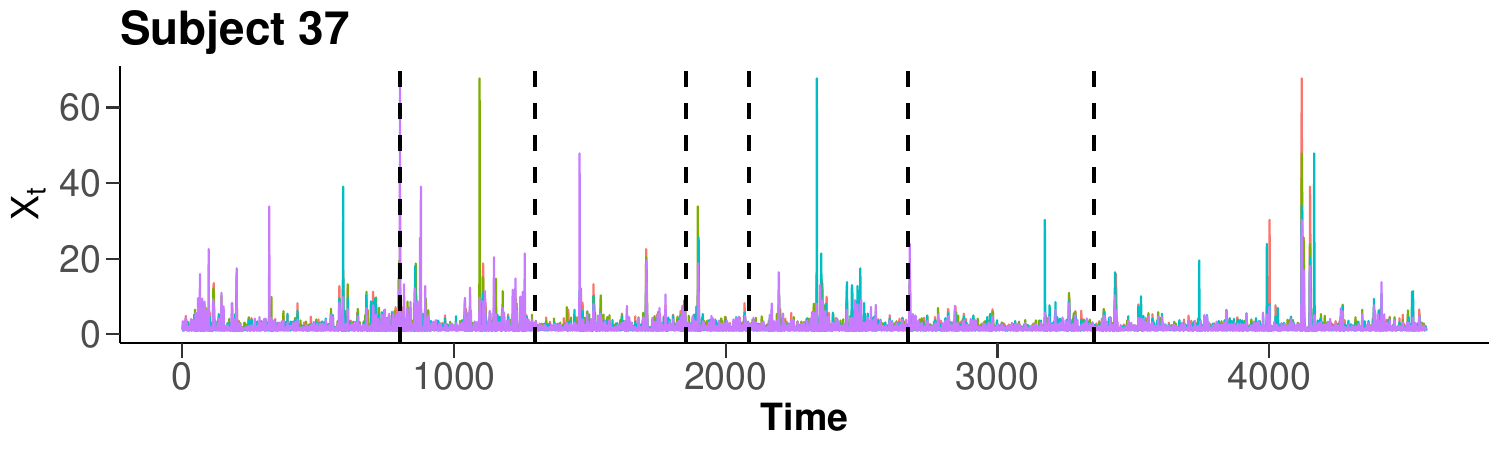}
     \includegraphics[width = 0.32\linewidth]{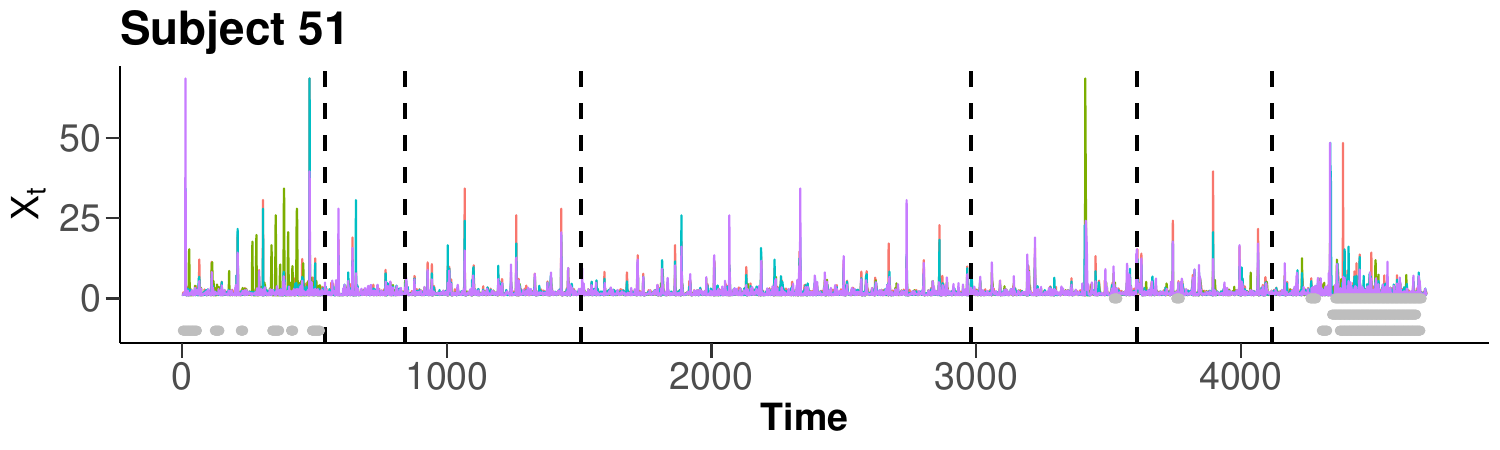}
     \includegraphics[width = 0.32\linewidth]{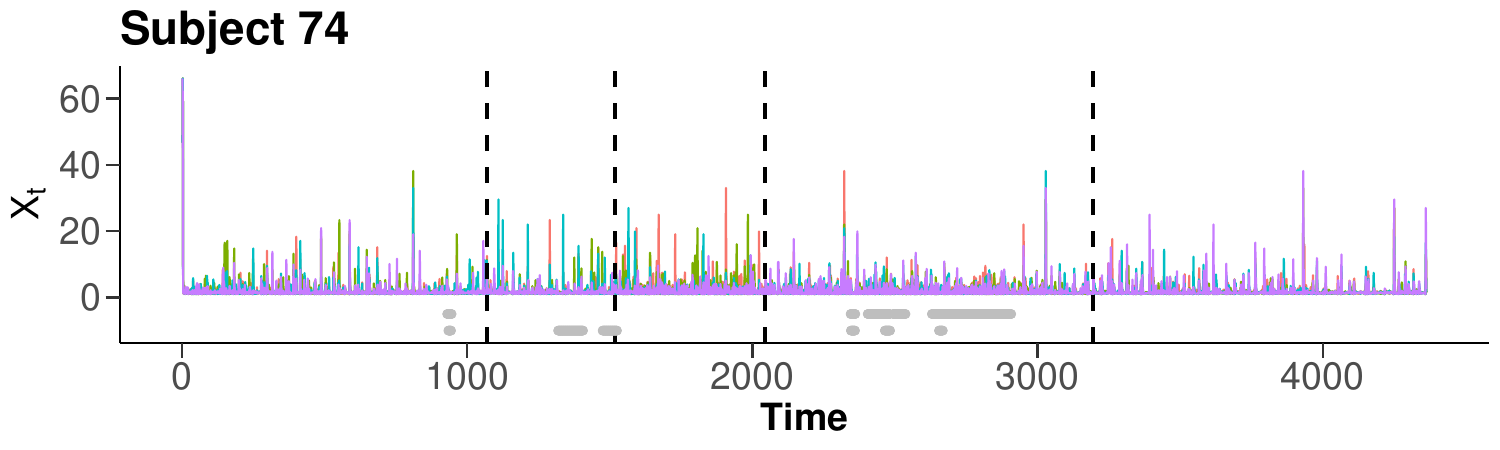}
    \includegraphics[width = 0.32\linewidth]{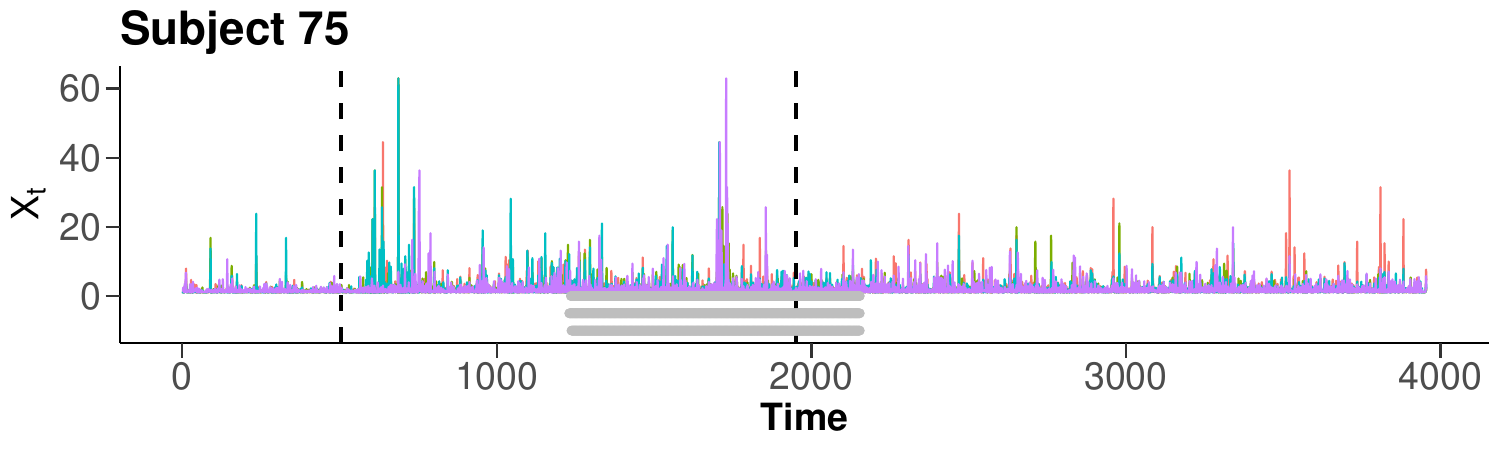}
    \caption{Change point estimates, using MMMOPED, are denoted by the vertical dotted lines. Observations of $\{X_{i,t}\}^n_{t=1}$ are plotted against time $t$, with the colour corresponding to the $i$-th channel, $i=1,\dots,d$, for $d=4$.}
    \label{fig:MMMOPED_results_sup1}
 \end{figure}
\end{landscape}

\begin{landscape}
 \begin{figure}
 \centering
 \includegraphics[width = 0.32\linewidth]{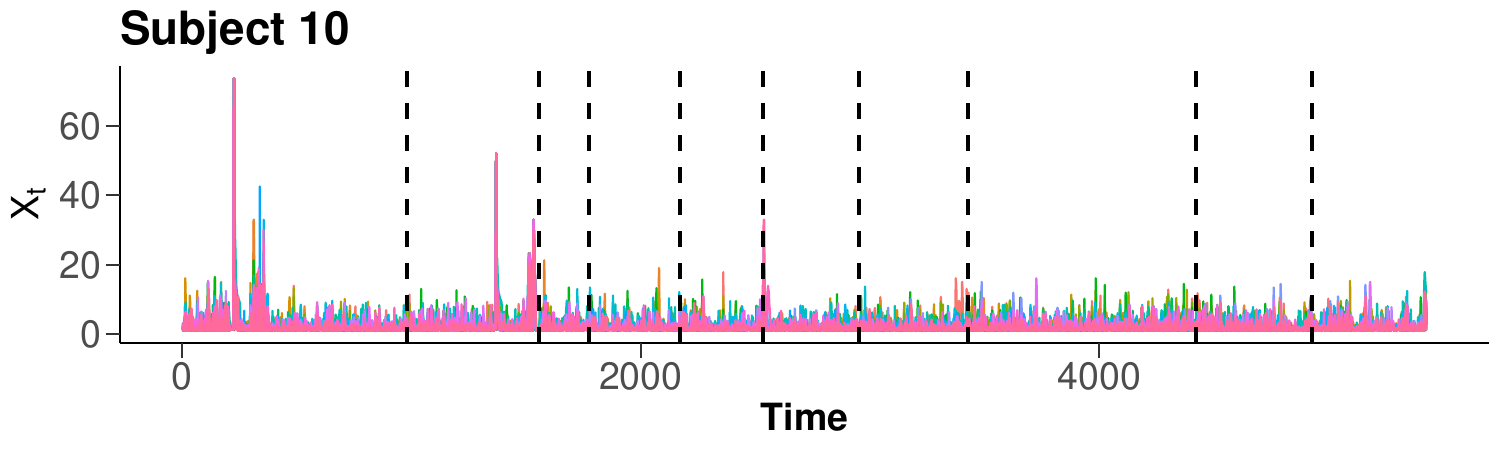}
   \includegraphics[width = 0.32\linewidth]{img/estimates_id22_EDMOSUM_multiscale_d19.pdf}
  \includegraphics[width = 0.32\linewidth]{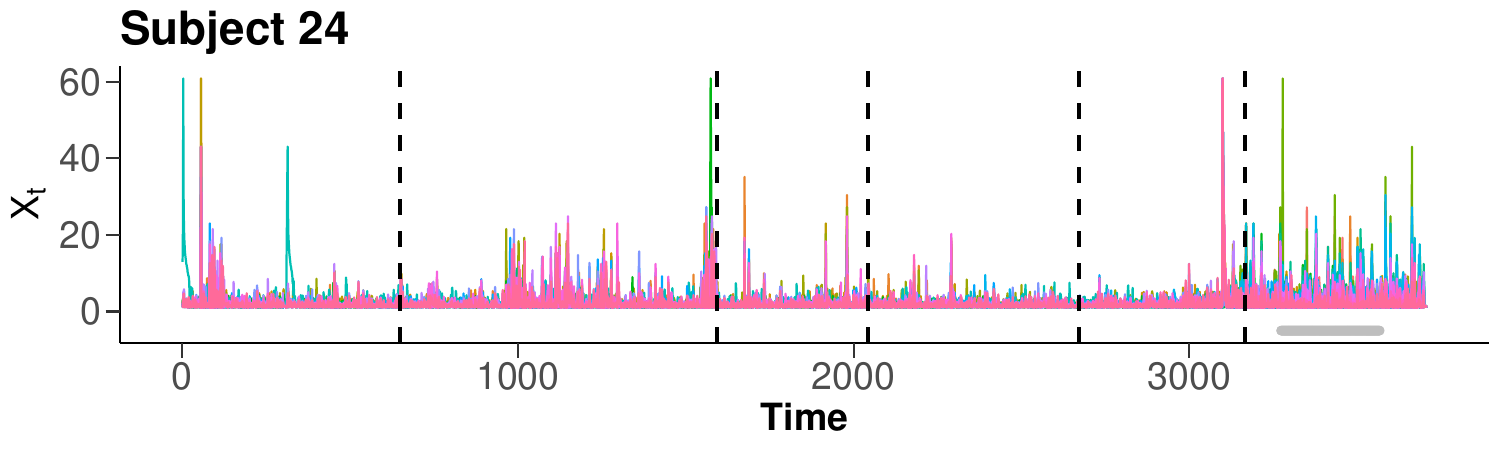}
 \includegraphics[width = 0.32\linewidth]{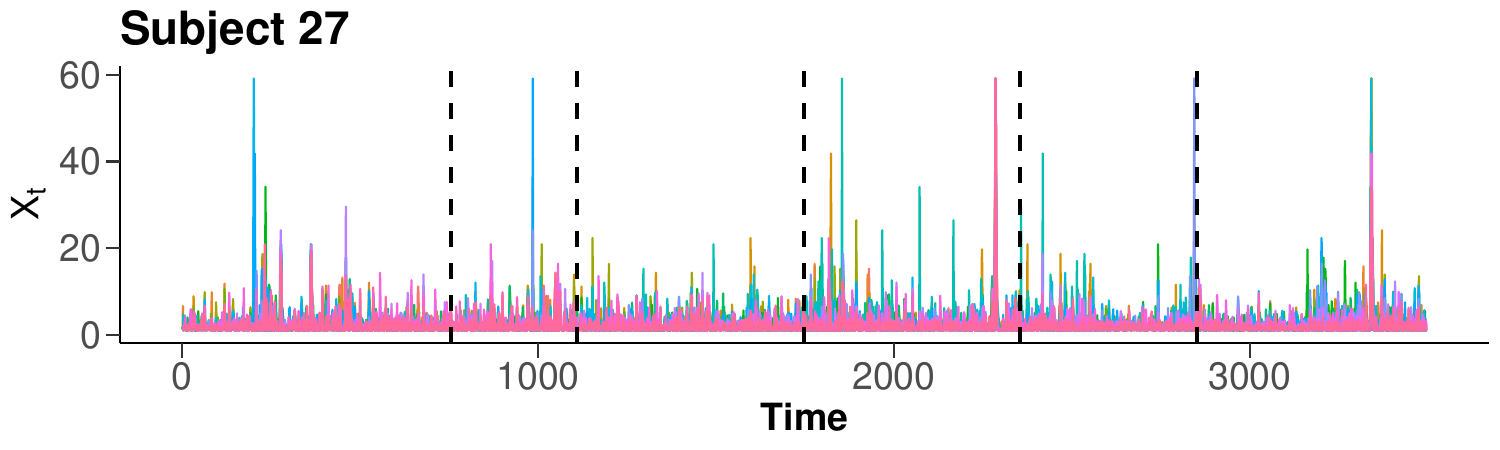}
  \includegraphics[width = 0.32\linewidth]{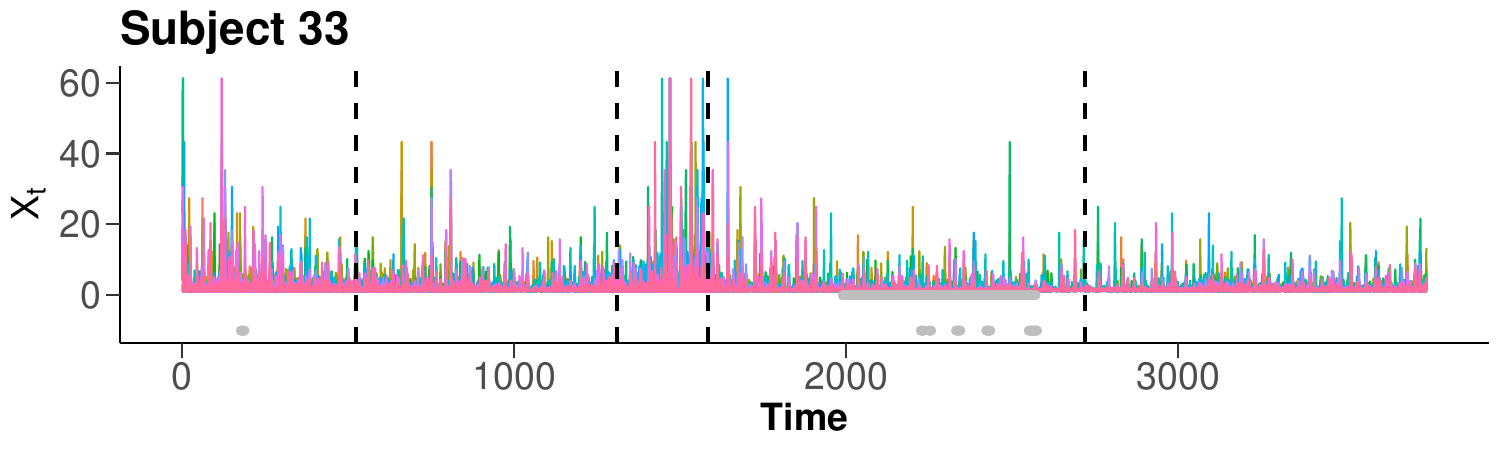}
   \includegraphics[width = 0.32\linewidth]{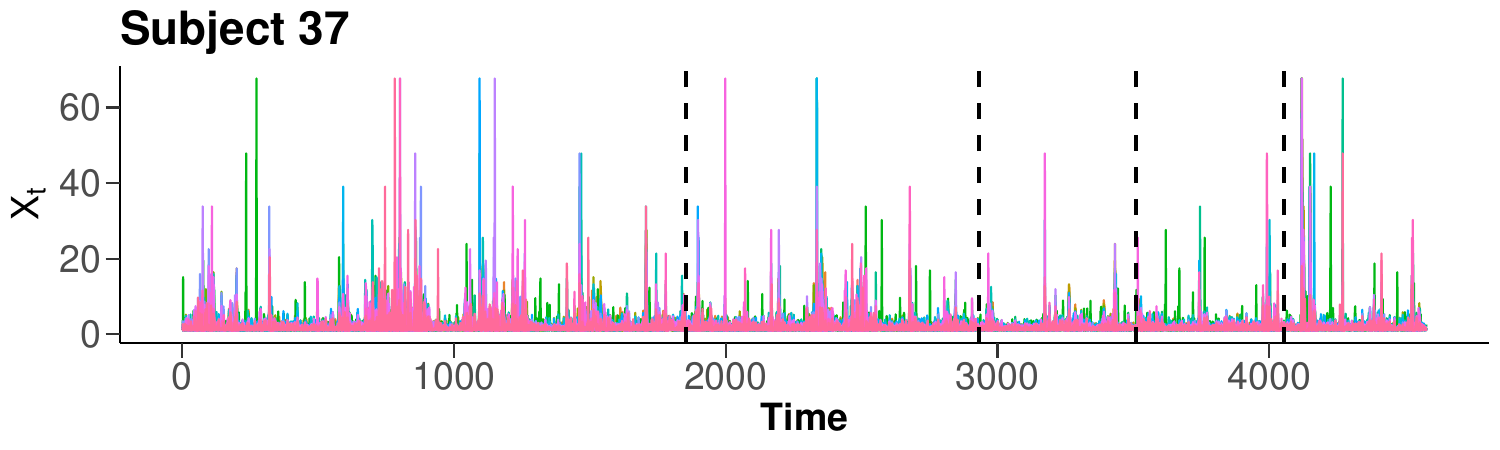}
     \includegraphics[width = 0.32\linewidth]{img/estimates_id51_EDMOSUM_multiscale_d19.pdf}
       \includegraphics[width = 0.32\linewidth]{img/estimates_id74_EDMOSUM_multiscale_d19.pdf}
    \includegraphics[width = 0.32\linewidth]{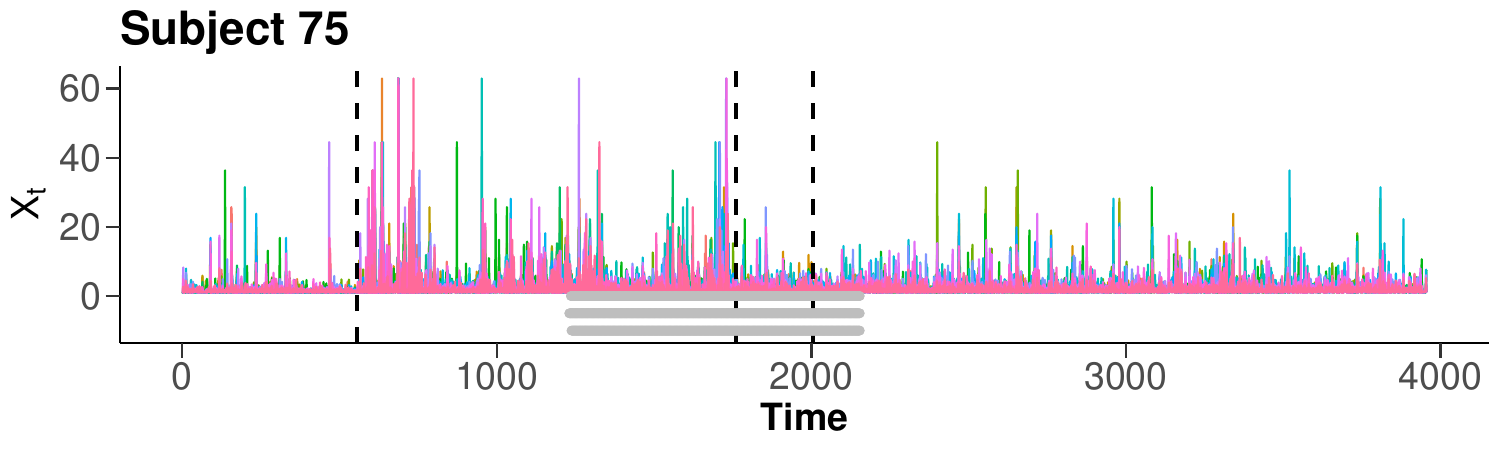}

    \caption{Change point estimates, using MMMOPED, are denoted by the vertical dotted lines. Observations of $\{X_{i,t}\}^n_{t=1}$ are plotted against time $t$, with the colour corresponding to the $i$-th channel, $i=1,\dots,d$, for $d=19$.}
    \label{fig:MMMOPED_results_sup2}
 \end{figure}
\end{landscape}
\end{appendix}

\baselineskip=14pt
\begingroup

\bibliographystyle{apalike}
\bibliography{ref}

\endgroup
\clearpage

\end{document}